\documentclass[aps,prl,superscriptaddress,twocolumn]{revtex4-1}

\usepackage{graphicx}
\usepackage{amsmath,amssymb,amsfonts}
\usepackage{xcolor}
\usepackage{soul}
\usepackage{bm}
\usepackage{tikz,tkz-euclide}
\usepackage[section]{placeins}

\usepackage{lipsum}

\usepackage[colorlinks=true,citecolor=blue,linkcolor=blue,urlcolor=blue]{hyperref}

\begin{document}

\title{Exact statistical transmutation of quantum mixtures on a 
ring}

\author{Wayne J. Chetcuti}
\affiliation{Université Grenoble-Alpes, CNRS, LPMMC, 38000 Grenoble, France}
\author{Nathan Goldman}
\affiliation{Laboratoire Kastler Brossel, Collège de France, CNRS, ENS-Université PSL,
Sorbonne Université, 11 Place Marcelin Berthelot, 75005 Paris, France}
\affiliation{International Solvay Institutes, 1050 Brussels, Belgium}
\affiliation{Center for Nonlinear Phenomena and Complex Systems, Université Libre de Bruxelles,
CP 231, Campus Plaine, 1050 Brussels, Belgium}

\author{Patrizia Vignolo}
\affiliation{Université Côte d’Azur, CNRS, Institut de Physique de Nice, 06200 Nice, France}
\affiliation{Institut Universitaire de France}
\author{Anna Minguzzi}
\affiliation{Université Grenoble-Alpes, CNRS, LPMMC, 38000 Grenoble, France}

\begin{abstract}
One-dimensional strongly repulsive quantum mixtures 
exhibit non-trivial exchange statistics arising from the interplay between orbital and spin degrees of freedom. Using an exact solution, we demonstrate that in the single-impurity limit the wavefunction of 
Fermi–Fermi and Bose–Bose mixtures on a ring threaded by an artificial gauge field display exact anyonic statistics under exchange of the impurity with the majority particles. The resulting fractional exchange
phase is fixed by the angular momentum sector selected through the applied flux. We find that the impurity momentum distribution coincide exactly with the anyonic one, independently of the  bosonic or fermionic nature of the mixture, with the large-momentum tails encoding a direct signature of the anyonic statistical angle. 
Finally,  we devise a quench protocol for  reversible dynamical anyonization. Our results provide  a path for realizing and manipulating anyonized states with ultracold atoms. 
\end{abstract}

\maketitle
Bosons and fermions are the two facets of the fundamental dichotomy of quantum statistics: exchanging two identical particles leaves the many-body wavefunction unchanged, or reverses its sign~\cite{dirac1926theory}. Anyons  escape 
this binary classification, with particle exchange imparting an arbitrary phase to the wavefunction~\cite{leinaas1977theory,wilczek1982quantum,khare2005fractional}. The prospect of continuously tunable exchange statistics has motivated the development of a broad class of anyonic models in continuum~\cite{kundu1999exact,girardeau2006anyon,calabrese2007correlation,batchelor2006one,batchelor2007bethe,santachiara2008one} and lattice~\cite{amico1998one,osterloh2000bethe,greschner2015anyon,bonkhoff2025anyonic} settings, providing a fertile framework for exploring the consequences beyond bosonic and fermionic statistics~\cite{forte1992quantum,stern2008anyons,greiter2024fractional}. 

Applications of anyons range from topology and strongly correlated matter to 
quantum information~\cite{wilczek1990fractional,nayak2008non}. In two dimensions, anyons are predicted to emerge as excitations of fractional quantum Hall states~\cite{arovas1984fractional}. Their exchange properties form the basis of braiding protocols for fault-tolerant quantum computation~\cite{kitaev2003fault}. Fractional exchange statistics was experimentally probed in nanoelectronic circuits~\cite{bartolomei2020fractional} and in ultracold atomic gases~\cite{kwan2024realization}. In one dimension (1D), anyons 
have been studied theoretically for decades~\cite{haldane1991fractional,ha1995fractional,greiter2024fractional} and are amenable to exact treatment, with Bethe Ansatz (BA) solutions available for interacting gases~\cite{batchelor2006one,batchelor2007bethe,scopa2020one} and closed expressions for the one-body density matrix in the hard-core limit~\cite{santachiara2007entanglement,santachiara2008one}. The associated momentum distribution provides a characteristic fingerprint of fractional statistics: as the statistical phase is varied, it interpolates between the bosonic and fermionic limits while developing a pronounced asymmetry at intermediate values~\cite{calabrese2007correlation,santachiara2007entanglement,patu2007correlation,hao2008ground,hao2009ground}.

A distinct route of engineering anyonic properties exploits a strongly interacting quantum impurity immersed in a quantum fluid~\cite{gamayun2020zero,gamayun2024emergence}. By setting the impurity into a momentum-carrying state, it was shown that in the thermodynamic limit its one-body density matrix coincides with that of an anyon. Such anyonization protocol was realized with ultracold bosonic atoms in a 1D optical lattice
~\cite{dhar2025observing} and examined via an effective swap model~\cite{wang2025anyonization}.

We propose an alternative protocol for 1D anyonization based on quantum impurities: by confining particles in a ring and applying an artificial gauge field, we demonstrate that in the strongly interacting regime the impurity acquires exact anyonic exchange statistics for arbitrary particle number. To obtain an explicit form for the exact many-body wavefunction we use the necklace Ansatz~\cite{aupetit2025necklace}, equivalent to BA in the strong-coupling limit, embedding the full permutation symmetry. This construction establishes that both the ring geometry and the single-impurity case are crucial for exact anyonization. We demonstrate that the large-momentum tails, owing to their sensitivity to nearest-neighbour exchanges, retain a direct imprint of fractional statistics and devise a gauge-field quench that prepares and reversibly connects distinct anyonized states dynamically. Our results provide a unified framework for realizing, probing, and controlling genuine anyonic exchange statistics in 1D.

{\textit{Model -- }} Consider a 1D two-component ($\alpha = \uparrow,\downarrow$) fermionic or bosonic mixture, where these internal degrees of freedom act as an effective spin. The system consists of $N_{p}$ particles of mass $m$ confined to a ring of circumference $L$ and subjected to strongly repulsive contact interactions, modeled by the Hamiltonian
\begin{align}\label{eq:Ham}
    \mathcal{H} \!\! = \!\!\!\!&\sum\limits_{\alpha = {\uparrow,\downarrow}}\sum\limits_{j=1}^{N_{\alpha}}\frac{\hbar^{2}}{2m}
    \bigg(\!-i\!\frac{\partial}{\partial x_{j,\alpha}}\!-\!\frac{2\pi}{L}\frac{\phi}{\phi_{0}}\bigg)^{2} \!+\! g_{\uparrow\downarrow}\!\sum\limits_{j,l}\delta (x_{j,\uparrow}-x_{l,\downarrow})\nonumber \\
&+ \sum\limits_{\alpha = \{\uparrow,\downarrow\}}\! g_{\alpha\alpha}\!\sum\limits_{j<l}^{N_{\alpha}}\delta(x_{j,\alpha}-x_{l,\alpha}).
\end{align}
Here, $g_{\uparrow\downarrow}$ is the inter-species interaction strength and  
$g_{\alpha\alpha}$ is the intra-species one, applying to the bosonic mixture. Hamiltonian~\eqref{eq:Ham} is SU(2) symmetric and BA integrable for arbitrary interactions~\cite{gaudin1967un,yang1967some,oelkers2006bethe} for fermions, and for bosons only if $g_{\uparrow\downarrow}\!=\!g_{\uparrow\uparrow}\!=\!g_{\downarrow\downarrow}$. The results established here for a continuous ring also hold for the lattice~\cite{suppmat}.

In the strongly repulsive limit, the spin and orbital degrees of freedom decouple, with the many-body wavefunction separating into an orbital and a spin part~\cite{ogata1990bethe,deuretzbacher2008exact,deuretzbacher2016momentum,osterloh2023exact}. The orbital sector is described by the corresponding spinless fermionic/Tonks-Girardeau (TG) wavefunction for fermionic/bosonic mixtures respectively. An effective Heisenberg Hamiltonian $H_{s} = \pm t\sum_{j}^{N_{p}}(P_{j,j+1}+\mathbf{I})$ governs the spin sector where $P_{j,j+1}$ is the permutation operator exchanging adjacent spins, $t$ sets the coupling strength, $\pm$ corresponds to fermions (bosons), and $\mathbf{I}$ is the identity matrix. The wavefunction then reads~\cite{ogata1990bethe,deuretzbacher2016momentum}
\begin{equation}
\Psi(X)=\sum\nolimits_{Q\in S_{N_{p}}}a_{Q}\theta_Q(X)\Psi_{F/A}(X),
\label{eq:wavef}
\end{equation}
where $X=\{x_{1},\ldots,x_{N_{p}}\}$, $\theta_{Q}(X)$ restricts the wavefunction to the coordinate sector $x_{Q(1)}\!<\! x_{Q(2)}\!<\!\cdots\!<\!x_{Q(N_{p})}$, and $a_{Q}$ is the spin amplitude associated with that sector. Here, $\Psi_{F}$ is the spinless fermionic wavefunction constructed as the Slater determinant of plane wave single-particle orbitals, $\varphi_j(x)=\frac{1}{\sqrt{L}}e^{i k_{j} x}$, where $k_{j}$ are the occupied wavevectors. The corresponding TG orbital wavefunction is defined through the Bose-Fermi mapping $\Psi_{A}(X)=\prod_{j<l}\mathrm{sgn}(x_{j}-x_{l})\Psi_F(X)$~\cite{girardeau1960relationship}.

Instead of relying directly on BA, the spin coefficients $a_{Q}$, defined over the particle-ordering sectors, are obtained through the necklace Ansatz~\cite{aupetit2025necklace}, which provides a reduced spin sectors description. Exploiting periodic boundary conditions, the different ordered spin configurations are organized into necklaces, i.e.~equivalence classes under cyclic rotations. Thus, a given necklace consists of all configurations generated from a representative spin ordering by successive applications of the cyclic permutation operator $P_{1\rightarrow N_{p}}$. Since Eq.~\eqref{eq:Ham} is invariant under these  rotations, $[\mathcal{H},P_{1\rightarrow N_{p}}]\!=\!0$ \cite{suppmat}, the spin states can be chosen as common eigenstates of $P_{1\rightarrow N_{p}}$ and $\mathcal{H}$. Consequently, the amplitudes  $a_{q,j}$  within a given necklace $q$ are not independently determined but related 
by phase factors fixed by the application of the cyclic operator (see Fig.~\ref{fig:schematic}), i.e. $a_{q,j} \!=\! e^{-i\frac{2\pi n}{N_{p}}j}c_{q}$ with $j\!=\! 0,\ldots,N_{q}\!-\!1$, $c_q$ being an overall coefficient,  $N_{q}$ the number of distinct configurations and $n$ labels the 
 integer quantum number allowed by the necklace structure.

{\textit{Impurity in a flux-threaded ring --}} Consider a ring-shaped system pierced by a synthetic flux $\phi$, which in cold atoms can be engineered through various implementations~\cite{amico2021roadmap,amico2022colloquium,dalibard2011colloquium,polo2025persistent} such as stirring or phase imprinting. For free particles, the coupling of the flux to the particle momenta shifts the spectrum, producing an energy landscape composed of piecewise parabolic branches labelled by the angular momentum $\ell$ per particle. Crossings between these branches allow the system to change $\ell$ to screen the applied flux, yielding a spectrum with period given by the flux quantum $\phi_{0}\!=\!\frac{\hbar}{m}$~\cite{leggett1991dephasing}. Entering the strongly repulsive regime, the ground-state energy exhibits a reduced periodicity  of $\phi_0/N_{p}$, originating from underlying spin correlations and  providing a spectral signature of angular momentum fractionalization~\cite{yu1992persistent,chetcuti2022persistent,pecci2023persistent}.

\begin{figure}[h!]
    \centering
    \includegraphics[width=0.85\linewidth]{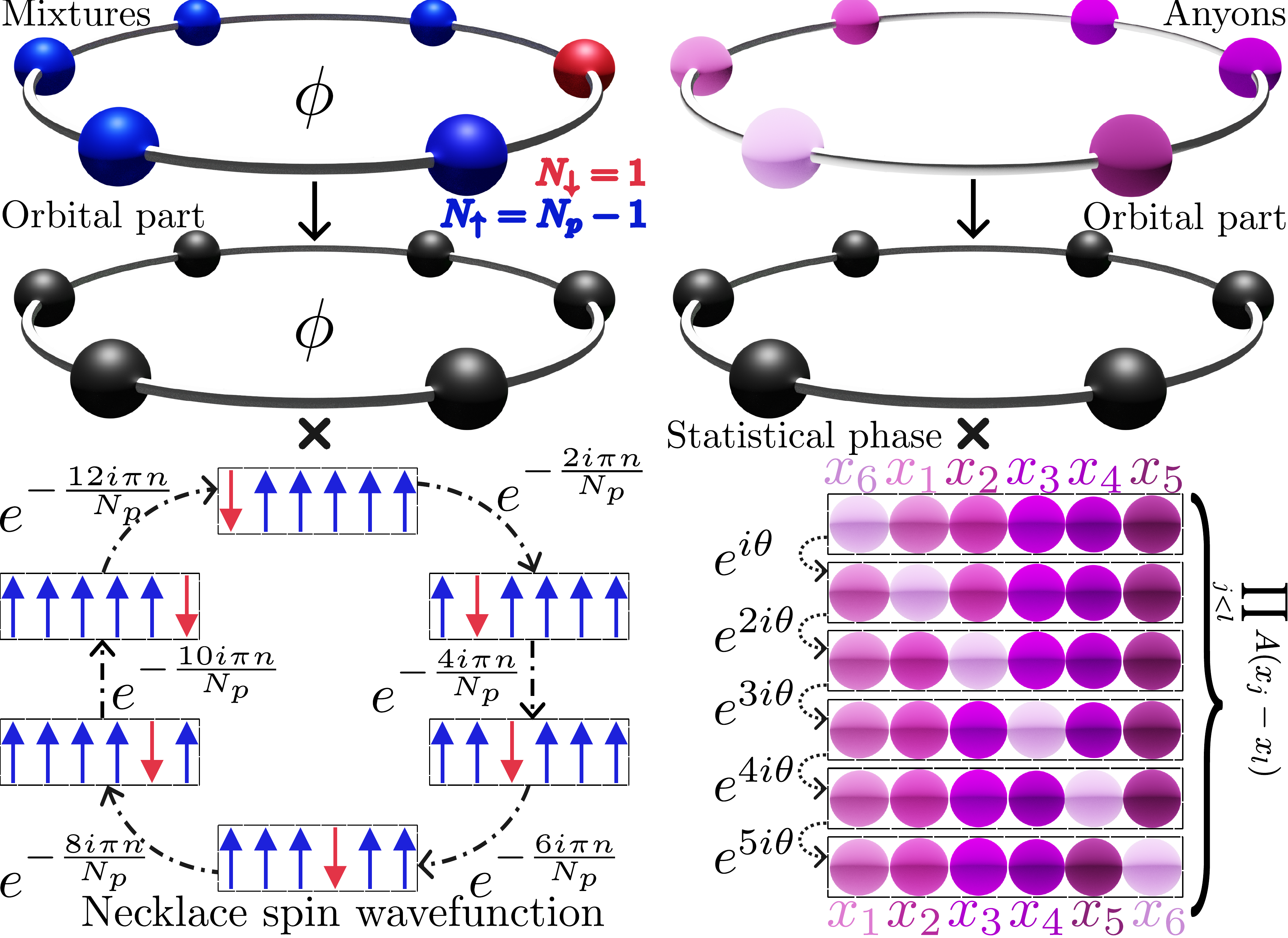}%
\caption{Scheme illustrating emerging anyonic exchange phases in single-impurity quantum mixtures. A repulsively interacting two-component mixture with $N_p =6$ particles pierced by an effective flux $\phi$ decouples into a spinless orbital part and a necklace spin wavefunction. The latter is constructed from cyclic translations of the impurity, with the $j$-th configuration acquiring a phase $e^{-i 2\pi n j/N_{p}}$.  In the anyonic case, the same orbital structure is instead dressed by the statistical factor $\prod_{j<l}A(x_{j}\!-\!x_{l})$, with $A(x_{j}-x_{l})=e^{i\theta}$ for $x_{j}\!<\!x_{l}$ and $A(x_{j}\!-\!x_{l})=1$ otherwise, where $\theta$ is the anyonic phase. Successive re-orderings acquire a phase $e^{i \theta}$, with the statistical phase playing the role of the necklace phase.}
    \label{fig:schematic}
\end{figure}
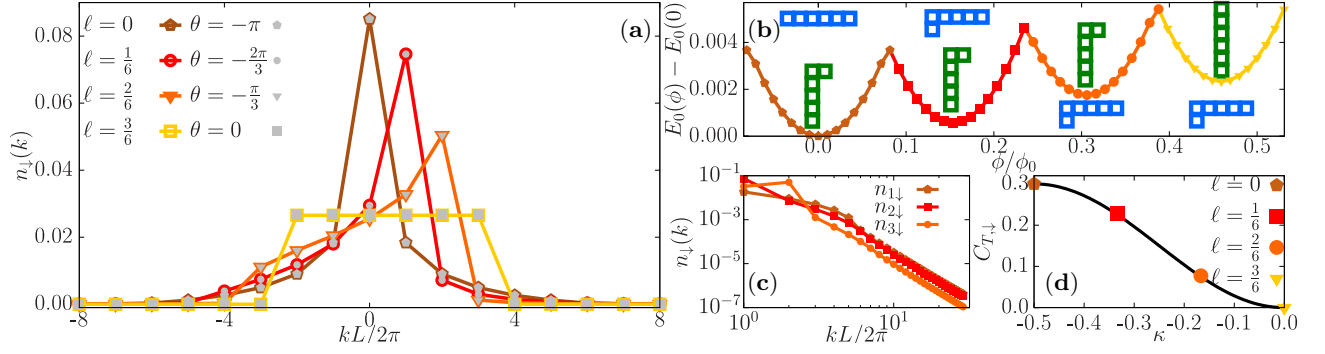
\begin{figure*}[ht!]
    \centering
    \includegraphics[width=0.95\linewidth]{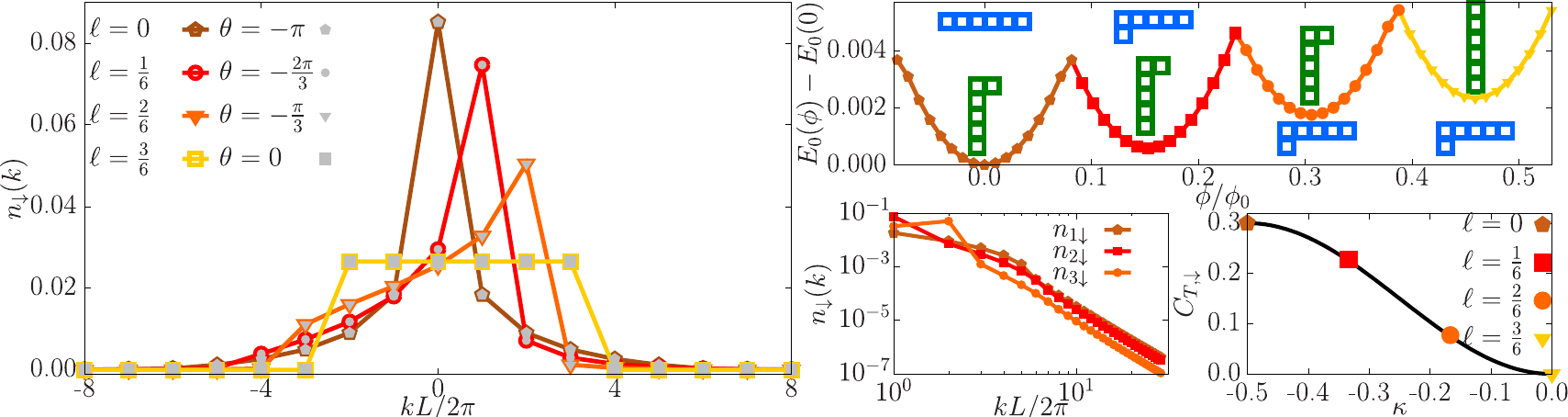}%
    \put(-255,115){(\textbf{a})}
    \put(-207,115){(\textbf{b})}
    \put(-207,20){(\textbf{c})}
    \put(-95,20){(\textbf{d})}
    \caption{Momentum distributions of mixtures and anyons. \textbf{(a)} Impurity momentum distribution $n_{\downarrow}(k)$ as a function of wavevector $k$, in units of $1/L$ and $2 \pi/L$ respectively, for fermionic mixtures with $N_{p}=6$ particles on a ring of length $L$. The colored curves show the exact results obtained from Eq.~\eqref{eq:dereutz} for different angular momenta $\ell$, while the gray symbols denote the corresponding anyonic distributions at the indicated statistical angles $\theta$, calculated from Eq.~\eqref{eq:santa}. \textbf{(b)} Ground-state energy $E_{0}(\phi)\!-\!E_{0}(0)$, in units of $J$, versus the flux $\phi/\phi_0$ obtained by exact diagonalization of the Fermi–Hubbard model with $U=500J$, where $U$ is the on-site interaction and $J$ is the hopping amplitude, together with the Young tableaux identifying the symmetry of each parabola for bosons (blue) and fermions (green). \textbf{(c}) Large-momentum tails of $n_{\downarrow}(k)$, shown on a log–log scale. \textbf{(d)} Impurity Tan's contact $C_{T,\downarrow}$ extracted from these tails as a function of $\kappa \!=\! \theta/2\pi$ (symbols) compared to Eq.~\eqref{eq:Tan_down} (solid line). }
    \label{fig:anyonized_mom}
\end{figure*}

At the level of the wavefunction, each fractionalized branch is characterized  by a modified set of momenta~\cite{suppmat} entering the orbital wavefunction $\Psi_{F\!/\!A}(X)$. Dictated by the winding number, this selects the total angular momentum sector and hence the corresponding parabola of the energy spectrum. 
Correspondingly, the spin sector is selected by retaining only the necklaces whose cyclic quantum number produces the required phase winding $e^{-i 2\pi n/N_p}$, namely $\ell \!=\! n/N_{p}$.  Thus, the integer $n$ links the fractional shift of the orbital sector to a definite spin state, encoded in the amplitudes  $a_{q,j}$ and determined by the corresponding Heisenberg Hamiltonian eigenstate. 

Focusing on the single-impurity limit $N_{\downarrow}=1$, the necklace Ansatz simplifies considerably: the spin configurations are generated by moving the spin-down particle through the spin-up background. Hence, there is a unique necklace, whose elements pertain to the impurity positions in the chain (see Fig.~\ref{fig:schematic}). Taking $|\!\downarrow\uparrow\ldots\uparrow\rangle$ as the initial configuration, the spin eigenstates read
\begin{equation}\label{eq:spinstate}
|\chi_n\rangle =\frac{1}{\sqrt{N_{p}}}
\sum\nolimits_{j=0}^{N_{p}-1}
e^{- \frac{2i\pi n j}{N{_p}}}
[P_{1\rightarrow N_{p}}]^{j}
|\!\downarrow\uparrow\cdots\uparrow\rangle .
\end{equation}
Inserting this spin amplitude in Eq.~(\ref{eq:wavef}), we obtain that the many-body wavefunction satisfies:
\begin{equation}\label{eq:parexchange}
\Psi(\ldots,x_{j},x_{l},\ldots) = e^{-i\frac{2\pi n}{N_{p}}}\Psi(\ldots,x_{l},x_{j},\ldots).
\end{equation}
Thus, by the necklace Ansatz we proved that the exchange of the impurity with a majority particle assigns a fractional phase to the wavefunction, fixed by the angular momentum sector and hence the necklace twist $2 \pi n/N_{p}$.

{\textit{Emergent anyonized one-body correlations --}} Anyon exchange statistics interpolate between fermions and bosons, satisfying anyonic anti-commutation relations~\cite{batchelor2006one}
\begin{align}
\hat{\psi}_{\theta}(x)\hat{\psi}_{\theta}^{\dagger}(y)+
e^{-i\theta \epsilon(x-y)}\hat{\psi}_{\theta}^{\dagger}(y)\hat{\psi}_{\theta}(x)&=\delta(x-y), \nonumber\\
\hat{\psi}_{\theta}(x)\hat{\psi}_{\theta}(y)+e^{i\theta\epsilon(x-y)}
\hat{\psi}_{\theta}(y)\hat{\psi}_{\theta}(x)&=0,
\end{align}
where $\epsilon(x\!-\!y)=\mathrm{sgn}(x\!-\!y)$, $\hat{\psi}_{\theta}(x)$ [$\hat{\psi}_{\theta}^{\dagger}(x)$] are the anyonic destruction (creation) field operators at position $x$. With this convention, the limits $\theta=0$ ($\theta=\pi$) describe fermions (bosons), while intermediate values $0\!<\!\theta\!<\!\pi$ correspond to anyons~\footnote{Equivalently, the algebra may be formulated utilizing bosonic anyonic commutation rules, which differ from the fermionic ones by a $\pi$ shift of the statistical parameter.}. The ground-state hard-core anyonic many-body wavefunction can be equivalently constructed through a fermion- or boson-anyon mapping 
\begin{equation}\label{eq:anyonwave}
\Psi^{\theta}_{0}(x_{1},...,x_{N_{p}})
\!=\!
\left[\prod\nolimits_{j<l} A(x_{j}\!-\!x_{l})\right]\Psi^{R}_{0}(x_{1},..,x_{N_{p}}),
\end{equation}
depending on the reference wavefunction $\Psi^{R}_{0}$ chosen to be the spinless fermionic or TG wavefunctions. The mapping factor is $A(x_{j}\!-\!x_{l}) = e^{i\theta}$ for $x_{j}\!<\!x_{l}$ and $A(x_{j}\!-\!x_{l})=1$ for $x_{j}\!>\!x_{l}$. From Eq.~\eqref{eq:anyonwave}, it is clear that the anyonic wavefunction acquires a spatial-ordering dependent phase, in close resemblance to the spin-dependent phases obtained for mixtures in Eq.~\eqref{eq:parexchange}. Such a connection was exploited to 
engineer anyonic correlations through spin waves  ~\cite{gamayun2020zero,gamayun2024emergence,dhar2025observing}.
The correspondence 
 applies only to exchanges between different spin components; same-species swaps leave the spin configuration unchanged and contribute only to $\pm$ sign change in the orbital wavefunction.

To illustrate the equivalence between the impurity and anyons, we define a conditional one-particle wavefunction by fixing $N_{p}\!-\!1$ spatial coordinates~\cite{chetcuti2025interferometric}. For the mixture, it is the  spin-down particle that remains mobile. Each crossing of the impurity with a fixed particle changes the spin ordering causing the spin wavefunction to acquire a phase. Likewise, the anyonic wavefunction acquires the same necklace phase once we select discrete values of $\theta = \frac{2\pi n}{N_{p}}$. Sticking to the fermion–anyon convention, $\theta\!=\!0$ yields a fully fermionized one-particle wavefunction with $N_{p}\!-\!1$ nodes, while $\theta\!=\!\pi$ bosonizes it, converting all the nodes into cusps. At intermediate $\theta$, the wavefunction is complex, with its amplitude traversing between real and imaginary planes across successive crossings, reflecting the ordering-dependent phases characterizing anyons.

Further insight follows from the one-body correlator. For hard-core anyons, it was previously shown that for both Fermi- and Bose-mapped anyons for ground-state wavefunctions with consecutively occupied momenta and odd $N_{p}$, the correlator can be represented exactly
as an $(N_{p}\!-\!1)\!\times\!(N_{p}\!-\!1)$ Toeplitz determinant~\cite{santachiara2007entanglement,santachiara2008one}
\begin{equation}\label{eq:santa}
    \rho_{N_{p}}^{\theta}(x) = e^{-i(\theta_{0}^{F} + \theta)\frac{x}{L}}\det[\varphi_{m}^{\sigma,\theta}(x)]_{1}^{N_{p}-1}.
\end{equation}
The matrix elements entering the determinant are given by $\varphi_{j,l}^{\sigma,\theta} \!=\!\int\nolimits_{0}^{2\pi}\frac{2}{\pi}e^{\imath (j-l)s}A\left(s\!-\!\frac{2\pi x}{L}\right)\mathcal{S}_{\sigma}\left(\frac{s}{2}\!-\!\frac{\pi x }{L}\right)\mathcal{S}_{\sigma}\left(\frac{s}{2}\right)\mathrm{d}s$ with $\mathcal{S}_{F}(u)\!=\! \sin(u)$ and $\mathcal{S}_{B}(u)\!=\! |\sin(u)|$ for fermions and bosons respectively. Here, we extend the analysis to even particle numbers and mesoscopic-sized systems. Our treatment introduces an extra shift $\theta_{0}^{F}=\pi$, specific to the fermion-anyon mapping with even particle numbers, which arises from the integer asymmetric momentum configuration. Moreover, we show that the statistical angle shifts the momenta entering the kernel $\varphi_{j,l}^{\sigma,\theta}(x)$ in accordance with the anyonic Bethe equations~\cite{batchelor2007bethe}. This shift ultimately contributes an extra phase to the determinant that exactly reproduces the effect of the respective angular momentum in the mixture.

For strongly repulsive mixtures, decoupling of the spin and orbital degrees of freedom allows the one-body correlator to be split into an orbital part and spin part~\cite{deuretzbacher2016momentum,ogata1990bethe}
\begin{equation}\label{eq:dereutz}
    \rho_{\alpha}(x,x') = \sum\nolimits_{j,l}^{N_{p}} (\pm 1)^{j+l}\rho^{j,l}(x,x')\omega^{j,l}_{\alpha}.
\end{equation}
Here, $\rho^{j,l}(x,x')\!=\!\int_{I_{jl}}\Psi^{*}(x_{1},\ldots,x_{j-1},x,x_{j+1},\ldots,x_{N_{p}})\times\Psi(x_{1},\ldots,x_{j-1},x',x_{j+1},\ldots,x_{N_{p}})\prod_{n\neq j}\!\mathrm{d}x_{n}$ is the orbital contribution with the $\pm$ signs corresponding to bosons/fermions~\cite{suppmat}. The spin matrix elements are given by $\omega_{\alpha}^{j,l}\!=\!\langle\chi|\delta_{\alpha_{j}}^{\alpha}\hat{P}_{j,...,l}|\chi\rangle$, where $|\chi\rangle$ is the  spin wavefunction, $\hat{P}_{j,\ldots,l}$ is the (anti-)cyclic permutation operator 
and $\delta_{\alpha_{j}}^{\alpha}$ chooses only particles with spin $\alpha_{j}\!=\!\alpha$.

In the case of hard-core anyons, we find that the one-body correlator can be written in an analogous form,
\begin{equation}
\rho_{\theta}(x,x')=\sum\nolimits_{j,l}^{N_{p}}(\pm1)^{j+l}\rho^{j,l}(x,x'){\color{black}e^{-i\theta_{B/F}(j-l)}} .
\end{equation}
\noindent Here, the orbital contribution $\rho^{j,l}(x,x')$ is identical to that of the mixture, while $e^{-i\theta_{B/F}(j-l)}$ represents the statistical phase accumulated by the anyonic string.  $\theta_{B}$ and $\theta_{F}$ correspond to the Bose- and Fermi-anyon mappings, respectively, and are related by $\theta_{F}=\theta_{B}+\pi$. Since this phase depends only on the change in ordering sector between $x$ and $x'$, labelled by $j$ and $l$, it plays the same role as the spin matrix element $\omega_{\alpha}^{j,l}$. For $\theta=2\pi n/N_p$, the anyonic factor coincides with the spin contribution obtained from Eq.~\eqref{eq:spinstate}. The anyonic and impurity correlators therefore share both the same orbital part and sector-dependent phase structure, establishing an exact mapping between hard-core anyons and the fractionalized single-impurity states of the strongly repulsive mixture.

{\textit{Momentum distributions of impurity versus anyons --}} To reveal the emergent anyonic character of the single impurity, we examine the momentum distribution $n_{\alpha}(k) \!=\!\int\nolimits_{0}^{L} \!\mathrm{d}x\!\int\nolimits_{0}^{L} \mathrm{d}x'\rho_{\alpha}(x,x') e^{-ik (x-x')}$ with $\rho_{\alpha}(x,x')$ taken from Eq.~\eqref{eq:dereutz-supp}. Comparing momentum distributions of hard-core anyons and the impurity in the strongly repulsive mixture, we show that the latter exhibits exact anyonic behaviour. 
Fig.~\ref{fig:anyonized_mom}\textbf{(a)} compares the impurity momentum distribution of fermionic mixtures with those of hard-core anyons for several fractionalized parabolas shown in Fig.~\ref{fig:anyonized_mom}\textbf{(b)}. Three characteristic regimes emerge: (i) TG or bosonized regime, marked by a single dominant peak; (ii) the fermionized regime, marked by a flat distribution without tails; and (iii) the genuinely anyonic regime, where the statistical phase produces a skewed profile. In all cases, the impurity momentum distribution exactly  matches the anyonic result for the corresponding value of the statistical angle shifted by $\theta_0^F$.
Signatures of anyonization stemming from the momentum distribution are manifested in spiral interferograms, and directly related to the anyonic phase \cite{suppmat}.

A peculiar parity effect uncovered in our study distinguishes between odd and even particle numbers. For odd $N_{p}$, bosonic and fermionic mixtures each reproduce the anyonic momentum distribution associated to their own mapping, yet they remain distinct from one another. By contrast, for even $N_{p}$, the impurity momentum distributions of the bosonic and fermionic mixtures become identical for every allowed value of $\theta$. Such equivalence arises only for even $N_{p}$,  because the discrete set of statistical angles contains $\theta\!=\!\pi$  where the two anyonic mappings exchange their bosonic and fermionic character. Hence, the two mappings interpolate in opposite directions: the bosonic reference evolves from bosonized at $\theta\!=\!0$ to fermionized at $\theta\!=\!\pi$, whereas the fermionic reference follows the reverse path. This interchange is impossible for odd $N_{p}$ since $\theta\!=\!\pi$ is excluded. In essence, the impurity effectively rescinds any signature of its original particle statistics and undergoes full \textit{statistical transmutation}, so that its bosonic or fermionic origin can no longer be inferred from the momentum distribution alone~\footnote{This loss of statistical identity is confined to the impurity, as the momentum distributions of the majority component remain distinguishable in bosonic and fermionic mixtures.}.

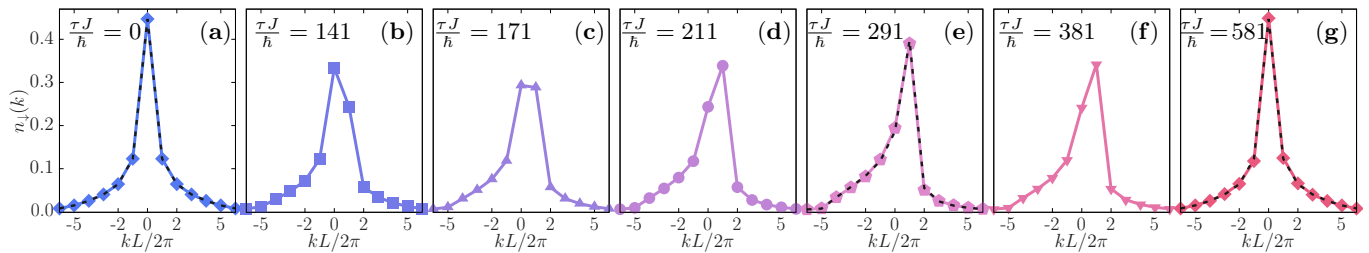
\begin{figure*}[ht!]
    \centering
    \includegraphics[width=\linewidth]{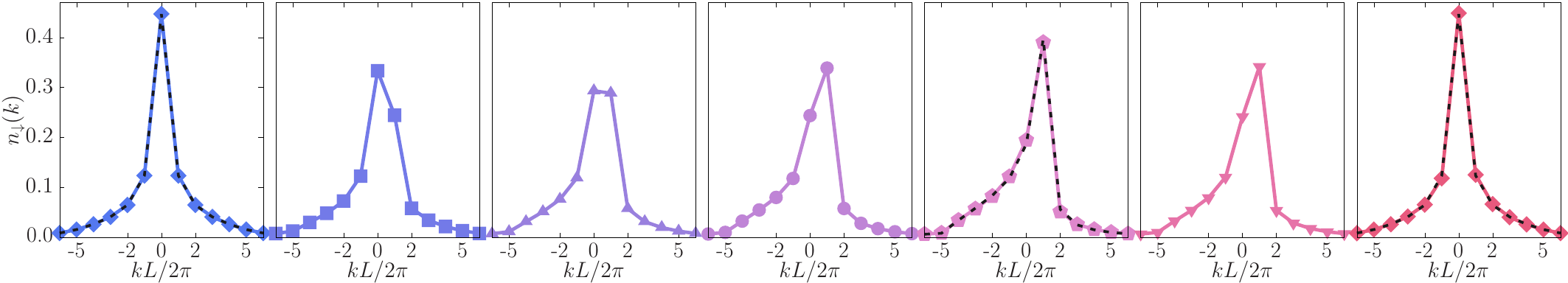}
    \put(-438,80){(\textbf{a})}
    \put(-488,80){$\frac{\tau J}{\hbar} = 0$}
    \put(-370,80){(\textbf{b})}
    \put(-418,80){$\frac{\tau J}{\hbar} = 141$}
    \put(-297,80){(\textbf{c})}
    \put(-350,80){$\frac{\tau J}{\hbar} = 171$}
    \put(-228,80){(\textbf{d})}
    \put(-280,80){$\frac{\tau J}{\hbar} = 211$}
    \put(-158,80){(\textbf{e})}
    \put(-210,80){$\frac{\tau J}{\hbar} = 291$}
    \put(-88,80){(\textbf{f})}
    \put(-138,80){$\frac{\tau J}{\hbar} = 381$}
    \put(-18,80){(\textbf{g})}
    \put(-70,80){$\frac{\tau J}{\hbar} \!=\! 581$}
    \caption{Dynamical anyonization. Time evolved momentum distribution $n_{\downarrow}(k)$ in units of $1/L$ as a function of the wavevector $k$ in units of  $2 \pi/L$  under the symmetry-breaking protocol, following a sudden quench of the flux from $\phi/\phi_{0}\!=\!0$ to $\phi/\phi_{0}\!=\!1/12$ at selected times $\tau$. During the dynamics, the momentum distribution oscillates from the bosonic one \textbf{(a)} with angular momentum $\ell \!=\! 0$  to the anyonic one \textbf{(e)}  with $\ell \!=\! 1/6$ , recovering its bosonized profile in \textbf{(g)}. Intermediate times \textbf{(b)}-\textbf{(d)} and  \textbf{(f)} depict hybridized contributions from both states. Black dotted lines are the ground-state momentum distributions profiles for $\ell=0$ in \textbf{(a)} and \textbf{(g)},  and $\ell=1/6$ in \textbf{(e)}. 
    Results obtained from exact diagonalization of the Fermi–Hubbard model with interactions $U/J$ = 1000, $N_{p}\!=\!6$ ,  on  $N_{s}\!=\!13$ sites, impurity barrier strength $\lambda_{\downarrow}/J \!=\! 0.1$ and hopping amplitude $J$.}
    \label{fig:dyna}
\end{figure*}

Next, we analyze the large-momentum tails through Tan's contact of the spin-down impurity, defined from the asymptotic behavior $C_{T,\downarrow}\!=\!\lim_{|k|\rightarrow\infty}k^{4} n_{\downarrow}(k)$. Since this regime is governed by the short-distance behavior of the correlator~\cite{musolino2024symmetry,musolino2025symmetry}, the contact is
$C_{T,\downarrow}\!=\!\frac{m^{2}tg_{\uparrow\downarrow}}{2 \pi\hbar^{4}}
\sum_{j}^{N_{p}-1}\langle\chi|\delta_{\downarrow}^{j}(P_{j-1,j}\!+P_{j,j+1}\!+ 2 \mathbf{I})|\chi\rangle$. Thus, Tan's contact isolates the local exchange encoded by $P_{j,j+1}$ providing an exact signature of the fractional statistics. By noting that $\sum_{j}^{N_{p}-1}\langle\chi|\delta_{\downarrow}^{j}(P_{j-1,j}\!+P_{j,j+1}\!+ 2 \mathbf{I})|\chi\rangle \!=\! 2[1\!\pm\!\cos(\theta)]$, Tan's contact $C_{T,\downarrow}$ can be re-expressed as 
\begin{equation}\label{eq:Tan_down}
 C_{T,\downarrow} = C_{0}[1\pm\cos(\theta)]/2
\end{equation}
where $C_{0}$ denotes  Tan's contact at $\theta\!=\!0$ ($\theta=\pi$) for bosonic (fermionic) mixtures. Fig.~\ref{fig:anyonized_mom}\textbf{(c)} shows the large-momentum tails for several values of $\theta$, while Fig.~\ref{fig:anyonized_mom}\textbf{(d)} compares the extracted values of $C_{T,\downarrow}$, with Eq.~\eqref{eq:Tan_down}, finding perfect agreement. Thus, the tails directly reflect the statistical phase through their analytic dependence on $\theta$. The same modulation is found in anyons  \cite{kundu1999exact,santachiara2008one}. While the anyonic correspondence here reported manifests at the single-impurity level, it also extends to global observables: the persistent current of anyons  coincides with that of the full quantum mixture \cite{suppmat}.

{\textit{Dynamical anyonization --}} Finally, we outline the protocol for dynamically reaching an anyonized state. The dynamics is induced by a quench in flux and computed by exact diagonalization of the lattice counterpart of Eq.~\eqref{eq:Ham}. The initial quench Hamiltonian is supplemented by a localized color-selective barrier $\lambda$ at site $j_{0}$: $\mathcal{H}_{b}\!=\!(\lambda_{\uparrow}n_{j_{0},\uparrow}  \!+\!\lambda_{\downarrow}n_{j_{0},\downarrow})$. The barrier's presence breaks translational invariance, inducing the opening of spectral gaps at the crossings between the energy parabolas. To drive oscillations between two selected branches, the flux is quenched to their avoided crossing. There, states with different angular momenta hybridize into superpositions, enabling the dynamics to connect distinct momentum distributions. In SU($N$)-invariant systems, the barrier couples only neighboring parabolas whose spin states belong to the same symmetry sector~\cite{polo2026static}, indicated by the Young diagrams in Fig.~\ref{fig:anyonized_mom}\textbf{(c)}. In order to access arbitrary states, we explicitly break SU($N$) invariance by applying the barrier only to the spin-down impurity~\cite{suppmat}. This lifts the symmetry-sector constraint. Using this symmetry-breaking quench protocol, dynamical anyonization is observed in Fig.~\ref{fig:dyna} through the time-evolved momentum distribution: its profile oscillates between those of the two coupled branches, while intermediate times show hybridized contributions from both, demonstrating coherent dynamics between distinct anyonized states.

{\textit{Conclusions --}} We have proven that the wavefunction of a strongly repulsive 1D imbalanced fermionic or bosonic mixture under an artificial gauge field displays flux-tunable anyonic exchange statistics. The fractionalized angular momentum of the mixture fixes the anyonic statistical angle, while the spin wavefunction generates the corresponding ordering-dependent phases as the impurity moves through the background. 
This exact correspondence, stemming from the single-necklace solution for the mixture,
necessitates
the single-impurity limit, ring geometry and same flux  for both species~\cite{suppmat}, and  holds for arbitrary particle numbers. While the anyonic character is inherited by the impurity momentum distribution regardless of parity, in the even case it establishes genuine anyonization, independently of the bosonic or fermionic nature of the mixture. We show that the impurity Tan's contact carries information on nearest-neighbor particle exchanges, making the large-momentum tails a direct experimental observable for the anyonic phase.
Finally, we find that a color-selective barrier can coherently couple states with different fractionalized angular momenta, enabling dynamical evolution between distinct anyonic sectors. Our anyonization scheme, where the fractional statistic  emerges naturally from angular momentum fractionalization,
is amenable to  experimental implementation~\cite{lunt2024engineering,lunt2024realization}, allowing anyonic features to be  exactly realized, detected, and coherently controlled. 

{\textit{Acknowledgements --}} We thank M. Albert, L. Amico, S. Jochim, J. Polo, A. Vashisht and B. Wang   for helpful and fruitful discussions, and G. Aupetit-Diallo, P. Garnier and  S. Musolino   for their help in the early stage of this work. WJC received funding from the European Union’s Horizon research and innovation programme under the Marie Sk\l odowska-Curie grant agreement \textit{SUN\textunderscore Atomtronics} (no. 101205763). 
NG acknowledges support by the ERC Starting Grant LATIS, the EOS project CHEQS, the Fondation ULB and the ANR PEPR Grant QUTISYM ANR-23-PETQ-0002. AM and PV acknowledge grants from projects Quantum-SOPHA ANR-21-CE47-0009 and Dyn1D ANR-23-PETQ-0001.


%

\onecolumngrid

\setcounter{equation}{0}
\renewcommand{\theequation}{S.\arabic{equation}}

\newpage

\section*{Supplemental Material for `Exact statistical transmutation of quantum mixtures on a ring}

\setcounter{section}{0}
\setcounter{secnumdepth}{2}

\setcounter{subsection}{0}
\setcounter{secnumdepth}{3}

\subsection{Many-body and conditional wavefunctions}

\noindent  In this section, we begin with the Bethe Ansatz many-body wavefunction and show how its spin sector is described by the necklace Ansatz in the strongly interacting limit. We then use the resulting spin amplitudes to construct a conditional one-particle wavefunction and establish the connection between the impurity in the mixture and a hard-core anyon.

\subsubsection{Bethe Ansatz wavefunction for imbalanced Bose-Bose and Fermi-Fermi mixtures}
\noindent To begin, we consider the Bethe Ansatz wavefunction for a mixture of $N_{p}$ bosons or fermions in a one-dimensional ring with two internal states, denoted by $\boldsymbol{\alpha}=(\alpha_{1},\ldots,\alpha_{N_{p}})$, which play the role of an effective spin~\cite{gaudin1967un,yang1967some,oelkers2006bethe,deguchi2000thermodynamics}. Within a coordinate sector $Q$, defined by $x_{Q1}<\ldots<x_{QN_{p}}$, the wavefunction is a superposition of plane waves weighted by the spin amplitudes $a_{Q}$:
\begin{equation}\label{eq:full_BA_wavefunction}
\Psi(X;\boldsymbol{\alpha})=\sum_{P\in S_{N_{p}}}
 a_{Q}(\bar{k}P|\bar{\lambda};\boldsymbol{\alpha})(-1)^{(1-\chi_{B})|P|}\exp\left(i\sum_{j=1}^{N_p}k_{Pj}x_{Qj}\right).
\end{equation}
Here $X = (x_{1},\ldots,x_{N_{p}})$, $P$ corresponds to the permutation forming the symmetric group $S_{N_{p}}$ acting on the charge quasimomenta $\bar{k} = (k_{1},\ldots,k_{N_{p}})$ with $|P|$ denoting the number of transpositions with respect to the sector $k_{1}\leq\ldots\leq k_{N_{p}}$ while $\chi_{B} = 0,1$ for fermions and bosons respectively. The spin amplitudes $a_{Q}(\bar{k}P|\bar{\lambda};\boldsymbol{\alpha})$ have the structure of the Bethe eigenstates of an  inhomogenous XXX Heisenberg Hamiltonian~\cite{deguchi2000thermodynamics,korepin1993quantum} and are dictated by a set of spin rapidities $\bar{\lambda} = (\lambda_{1},\ldots,\lambda_{N_{p}})$. These amplitudes also include the factor $(-1)^{|Q|}$ associated with the permutation of the coordinate ordering. \\

\noindent In the limit of strong repulsive interactions, the Bethe Ansatz equations for charge and spin degrees of freedom decouple~\cite{ogata1990bethe,osterloh2023exact,deuretzbacher2016momentum}. The orbital sector is described by a spinless model, while the spin sector is governed by the Heisenberg Hamiltonian. Consequently, the many-body wavefunction factorizes as 
\begin{equation}\label{eq:oneparwave}
    \Psi(X;\boldsymbol{\alpha}) = a_{Q}(\bar{\lambda};\boldsymbol{\alpha})\Psi_{F/A}(X),
\end{equation}
where $\Psi_{F/A}(X)$, hold for fermionic (bosonic) mixtures respectively, and 
$\Psi_{F}$ is the spinless fermionic wavefunction constructed as the Slater determinant of plane wave single-particle orbitals $\varphi_{j}(x) = \frac{1}{\sqrt{L}}e^{i k_{j}x}$ whilst $\Psi_{A} = \prod_{j<l}\mathrm{sgn}(x_{j}-x_{l})\Psi_{F}(X)$ is the corresponding Tonks-Girardeau wavefunction. The momenta $k_{j}$ are not the bare lattice momenta, but the charge rapidities obtained from the Bethe-Ansatz equations in the limit $U\rightarrow + \infty$, $k_{j} = \frac{2\pi}{L}\left[I_{j}\pm\frac{1}{N_{p}}\sum_{l}^{M}J_{l}\right]$ for $j=1,\ldots,N_{p}$
where the upper and lower signs correspond to fermions and bosons, respectively. Here, $M$ is the number of spin-down particles, $L$ is the circumference of the ring, and $I_{j}$ and $J_{l}$ are the charge and spin quantum numbers characterizing the spectrum~\cite{essler2005one,chetcuti2022persistent,pecci2023persistent}. These quantum numbers are integers or half-odd-integers depending on the parities of $N_{p}$ and $M$ (see Table~\ref{tab:quantum_number_parity}). 

\begin{table}[h!]
\centering
\caption{Parity of the charge and spin quantum numbers for SU(2) bosonic and fermionic mixtures. Here, $N_{p}$ is the total particle number and $M=N_{\downarrow}$, such that $N_{p}-M=N_{\uparrow}$.}
\label{tab:quantum_number_parity}
\begin{tabular}{|c| c| c| c|}
\hline
Statistics & Quantum numbers & Integer & Half-odd integer \\
\hline
Bosons
& Charge $\{I_j\}$ and spin $\{J_\alpha\}$
& $N_{p}-M$ odd
& $N_{p}-M$ even
\\
Fermions
& Charge $\{I_{j}\}$
& $M$ even
& $M$ odd
\\
Fermions
& Spin $\{J_\alpha\}$
& $N_{p}-M-1$ even
& $N_{p}-M-1$ odd
\\
\hline
\end{tabular}
\end{table}

\noindent Owing to the spin-charge decoupling, the spin wavefunction loses its dependence on the charge quasimomenta, which acted as inhomogeneities in the XXX model. Therefore, the spin amplitudes are determined solely by the spin rapidities $\bar{\lambda}$, which satisfy the Bethe equations of the homogeneous XXX Heisenberg chain and are labelled by the same spin quantum numbers $\{J_{l}\}$ as the original mixtures.

\subsubsection{Necklace Ansatz wavefunction for imbalanced Fermi-Fermi and Bose-Bose mixtures}

\noindent An equivalent but alternative construction to Bethe Ansatz is provided by the necklace Ansatz~\cite{aupetit2025necklace}, which groups together all snippets connected by successive application of the full cyclic permutation operator $P_{1\rightarrow N_{p}}$. For fixed magnetization $M$, the spin Hilbert space contains $N_{q}=\binom{N_{p}}{M}$ snippets (sectors). These are partitioned into $q$ distinct necklaces that are disjoint and do not map onto one another. The number of necklaces is given by $N_{neck} = \left(N_{q}-\sum_{j=1}^{R}p_{j}\right)/N_{p}+R$ with $N_{q}$ being the snippet number and $R$ corresponding to the number of high-symmetry configurations with period $p_{j}<N_{p}$. \\

\noindent Within a given necklace, snippets are connected by the successive application of the cyclic permutation operator. Together with the ring boundary conditions, this fixes their relative amplitudes to differ by a constant phase: $a_{q,j} = c_{q}e^{-i\frac{q2\pi n}{N_{p}}j}$ with $j = 0,\ldots,N_{q}-1$  
where $c_{q}$ is the overall coefficient of necklace $q$. The integer $n$ labels the cyclic quantum numbers allowed by the necklace structure and identifies the corresponding Heisenberg eigenstate, characterized by the spin quantum numbers $\sum_{l}J_{l} = n$. Since the relative amplitudes within each necklace are fixed by cyclic symmetry, only $N_{neck}$ independent coefficients must be determined, rather than the full set of $N_{q}$ snippet amplitudes. The necklace Ansatz therefore provides a more economical description than a direct Bethe Ansatz construction, substantially reducing the complexity of the spin problem. \\

\noindent For mixtures in the single-impurity limit, i.e. $N_{\downarrow}=1$ and $N_{\uparrow}=N_{p}-1$, which is the regime considered in this work, the entire spin basis forms a single necklace of length $N_{p}$. Each snippet corresponds to a different position of the impurity along the spin chain. Starting from the configuration $|\downarrow\uparrow\ldots\uparrow\rangle$, the spin eigenstates are
\begin{equation}\label{eq:spinwaveapp}
|\chi_n\rangle =\frac{1}{\sqrt{N_{p}}}
\sum\nolimits_{j=0}^{N_{p}-1}
e^{- \frac{2i\pi n j}{N{_p}}}
[P_{1\rightarrow N_{p}}]^{j}
|\!\downarrow\uparrow\cdots\uparrow\rangle .
\end{equation}
Each translation of the impurity by one position through the background therefore causes the spin amplitude to acquire a phase of $e^{-i\frac{2\pi n}{N_{p}}}$. The resulting state is a delocalized spin-wave with winding number $\ell=n/N_{p}$, corresponding to the angular momentum per particle. Equivalently, exchanging the impurity with an adjacent background particle gives $\Psi(\ldots,x_{\downarrow},x_{\uparrow},\ldots) = e^{-i\frac{2\pi n}{N_p}}\Psi(\ldots,x_{\uparrow},x_{\downarrow},\ldots)$, which mirrors the exchange rule characteristic of anyonic particles. 

\subsubsection{Link of the hard-core anyons wavefunction to the necklace Ansatz one}

\noindent The ground-state wavefunction of hard-core anyons~\cite{santachiara2007entanglement,santachiara2008one,batchelor2006one}, constructed through either a Fermi- or Bose-anyon mapping, can be written as
\begin{equation}\label{eq:hardamap0}
\Psi^\theta_0(x_1,\ldots,x_{N_p}) = \left[\prod_{j<l}A_\theta(x_j-x_l)\right]\Psi^R_0(x_1,\ldots,x_{N_p}),
\end{equation}
where $\Psi^{R}_{0}$ is the reference spinless-fermion or Tonks--Girardeau wavefunction and the mapping factor reads
\begin{equation}
A_\theta(x_j-x_l)
\begin{cases}
e^{i\theta}, & x_j<x_l,\\
1, & x_j>x_l.
\end{cases}
\end{equation}
In order to establish a direct correspondence between the impurity wavefunction of the mixture and the anyonic one, we introduce the conditional one-particle wavefunction~\cite{chetcuti2025interferometric}, where $N_{p}-1$ coordinates are fixed to a given spatial position and the other particle is left to be free. Starting from an ordered configuration of anyons $x_{1}<\ldots<x_{N_{p}}$, each pair contributes $e^{i\theta}$ to the mapping factor, yielding
\begin{equation}
    \prod_{j<l}A_{\theta}(x_{j}-x_{l}) =e^{i\theta\frac{N_{p}(N_{p}-1)}{2}}, \quad \mathrm{for} \hspace{1mm} x_{j}<x_{l}.
\end{equation}
Each time the mobile particle crosses one of the fixed particles, one pair changes its ordering and the mapping factor is multiplied by $e^{-i\theta}$. After $j$ such exchanges of which there are only $N_{p}-1$, we have that
\begin{equation}
\prod_{r<s}A_\theta(x_{r}-x_{s})
e^{i\theta\left[\frac{N_{p}(N_{p}-1)}{2}-j\right]} = e^{i\theta\frac{N_{p}(N_{p}-1)}{2}}e^{-i\theta j},\qquad j=0,\ldots,N_{p}-1.
\end{equation}
By setting $\theta = \frac{2\pi n}{N_{p}}$, we can define $A_{n} \equiv e^{i\frac{2\pi n}{N_{p}}[\frac{N_{p}(N_{p}-1)}{2}]}$ such that the statistical phase after $j$ exchanges becomes $A_{n}e^{-i\frac{2\pi n}{N_{p}}j}$. This is exactly the  phase acquired by the impurity spin amplitude as it is successively exchanged with the spin-up background. Thus, once the corresponding spinless fermionic or Tonks-Girardeau orbital reference state is chosen, the conditional impurity wavefunction coincides with that of a hard-core anyon with statistical angle $\theta = \frac{2\pi n}{N_{p}}$. 

\subsubsection{Conditional one-particle wavefunctions}
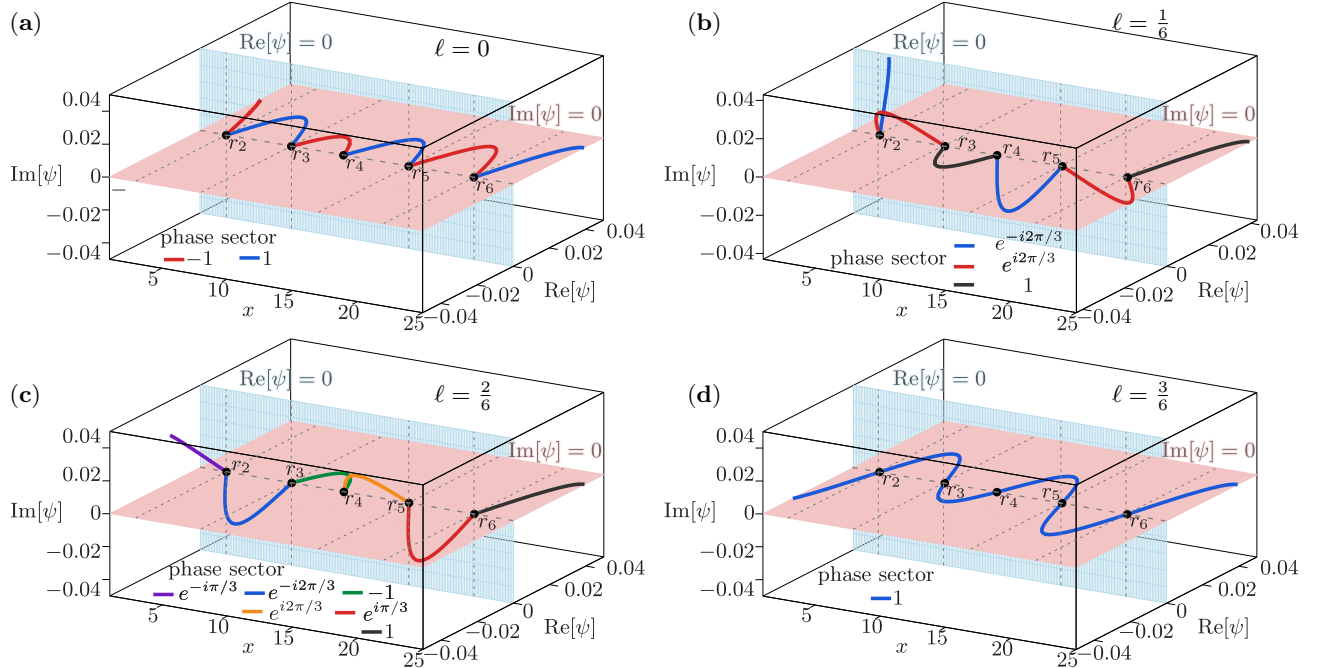
\begin{figure}[h!]
    \centering
    \includegraphics[width=0.95\linewidth]{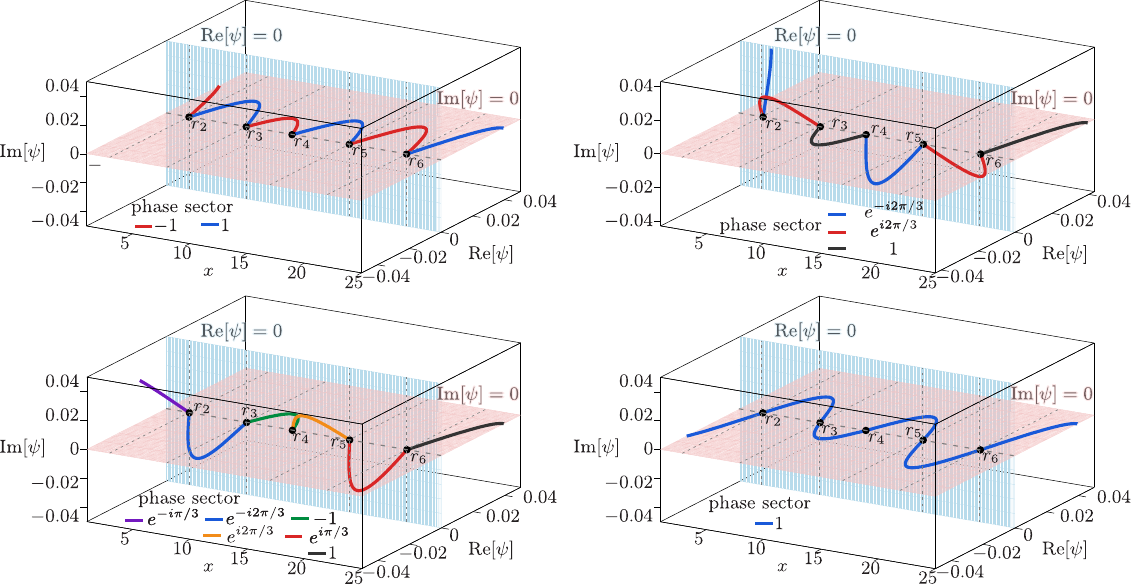}%
    \put(-485,240){(\textbf{a})}
    \put(-325,230){$\ell = 0$}
    \put(-230,240){(\textbf{b})}
    \put(-70,240){$\ell = \frac{1}{6}$}
    \put(-485,100){(\textbf{c})}
    \put(-325,100){$\ell = \frac{2}{6}$}
    \put(-230,100){(\textbf{d})}
    \put(-70,100){$\ell = \frac{3}{6}$}
\caption{One-particle wavefunction for the impurity in the fermionic mixture. Real and imaginary parts of the one-particle wavefunction $\Psi(x)$ plotted as functions of the impurity position $x$ around the ring for different angular momenta per particle $\ell$, with the $N_{p}-1$ spin-up particle coordinates fixed to $\{r2,r3,r4,r5,r6\} = \{3,8,12,17,22\}$. The wavefunctions are obtained from Eq.~\eqref{eq:oneparwave} using the spin amplitudes in Eq.~\eqref{eq:spinwaveapp} for the Heisenberg eigenstate selected at each $\ell$. Panels \textbf{(a)}-\textbf{(d)} correspond to $(\ell,n)$=(0,3),(1/6,2), (2/6,1), and (3/6,0), respectively and hence to the Fermi-mapped statistical angles $\theta= -\pi$, $-2\pi/3$, $-\pi/3$ and 0. Panels \textbf{(a)} and \textbf{(d)} display the bosonized and fermionized limits, whereas \textbf{(b)} and \textbf{(c)} show fractional anyonic states. The differently colored branches identify distinct spin-configuration sectors, or snippets, generated as the spin-down impurity successively exchanges position with the spin-up background particles. Each crossing moves the wavefunction into the next sector and imparts the corresponding statistical phase. Note that the same wavefunctions follow from the Bose-anyon mapping after a $\pi$ shift, for which the corresponding assignments are $n=0$, $-1$, $-2$, and 3, respectively. The phase sectors the impurity traverses upon each swap are illustrated in the figure. The grey dotted lines, as well as the colored planes are to guide the eye.}
    \label{fig:oneparwave}
\end{figure}

\noindent Further insight is provided by the conditional one-particle wavefunction shown in Fig.~\ref{fig:oneparwave}. Focusing on Fermi-mapped anyons, at $\theta=0$ the wavefunction exhibits the characteristic $N_{p}-1$ nodes associated with fermionic statistics. At $\theta=\pi$, corresponding to the bosonic limit, these nodes are converted into cusps. The opposite behavior occurs for Bose-mapped anyons. For mixtures with an even number of particles, the bosonized and fermionized limits occur at winding numbers $\ell=0$ and $\ell=\frac{1}{2}$, respectively, irrespective of the underlying bosonic or fermionic statistics~\footnote{This correspondence depends crucially on particle-number parity. Here, we restrict the discussion to even $N_{p}$ that will be addressed later.}. At intermediate values of $\theta$, corresponding to fractional winding numbers $\ell$, each crossing rotates the wavefunction by the statistical phase, producing a complex profile whose successive phase changes directly reveal its anyonic character. \\

\noindent For mixtures confined to a ring pierced by an effective flux, different configurations of the spin quantum numbers $\{J_{m}\}$ are selected to compensate the flux-induced shift of the charge sector. With the convention adopted above, their sum is related to the winding number by
\begin{equation}
\frac{1}{N_p}\sum_{m=1}^{M}J_{m} =  \begin{cases}
-\ell, & \text{fermions},\\
+\ell, & \text{bosons}.
\end{cases}
\end{equation}
The chosen configuration modifies the charge quasimomenta $k_{j}$ entering the orbital wavefunction while simultaneously selecting the corresponding Heisenberg eigenstate through its spin amplitudes. The allowed arrangements of $\{J_{m}\}$ depend on the particle statistics and on the parities of $N_{p}$ and $M$. For the even particle numbers considered here, increasing the winding from $\ell = 0$ to $\ell=\frac{1}{2}$ modifies $\sum_{m}J_{m}$ from 0 to $-\frac{N_{p}}{2}$ for fermions and from $+\frac{N_{p}}2$ to 0 for bosons. This opposite evolution reflects the relative $\pi$-shift between the Fermi- and Bose-mapped anyonic descriptions.

\subsection{Exact one-body density matrix for strongly repulsive quantum mixtures and hard-core anyons}\label{sec:onebodcorr}

\noindent In this section, we present compact derivations of the one-body correlator for hard-core anyons starting from the Fermi- and Bose-anyon mappings following the approach in~\cite{santachiara2007entanglement,santachiara2008one}. We then express the correlator as separate orbital and statistical contributions, directly mirroring the decomposition into orbital and spin degrees of freedom in the mixture. \\

\noindent Starting from the ground-state hard-core Fermi-anyon mapping in Eq.~\eqref{eq:hardamap}
\begin{equation}
  \label{eq:hardamap}
\Psi^{\theta}_{0}(x_{1},\ldots,x_{N_p}) = \left[\prod_{j<l}A_{\theta}(x_{j}-x_{l})\right]\Psi^{F}_0(x_{1},\ldots,x_{N_p}),
\end{equation}
with $A_{\theta}(x_{j}-x_{l})=e^{i\theta}$ for $x_{j}<x_{l}$ and 1 for $x_{j}>x_{l}$, the one-body correlator $\rho^{\theta}_{N_{p}}(x,x')$ can be written as
\begin{equation}
  \label{eq:anyonbodcorr}
    \rho^{\theta}_{N_{p}}(x-x') = N_{p}\int\limits^{L}_{0}\mathrm{d}x_{2}\ldots\int\limits_{0}^{L}\mathrm{d}x_{N_{p}}\bar{\Psi}^{\theta}_{0}(x,x_{2},\ldots,x_{N_{p}})\Psi_{0}^{\theta}(x',x_{2},\ldots,x_{N_{p}}).
\end{equation}
Translational invariance implies that the correlator depends only on
$x-x'$. We may therefore set $x'=0$ and define
$\rho^{\theta}_{N_{p}}(x)\equiv\rho^{\theta}_{N_{p}}(x,0)$. Focusing first on odd $N_{p}$, the free-fermion reference ground-state occupies the $N_{p}$ consecutive wavevectors $k_{j}=\frac{2\pi}{L}j$ for $j=-\frac{N_{p}-1}{2},\ldots,\frac{N_{p}-1}{2}$.  The corresponding spinless-fermion wavefunction is the Slater determinant
\begin{equation}
\Psi_{0}^{F}(x_{1},\ldots,x_{N_{p}}) = \frac{1}{\sqrt{N_{p}!}L^{N_{p}/2}}\mathrm{det}[e^{ik_{j}x_{l}}].
\end{equation}
Using the Vandermonde identity  $\det\!\left[x_{j}^{\,l-1}\right]_{j,l=1}^{N_{p}}=
\prod\limits_{1\leq j<l\leq N_{p}}(x_{l}-x_{j})$, one can show that
\begin{equation}
    \Psi_{0}^{F}(x_{1},\ldots,x_{N_{p}}) = C_{N_{p}}\prod\limits_{1\leq j<l\leq N_{p}}\sin\left[\frac{\pi (x_{j}-x_{l})}{L}\right],
\end{equation}
where $C_{N_p}$ is an overall complex constant satisfying $|C_{N_{p}}|^{2} = \frac{2^{N_p(N_p-1)}}{N_{p}!L^{N_{p}}}$.
Substituting this expression into the anyonic mapping and then into the one-body correlator gives
\begin{align}
    \bar{\Psi}^{\theta}_{0}(x,x_{2},\ldots,x_{N_{p}})\Psi^{\theta}_{0}(0,x_{2},\ldots,x_{N_{p}}) = |C_{N_{p}}|^{2}\left(\prod\limits_{j=2}^{N_{p}}\bar{A}_{\theta}(x-x_{j})A_{\theta}(-x_{j})\right)\left[\prod\limits_{l=2}^{N_{p}}\sin\left(\frac{\pi (x_{l}-x)}{L}\right)\sin\left(\frac{\pi x_{l}}{L}\right)\right]\nonumber\\
    \times\prod\limits_{2\leq p<q\leq N_{p}}\sin^{2}\left[\frac{\pi (x_{p}-x_{q})}{L}\right],
\end{align}
with $\bar{A}_{\theta}$ being the conjugate of $A_{\theta}$. The anyonic factors associated with pairs among the integrated coordinates $x_2,\ldots,x_{N_p}$ cancel between the two wavefunctions, leaving only those involving the external coordinates $x$ and $0$. Noting that $\bar{A}_{\theta}(x-x_{j})A_{\theta}(-x_{j}) = A_{\theta}(x_{j}-x)$ and introducing the variables
\begin{equation}
    s_j=\frac{2\pi x_j}{L},
    \qquad
    \gamma=\frac{2\pi x}{L},
    \qquad
    \mathrm{d}x_j=\frac{L}{2\pi}\,\mathrm{d}s_j,
\end{equation}
we can express the hard-core anyonic one-body correlator as
\begin{equation}
    \rho_{N_{p}}^{\theta}(x) = N_{p}|C_{N_{p}}|^{2}\left(\frac{L}{2\pi}\right)^{(N_{p}-1)}\left(\int_{0}^{2\pi}\mathrm{d}s_j\right) \left[\prod\limits_{j=2}^{N_{p}}A_{\theta}(s_{j}-\gamma)\sin\left(\frac{s_{j}-\gamma}{2}\right)\sin\left(\frac{s_{j}}{2}\right)\right]\prod\limits_{2\leq p<q\leq N_{p}}\sin^{2}\left(\frac{s_{p}-s_{q}}{2}\right).
\end{equation}
This can be further simplified by realizing that 
$\sin^{2}\left(\frac{s_{p}-s_{q}}{2}\right) = \frac{1}{4}|e^{is_{p}}-e^{is_{q}}|^{2}.$
Since the product contains $\frac{(N_p-1)(N_p-2)}{2}$ distinct pairs, it follows that
\begin{equation}
    \prod\limits_{2\leq p<q\leq N_{p}}\sin^{2}\left(\frac{s_{p}-s_{q}}{2}\right) = 4^{-\frac{(N_{p}-1)(N_{p}-2)}{2}}\prod\limits_{2\leq p<q\leq N_{p}}|e^{is_{p}}-e^{is_{q}}|^{2}.
\end{equation}
Combining this factor with $|C_{N_p}|^{2}$ and the Jacobian from the coordinate transformation gives an overall constant of $\frac{4^{(N_{p}-1)}}{N_{p}!L(2\pi)^{N_{p}-1}}$. Therefore, defining the single-variable weight
\begin{equation}
g_{\theta}(s;\gamma)=A_{\theta}(s-\gamma)\sin\left(\frac{s-\gamma}{2}\right)\sin\left(\frac{s}{2}\right),
    \qquad
    \gamma=\frac{2\pi x}{L},
\end{equation}
the correlator can be written as
\begin{align}
\rho_{N_p}^{\theta}(x) = N_{p}\frac{4^{N_{p}-1}}{N_{p}!\,L}
    \prod_{j=1}^{N_{p}-1}
    \int_{0}^{2\pi}\frac{\mathrm{d}s_{j}}{2\pi}
    \prod_{j=1}^{N_{p}-1}
    g_{\theta}(s_{j};\gamma)\prod_{1\leq p<q\leq N_{p}-1}
    \left|e^{is_{p}}-e^{is_{q}}\right|^{2}.
\end{align}
For a general weight $g(s)$, Heine's identity reads
\begin{align}
    \prod_{j=1}^{n}
    \int_{0}^{2\pi}\frac{\mathrm{d}s_{j}}{2\pi}
    \prod_{j=1}^{n}g(s_{j})
    \prod_{1\leq p<q\leq n}
    \left|e^{is_{p}}-e^{is_{q}}\right|^{2}=
    n!\det_{1\leq k,l\leq n}
    \left[\mu_{k,l}\right],
\end{align}
where $\mu_{k,l}=\int_{0}^{2\pi}\frac{\mathrm{d}s}{2\pi}\,g(s)e^{i(k-l)s}$.
Applying this identity with $n=N_{p}-1$ and
$g(s)=g_{\theta}(s;\gamma)$ gives
\begin{align}
\rho_{N_{p}}^{\theta}(x)=N_{p}\frac{4^{N_{p}-1}(N_{p}-1)!}{N_{p}!\,L}\det_{1\leq k,l\leq N_p-1}\left[\mu_{k,l}\right]=\frac{4^{N_{p}-1}}{N_{p}L}
\det_{1\leq k,l\leq N_{p}-1}\left[\mu_{k,l}\right].
\end{align}
Introducing the rescaled matrix elements
\begin{equation}
\varphi_{k,l}^{\theta}(x)\equiv4\mu_{k,l}=4\int_{0}^{2\pi}\frac{\mathrm{d}s}{2\pi}\,g_{\theta}(s;\gamma)e^{i(k-l)s},
\end{equation}
and using the fact that the determinant has dimension $N_{p}-1$, $\det\left[\varphi_{k,l}^{\theta}(x)\right]=4^{N_{p}-1}\det\left[\mu_{k,l}\right]$,
we obtain
\begin{equation}
\rho_{N_{p}}^{\theta}(x)=\frac{1}{L}
\det_{1\leq k,l\leq N_{p}-1}\left[\varphi_{k,l}^{\theta}(x)\right].
\label{eq:anynumb}
\end{equation}
The Toeplitz matrix elements are explicitly of the form
\begin{align}
    \varphi_{k,l}^{\theta}(x)&=\frac{2}{\pi}
    \int_{0}^{2\pi}\mathrm{d}s\,
    e^{i(k-l)s}
    A_{\theta}(s-\gamma)
    \sin\left(\frac{s-\gamma}{2}\right)
    \sin\left(\frac{s}{2}\right)
    \nonumber\\
    &=\frac{2}{\pi}
    \int_{0}^{2\pi}\mathrm{d}s\,
    e^{i(k-l)s}
    A_{\theta}(s-\gamma)
    \sin\left(\frac{s}{2}-\frac{\pi x}{L}\right)
    \sin\left(\frac{s}{2}\right).
\end{align}
In the limiting cases $\theta=0$ and $\theta=\pi$, one recovers the well-known results for spinless fermions and hard-core bosons, respectively:
\begin{align}\label{eq:santaeqs}
   &\varphi_{k,l}^{0}(x)= \frac{2}{\pi}\int_{0}^{2\pi}\mathrm{d}s\,
    e^{i(k-l)s}\sin\left(\frac{s}{2}-\frac{\pi x}{L}\right)
    \sin\left(\frac{s}{2}\right),\\
    &\varphi_{k,l}^{\pi}(x)= \frac{2}{\pi}\int_{0}^{2\pi}\mathrm{d}s\,
    e^{i(k-l)s}\left|\sin\left(\frac{s}{2}-\frac{\pi x}{L}\right)\right|
    \left|\sin\left(\frac{s}{2}\right)\right|.
\end{align}
The same derivation can be carried out starting from the Tonks-Girardeau wavefunction by following the approach of Ref.~\cite{forrester2003finite}. This yields the corresponding Bose-anyon expressions, with the statistical angle shifted by $\pi$ relative to the Fermi-anyon mapping. Consequently, the bosonic and fermionic limits are interchanged between the two constructions. \\

\noindent The kernels entering the Toeplitz determinant for the fermionic construction are
\begin{align}
    &\varphi^{\theta}_{r=0} = \frac{2}{\pi}\left(\pi\cos(a) + \left(e^{i\theta}-1\right)[a\cos(a)-\sin(a)]\right) \\
    &\varphi^{\theta}_{r=\pm1} = \frac{e^{\pm ia}}{\pi}\left[-\pi + \left(e^{i\theta}-1\right)(\sin(a)\cos(a)-a)\right] \\
    &\varphi^{\theta}_{r\neq 0,\pm1} = \left(\frac{e^{i\theta}-1}{\pi}\right)\left[\frac{r\sin(a)\left(1+e^{2ira}\right)+i\cos(a)\left(e^{2ira}-1\right)}{r(r^{2}-1)}\right]
\end{align}
where $a = \frac{\pi x}{L}$ and $r = k-l$. Whilst for the bosonic case, we have that
\begin{align}
\varphi_{r=0}^{\theta}&=\frac{2}{\pi}
\left[\pi\cos(a)+\left(e^{i\theta}+1\right)\left(a\cos(a)-\sin(a)\right)\right],\\
\varphi_{r=\pm1}^{\theta}&=
\frac{e^{\pm ia}}{\pi}
\left[
-\pi + 
\left(e^{i\theta}+1\right)
\left(a-\sin(a)\cos(a)\right)
\right],\\
\varphi_{r\neq0,\pm 1}^{\theta}
&=\left(\frac{e^{i\theta}+1}{\pi}\right)
\left[\frac{
r\sin(a)\left(1+e^{2ira}\right)
+i\cos(a)\left(e^{2ira}-1\right)
}{r\left(r^{2}-1\right)
}\right].
\end{align}

\noindent Now consider a uniform dimensionless shift $Z$ of the occupied momenta, $k_{j}=\frac{2\pi}{L}(j+Z)$, such that the momentum configuration is no longer symmetric around zero. The spinless-fermion wavefunction then becomes
\begin{equation}
    \Psi_{0,Z}^{F}(x_{1},\ldots,x_{N_p})
    =C_{N_{p}}\exp\left(i\frac{2\pi Z}{L}\sum_{p=1}^{N_{p}}x_{p}\right)
    \prod_{1\leq j<l\leq N_{p}}\sin\left[\frac{\pi(x_{j}-x_{l})}{L}
    \right].
\end{equation}
The exponential factor originates from the center-of-mass momentum introduced by the non-symmetric set. In the one-body correlator, the phases associated with the integrated coordinates cancel between the two wavefunctions, leaving only the contribution from the external coordinate:
\begin{equation}
    \rho_{N_{p},Z}^{\theta}(x)=
    e^{i\frac{2\pi Z}{L}x}\rho_{N_{p},0}^{\theta}(x).
\end{equation}
Following the same steps as above, one therefore obtains
\begin{equation}\label{eq:eveneq}
\rho_{N_{p},Z}^{\theta}(x)=\frac{e^{i\frac{2\pi Z}{L}x}}{L}\det_{1\leq k,l\leq N_{p}-1}\left[\varphi_{k,l}^{\theta}(x)\right].
\end{equation}

\noindent For even $N_p$, the lowest-energy configuration of spinless fermions is formed from an integer-valued set of momenta that cannot be symmetric about zero. Instead, the occupied set is centered at $Z=\pm1/2$, with the sign selecting one of the two degenerate ground-state configurations. By contrast, the momenta entering the bosonic construction remain symmetric: they are integer valued for odd $N_{p}$ and half-odd-integer valued for even $N_{p}$. Equation~\eqref{eq:anynumb} can therefore be used for bosons at either parity, whereas even $N_{p}$ fermions require Eq.~\eqref{eq:eveneq} with $Z=\pm1/2$, corresponding to a center-of-mass phase shift of $\pm\pi$. \\

\noindent A further role of $Z$ is to account for the shift of the quasimomenta induced by the statistical angle $\theta$. In the strongly repulsive limit, the anyonic Bethe Ansatz~\cite{batchelor2006one,batchelor2007bethe} shows that the bosonic Bethe equations acquire a uniform statistical shift,
\begin{equation}
    k_{j}L=
    2\pi\left[I_j+\kappa(N_{p}-1)\right],
    \qquad
    \theta=2\pi\kappa,
\end{equation}
where $\{I_j\}$ are the charge quantum numbers. The limits $\theta=0$ and $\theta=\pi$ recover the usual hard-core bosonic and spinless-fermionic momentum configurations, respectively. For the discrete angles $\theta=2\pi n/N_p$ considered here, this shift is equivalent to the one generated in the mixture by the spin quantum numbers $X=\sum_{\alpha}J_{\alpha}$, which enter the charge quasimomenta through $X/N_{p}$ and determine the angular-momentum sector. Consequently, this contribution must also be included in the anyonic wavefunction and its one-body correlator. In the convention adopted here, odd-$N_{p}$ fermionic and bosonic mixtures of either parity therefore correspond to $Z=-\frac{X}{N_{p}}=\frac{\theta}{2\pi}$, whereas even-$N_{p}$ fermions contain the additional half-integer shift of the reference momentum configuration, giving $2\pi Z=(\pi+\theta)$. \\

\noindent For strongly repulsive mixtures, spin--charge decoupling allows the one-body correlator to be separated into orbital and spin contributions~\cite{deuretzbacher2016momentum} (see Refs.~\cite{ogata1990bethe,osterloh2023exact} for the lattice counterpart),
\begin{equation}\label{eq:dereutz-supp}
\rho_{\alpha}(x,x')=\sum_{j,l=1}^{N_{p}}
(\pm1)^{j+l}\,
\rho^{j,l}(x,x')\,
\omega_{\alpha}^{j,l},
\end{equation}
where the positive and negative signs correspond to bosons and fermions, respectively. Introducing the ordered spectator coordinates $\mathbf{y}=(y_{1},\ldots,y_{N_{p}-1})$, the orbital contribution is
\begin{equation}
\rho^{j,l}(x,x')=\int_{I_{jl}}
\prod_{n\neq j}^{N_{p}-1}\mathrm{d}y_{n}\,\bar{\Psi}_{F}
(y_{1},\ldots,y_{j-1},x,y_{j},\ldots,y_{N_{p}-1})\Psi_{F}
(y_{1},\ldots,y_{l-1},x',y_{l},\ldots,y_{N_{p}-1}).
\end{equation}
For $x<x'$ and $j<l$, the integration domain $I_{jl}$ is defined by
$y_{1}<\cdots<y_{j-1}<x<y_{j}<\cdots<y_{l-1}<x'<y_{l}<\cdots<y_{N_{p}}$.
The spin contribution is
\begin{equation}\label{eq:omegajl}
\omega_{\alpha}^{j,l}=\langle\chi|\delta^{\alpha}_{\alpha_{j}}\hat{P}_{j,\ldots,l}|\chi\rangle,
\end{equation}
where $|\chi\rangle$ is the spin wavefunction and
$\delta_{\alpha_{j}}^{\alpha}$ projects the spin at site $j$ onto component $\alpha$. The loop-permutation operator $\hat{P}_{j,\ldots,l}$ acts as cyclic
\begin{equation}
j\rightarrow j+1\rightarrow\cdots\rightarrow l\rightarrow j
\end{equation}
for $j<l$, and as anti-cylic
\begin{equation}
j\rightarrow j-1\rightarrow\cdots\rightarrow l\rightarrow j
\end{equation}
for $j>l$, while $\hat{P}_{j,j}$ is the identity. By inserting  the necklace Ansatz expression (\ref{eq:spinwaveapp}) for the spin eigenstate into Eq.(\ref{eq:omegajl}), one readily obtains for the spin-down component
$\omega_{\downarrow}^{j,l}=e^{-i \frac{2 \pi n}{N_p}(l-j)}$.\\

\noindent Next, we express the anyonic correlator in an analogous form, making the correspondence between the impurity and anyons explicit. The anyonic correlator introduced in Eq.~\eqref{eq:anyonbodcorr} can be written as
\begin{align}
\rho^{\theta}_{N_{p}}(x,x')
=
N_p\int\mathrm{d}x_{2}\cdots\mathrm{d}x_{N_{p}}\,
\bar{\Psi}_{0}^{F}(x,x_{2},\ldots,x_{N_p})
\Psi_{0}^{F}(x',x_{2},\ldots,x_{N_p})
\bar{\mathcal{A}}_{\theta}(x,x_{2},\ldots,x_{N_p})
\mathcal{A}_{\theta}(x',x_{2},\ldots,x_{N_p}),
\end{align}
where 
$\mathcal{A}_{\theta}(x_1,x_{2},\ldots,x_{N_p})=\prod_{p<q}A_{\theta}(x_{p}-x_{q})$
is the anyonic mapping factor. This expression already separates the fermionic orbital contribution from the statistical phase. To make this separation explicit, we partition the integration domain into ordered sectors. Consider $x<x'$, and denote the ordered spectator coordinates (i.e. those which are integrated over) by
\(y_{1}<\cdots<y_{N_p-1}\). If \(x\) and \(x'\) occupy positions \(j\) and \(l\), respectively, in the ordered coordinate chain, then \(j\leq l\), and the corresponding sector is $y_{1}<\cdots<y_{j-1}<x<y_{j}<\cdots<y_{l-1}<x'<y_{l}<\cdots<y_{N_{p}}$.
Therefore, the number of particles between $x$ and $x'$ is $l-j$. Within this sector, all contributions to the two anyonic strings cancel except those associated with the $l-j$ particles between $x$ and $x'$, giving
\begin{equation}
\bar{\mathcal{A}}_{\theta}(x,\mathbf{y})
\mathcal{A}_{\theta}(x',\mathbf{y})
=
e^{-i\theta(l-j)},
\end{equation}
where $\mathbf{y}=(y_{1},\ldots,y_{N_p-1})$. Consequently, the anyonic correlator  becomes
\begin{equation}
\rho^{\theta}_{N_{p}}(x,x')
=
\sum_{1\leq j\leq l\leq N_{p}}
(\pm)^{(j+l)}\rho^{j,l}(x,x')\,e^{-i\theta(l-j)},
\qquad x<x'.
\end{equation}
Thus, the statistical factor $e^{-i\theta(l-j)}$ coincides exactly with the spin matrix element $\omega_{\downarrow}^{j,l}$ appearing in the corresponding decomposition for the mixture. The anyonic correlator has the same orbital contribution as the mixture, $(\pm1)^{j+l}\rho^{j,l}(x,x')$, where the positive and negative signs correspond to the Bose- and Fermi-anyon mappings, respectively. The sign accounts for the exchange symmetry of the reference wavefunction between ordered sectors. Moving the particle from position $j$ to position $l$ requires $l-j$ nearest-neighbour exchanges. For the fermionic reference wavefunction, this produces the factor $(-1)^{l-j}=(-1)^{j+l}$ whereas no additional sign is acquired for the bosonic reference wavefunction. For $x>x'$, the correlator follows from Hermiticity: $\rho^{\theta}_{N_{p}}(x,x')=\left[\rho^{\theta}_{N_{p}}(x',x)\right]^{*}$.

\subsection{Momentum distribution profiles}
\subsubsection{Comparison between mixtures and anyons}

\noindent In this section, we present the momentum distributions of the spin-up and spin-down components, calculated exactly for both bosonic and fermionic mixtures using Eq.~\eqref{eq:dereutz-supp}. We compare these results with the anyonic momentum distribution obtained from Eq.~\eqref{eq:anyonbodcorr}, allowing us to identify which component-resolved distributions coincide with their anyonic counterparts and where this correspondence no longer holds.
\begin{figure}[h!]
    \centering
    \includegraphics[width=0.95\linewidth]{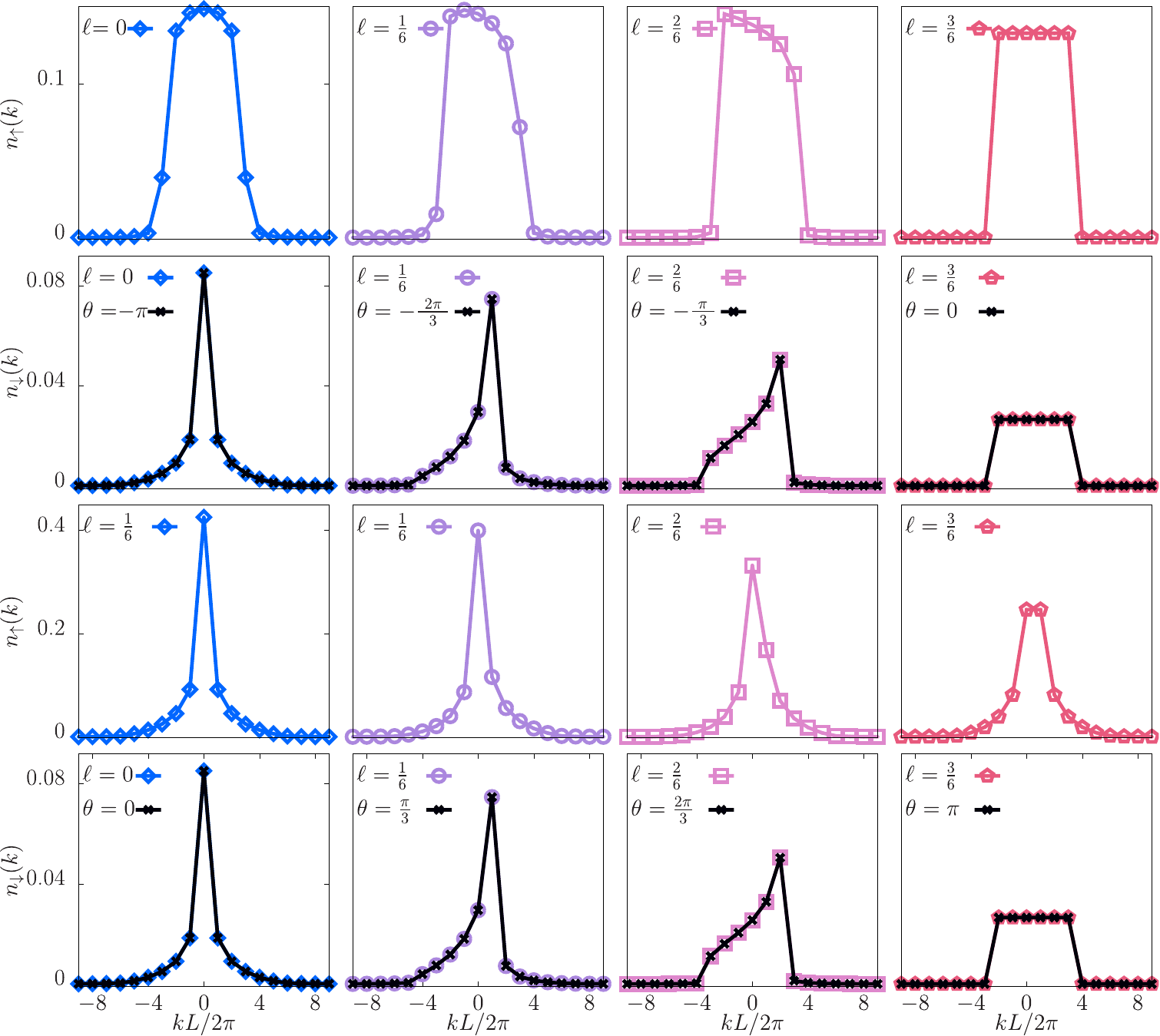}%
\caption{Anyonization of multi-component mixtures with an even particle number. Component-resolved momentum distributions $n_{\uparrow}(k)$ and $n_{\downarrow}(k)$, in units of $1/L$, for a mixture of $N_{\uparrow}=5$ spin-up and $N_{\downarrow}=1$ spin-down particles. The wavevector $k$ is expressed in units of $2\pi/L$. The top two rows correspond to fermionic mixtures and the bottom two to bosonic mixtures; within each pair, the upper and lower rows show $n_{\uparrow}(k)$ and $n_{\downarrow}(k)$, respectively. From left to right, the columns correspond to angular momenta per particle $\ell=0$, $1/6$, $2/6$, and $3/6$, calculated using Eq.~\eqref{eq:dereutz-supp}. The black curves show the corresponding anyonic momentum distributions, obtained via Eq.~\eqref{eq:anyonbodcorr} using the respective Fermi- or Bose-anyon mapping and the appropriate statistical angle $\theta$.}
    \label{fig:compar_odd}
\end{figure}

\begin{figure}[h!]
    \centering
    \includegraphics[width=0.85\linewidth]{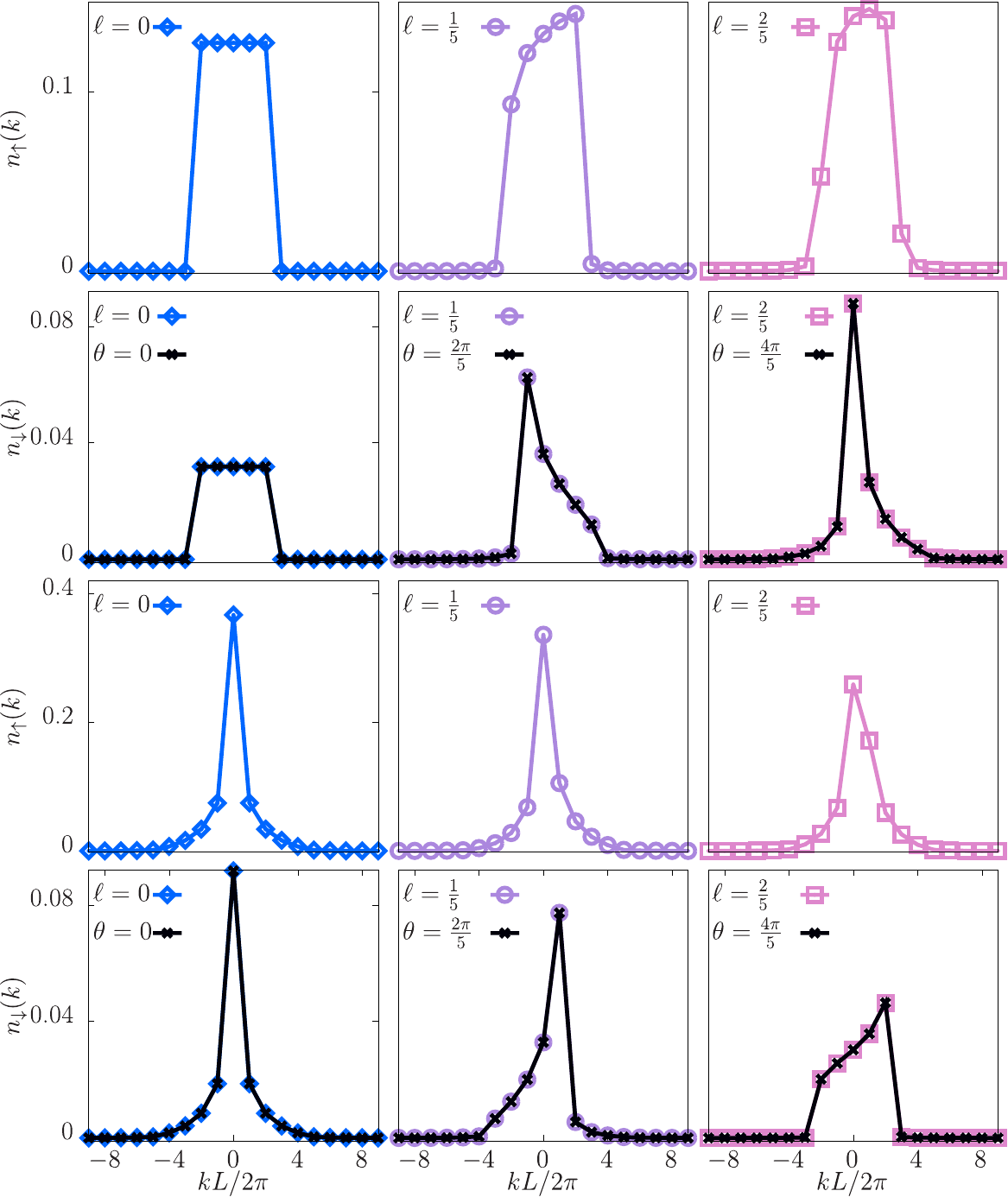}%
\caption{Anyonization of multi-component mixtures with an odd particle number. Component-resolved momentum distributions $n_{\uparrow}(k)$ and $n_{\downarrow}(k)$, in units of $1/L$, for a mixture of $N_{\uparrow}=4$ spin-up and $N_{\downarrow}=1$ spin-down particles. The wavevector $k$ is expressed in units of $2\pi/L$. The top two rows correspond to fermionic mixtures and the bottom two to bosonic mixtures; within each pair, the upper and lower rows show $n_{\uparrow}(k)$ and $n_{\downarrow}(k)$, respectively. From left to right, the columns correspond to angular momenta per particle $\ell=0$, $1/5$, $2/5$, and $3/5$, calculated using Eq.~\eqref{eq:dereutz-supp}. The black curves show the corresponding anyonic momentum distributions, obtained via Eq.~\eqref{eq:anyonbodcorr} using the respective Fermi- or Bose-anyon mapping and the appropriate statistical angle $\theta$.}
    \label{fig:compar_even}
\end{figure}

\noindent First, we observe that the impurity momentum distribution coincides exactly with that of the corresponding anyonic system for both odd and even particle numbers --Figs.~\ref{fig:compar_odd}and~\ref{fig:compar_even}. Comparing the bosonic and fermionic mixtures directly, however, reveals a parity-dependent distinction: for even $N_p$, the impurity momentum distributions of bosonic and fermionic mixtures are identical, whereas for odd $N_p$ they remain different.  Only for even $N_{p}$, do the two constructions become indistinguishable at the level of the $n_{\downarrow}(k)$. We refer to the exact correspondence between the bosonic and fermionic impurity distributions in the even case as \textit{complete statistical transmutation}, which will be discussed in detail below. By contrast, the momentum distributions of the spin-up background particles do not coincide between the bosonic and fermionic mixtures.

\subsubsection{Anyonization at finite interactions and lattice fillings}

\begin{figure}[h!]
    \centering
    \includegraphics[width=\linewidth]{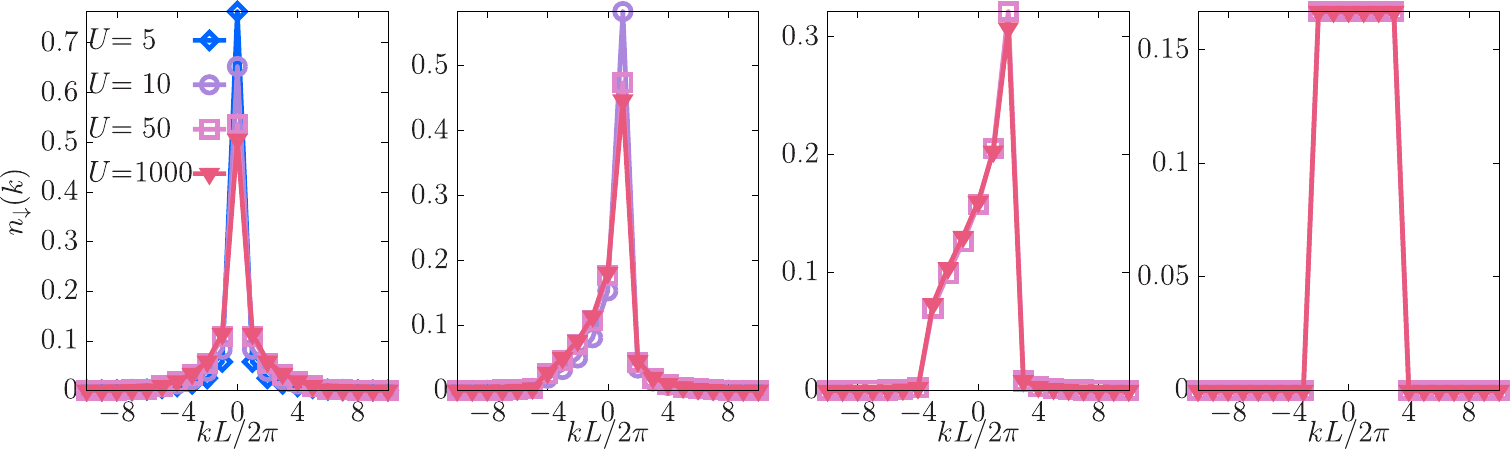}%
    \put(-395,140){(\textbf{a})}
    \put(-269,140){(\textbf{b})}
    \put(-144,140){(\textbf{c})}
    \put(-17,140){(\textbf{d})}
\caption{Anyonization at finite interactions. Impurity momentum distribution $n_{\downarrow}(k)$ as a function of the wavevector $k$, in units of $1/L$ and $2\pi/L$ respectively, for $N_{p}=6$ two-component fermions, on a ring of $N_{s}=20$ sites. From left to right, the panels correspond to angular momenta per particle $\ell=0$, $1/6$, $2/6$, and $3/6$, respectively, for different interaction strengths $U/J$. Curves are omitted when the fractionalized branch associated with a given $\ell$ has not yet become the ground-state at that interaction strength. The results were obtained by exact diagonalization of the Fermi-Hubbard model, with the interaction expressed in units of the hopping amplitude $J$.}
    \label{fig:mominter}
\end{figure}

\noindent The comparison presented so far between the impurity momentum distribution and that of hard-core anyons has focused on the strongly interacting regime, where the correspondence becomes exact. At weak and intermediate interactions, the two distributions no longer coincide quantitatively; nevertheless, their overall profiles already display the characteristic bosonized, fermionized, or anyonized behaviour --see Fig.~\ref{fig:mominter}. The essential requirement is the presence of crossings between ground- and excited-state branches, which gives rise to angular-momentum fractionalization and allows the different statistical regimes to be accessed. The interaction strength at which fractionalization first emerges is not universal, but depends on the system parameters, particularly the particle and site numbers denoted by $N_{p}$ and $N_{s}$ respectively~\cite{chetcuti2022persistent}. As the repulsive interaction is increased, the momentum distribution broadens and progressively approaches that of a Tonks-Girardeau gas, spinless fermions, or hard-core anyons, depending on the statistical character of the selected branch.

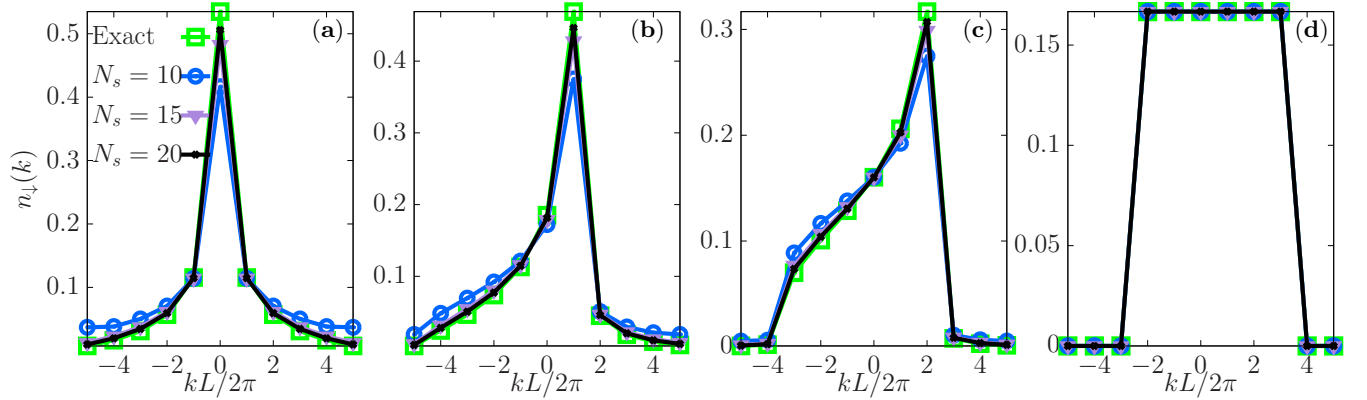
\begin{figure}[h!]
    \centering
    \includegraphics[width=\linewidth]{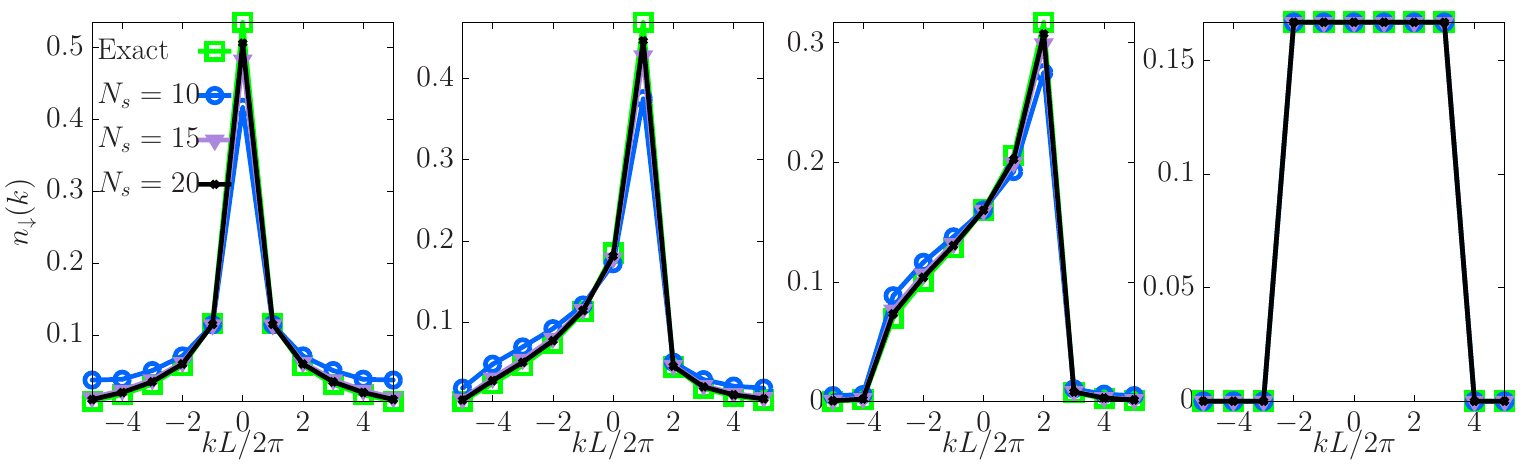}%
    \put(-395,140){(\textbf{a})}
    \put(-275,140){(\textbf{b})}
    \put(-150,140){(\textbf{c})}
    \put(-25,140){(\textbf{d})}
\caption{Anyonization in lattice rings. Impurity momentum distribution $n_{\downarrow}(k)$ as a function of the wavevector $k$, in units of $1/L$ and $2\pi/L$ respectively,  for $N_p=6$ two-component fermions,  at strong repulsion $U/J=1000$. From left to right, the panels correspond to angular momenta per particle $\ell=0$, $1/6$, $2/6$, and $3/6$, respectively, for rings with different numbers of sites $N_s$. The lattice results are compared with the exact continuum predictions shown by the green squares. The numerical data were obtained by exact diagonalization of the Fermi-Hubbard model, with energies expressed in units of the hopping amplitude $J$.}
    \label{fig:mom_lattice}
\end{figure}
\noindent Although most of the results presented in this work are obtained in the continuum, the same behaviour also arises on a lattice. By describing the system with the SU(2) Fermi--Hubbard and SU(2) Bose-Hubbard models, we recover the same bosonized, fermionized, and anyonized regimes --see Fig.~\ref{fig:mom_lattice}. In the dilute limit, obtained by increasing the number of ring sites $N_s$ at fixed particle number $N_p$, the lattice results converge to and eventually overlap with the exact continuum results derived here. Away from this limit, lattice effects prevent a quantitative match, but the momentum-distribution profiles and their evolution across the different statistical regimes remain qualitatively the same.

\subsubsection{Large momentum tails and Tan's relation}

\noindent Momentum-distribution tails provide a direct probe of the anyonic phase acquired under particle exchange. The following derivation follows Ref.~\cite{musolino2024symmetry}, where these arguments are presented in greater detail. After integrating out the orbital degrees of freedom, the spin-resolved  momentum distribution can be written as $n_\alpha(k)=\langle\chi|\hat{n}_{\alpha }(k)|\chi\rangle$,
where $|\chi\rangle$ is the spin state and
\begin{equation}
\hat{n}_\alpha(k)
=
\sum_{j,l=1}^{N_p} \delta_\alpha^j
\hat{P}_{j,\ldots,l}R^{(j,l)}(k),
\label{eq:hatn}
\end{equation}
with $\hat{P}_{j,\ldots, l}$ denoting the aforementioned cyclic  (anti-cyclic) permutation operator for $j<l$ ($j>l$) respectively. The orbital kernel is given by
\begin{align}
R^{(j,l)}(k)
=
\frac{N_{p} !}{2\pi}
\int\mathrm{d}x\,\mathrm{d}x'e^{-ik(x-x')}\int_{I_{jl}}
\left(\prod_{n\neq j}\mathrm{d}x_n\right)
\Psi_A^{*}
(x_1,\ldots,x_{j-1},x,x_{j+1},\ldots,x_{N_{p}})
\Psi_A
(x_1,\ldots,x_{j-1},x',x_{j+1},\ldots,x_{N_{p}}),
\end{align}
where, for $j<l$, the integration region is $I_{jl}=\left\{
x_{1}<\cdots<x_{j-1}<x<x_{j+1}<\cdots
<x_{l}<x'<x_{l+1}<\cdots<x_{N_{p}}
\right\}$.
Since the permutation operators are independent of $k$ and the sum contains finitely many terms,
\begin{equation}
\lim_{|k|\rightarrow\infty}k^4\hat{n}_{\alpha}(k)
=
\sum_{i,j=1}^{N_p} \delta_\alpha^j
\hat{P}_{i,\ldots,j}
\left[
\lim_{|k|\rightarrow\infty}k^4R^{(i,j)}(k)
\right].
\end{equation}

\noindent Although all orbital kernels $R^{(i,j)}(k)$ vanish as $|k|\rightarrow\infty$, the leading
terms decay as $k^{-4}$, so that $k^{4}R^{(i,j)}(k)$ remains finite, yielding the coefficient of the universal momentum-distribution tail,
\begin{equation}
n_{\alpha}(k)
\underset{|k|\rightarrow\infty}{\sim}
\frac{C_{T,\alpha}}{k^{4}},
\qquad
C_{T,\alpha}
=
\lim_{|k|\rightarrow\infty}
k^{4}n_{\alpha}(k),
\end{equation}
where $C_{T,\alpha}$ is Tan's contact for component $\alpha$. 
Only the diagonal kernels $R^{(j,j)}(k)$ and the nearest-neighbour kernels $R^{(j,j\pm1)}(k)$ contribute at order $k^{-4}$; all longer permutation cycles vanish more rapidly. The diagonal terms correspond to the identity, while the nearest-neighbour terms correspond to a single exchange. Consequently, Tan's contact for the impurity takes the form
\begin{equation}
C_{T,\downarrow}\!=\!\frac{m^{2}tg_{\uparrow\downarrow}}{2 \pi\hbar^{4}}
\sum_{j}^{N_{p}-1}\langle\chi|\delta_{\downarrow}^{j}(P_{j-1,j}\!+P_{j,j+1}\!+ 2 \mathbf{I})|\chi\rangle
\label{eq:tan-perm}
\end{equation}
where $t$ is the spin-exchange coupling and $g_{\uparrow\downarrow}$ is the inter-component interaction.  In the species-resolved impurity distribution, the permutation term describes the exchange of the spin-down particle with a neighbouring spin-up particle, thereby making the large-momentum tail sensitive to the anyonic exchange phase. \\

\noindent We evaluate the spin contribution entering the Tan contact of the impurity in Eq.(\ref{eq:tan-perm}) on the necklace state $|\chi_\theta\rangle= 1/\sqrt{N_{p}}\sum_l e^{-i \theta l}| l \rangle$, where $\theta=2 \pi n /N_p$ and
$| l \rangle$ is the single-spin-down basis 
\begin{equation}
    |l\rangle
    =
    |\uparrow_1\cdots\uparrow_{l-1}
    \downarrow_l
    \uparrow_{l+1}\cdots\uparrow_{N_p}\rangle ,
\end{equation}
with all site indices understood modulo $N_p$.
By introducing   $ u_j^\downarrow
=  \delta_{j,\downarrow}+\delta_{j+1,\downarrow}$
the Tan's contact in Eq.(\ref{eq:tan-perm}) reads
\begin{equation}
C_{T,\downarrow}\!=\!\frac{m^{2}tg_{\uparrow\downarrow}}{2 \pi\hbar^{4}}
\sum_{j}^{N_{p}} \sum_{l,l'} e^{-i \theta (l-l')} \langle l' |u_j^\downarrow
        \left(P_{j,j+1}+\mathbf{ I}\right)|l\rangle.
\label{eq:tan-chi-l}
\end{equation}

\noindent We proceed in evaluating Eq.(\ref{eq:tan-chi-l}),
defining for short-hand notation
\begin{equation}
    \hat{\mathcal O}_{\downarrow}
    =
    \sum_{j=1}^{N_p}
    u_j^\downarrow
    \left(P_{j,j+1}+\mathbf{ I}\right).
\end{equation}
In the single-impurity sector, for a fixed bond $(j,j+1)$, there are
three cases:
\begin{enumerate}
    \item The down spin is neither at site $j$ nor at site $j+1$: $|\cdots\uparrow_j\uparrow_{j+1}\cdots\rangle$. Hence,
    \begin{equation}
        u_j^\downarrow
        \left(P_{j,j+1}+\mathbf{ I}\right)|l\rangle
        =
        0,
        \qquad
        l\neq j,\quad l\neq j+1 .
    \end{equation}

    \item The down spin is at site $j$: $|\cdots\downarrow_j\uparrow_{j+1}\cdots\rangle$.
    Therefore,
    \begin{equation}
        u_j^\downarrow
        \left(P_{j,j+1}+\mathbf{ I}\right)|j\rangle
        =
        |j+1\rangle+|j\rangle .
    \end{equation}

    \item The down spin is at site \(j+1\): $|\cdots\uparrow_j\downarrow_{j+1}\cdots\rangle$. Therefore,
    \begin{equation}
        u_{j}^\downarrow
        \left(P_{j,j+1}+\mathbf{ I}\right)|j+1\rangle
        =
        |j\rangle+|j+1\rangle .
    \end{equation}
\end{enumerate}
Combining these three cases,
\begin{equation}
u_j^\downarrow
\left(P_{j,j+1}+\mathbf{ I}\right)|l\rangle
=
\begin{cases}
    0,
    & l\neq j,\quad l\neq j+1, \\[2mm]
    |j\rangle+|j+1\rangle,
    & l=j, \\[2mm]
    |j\rangle+|j+1\rangle,
    & l=j+1 .
\end{cases}
\end{equation}
Equivalently, for a fixed impurity position $l$, only the two adjacent
bonds $j=l-1$ and $j=l$ contribute. Therefore,
\begin{equation}
    \hat{\mathcal O}_{\downarrow}|l\rangle
    =
    2|l\rangle+|l-1\rangle+|l+1\rangle,
    \label{{eq:tan-pj}}
\end{equation}
giving us that, 
\begin{align}
    \hat{\mathcal O}_{\downarrow}|\chi_\theta\rangle=\left(2+e^{i\theta}+e^{-i\theta}\right)|\chi_\theta\rangle=
    2\left[1+\cos(\theta)\right]|\chi_\theta\rangle .
\end{align}
As a result, $\langle\chi_\theta|
    \hat{\mathcal O}_{\downarrow}
    |\chi_\theta\rangle
    =
    2\left[1+\cos(\theta)\right]$.
leading to
\begin{equation}
C_{T,\downarrow}\!=\!\frac{m^{2}tg_{\uparrow\downarrow}}{\hbar^{4}\pi}   (2 +  e^{i \theta} + e^{-i\theta})= \frac{C_{0}}{2} (1 + \cos(\theta)),
\end{equation}
with $C_0$  the impurity Tan's contact at $\theta=0$,
which coincides with Eq.(10) of the main text.

\subsubsection{Dynamical anyonization}

\begin{figure}[ht!]
    \centering
    \includegraphics[width=\linewidth]{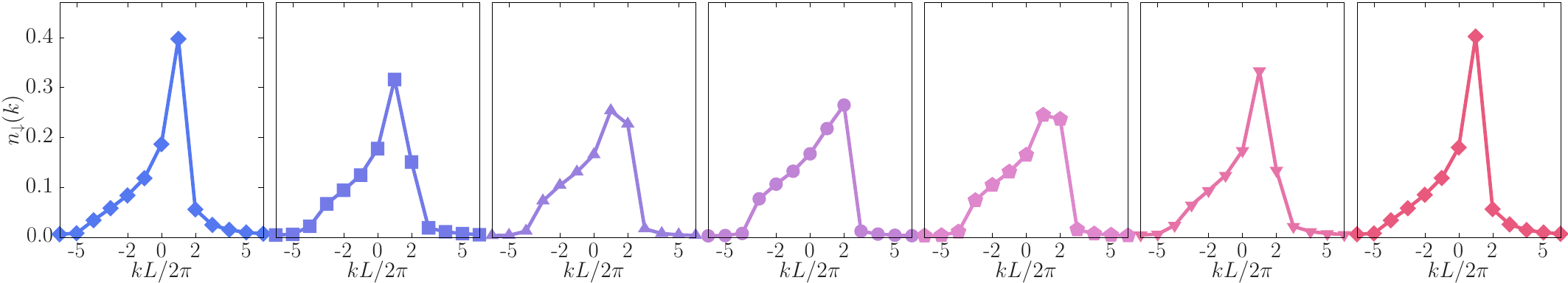}
    \put(-438,80){(\textbf{a})}
    \put(-488,80){$\frac{\tau J}{\hbar} = 0$}
    \put(-370,80){(\textbf{b})}
    \put(-418,80){$\frac{\tau J}{\hbar} = 141$}
    \put(-297,80){(\textbf{c})}
    \put(-350,80){$\frac{\tau J}{\hbar} = 211$}
    \put(-228,80){(\textbf{d})}
    \put(-280,80){$\frac{\tau J}{\hbar} = 291$}
    \put(-158,80){(\textbf{e})}
    \put(-210,80){$\frac{\tau J}{\hbar} = 370$}
    \put(-88,80){(\textbf{f})}
    \put(-138,80){$\frac{\tau J}{\hbar} = 470$}
    \put(-18,80){(\textbf{g})}
    \put(-70,80){$\frac{\tau J}{\hbar} \!=\! 600$}
    \caption{Dynamical anyonization from $\ell = 1/6$ to $\ell = 2/6$. Time evolved momentum distribution $n_{\downarrow}(k)$  as a function of the wavevector $k$, in units of $1/L$ and $2 \pi/L$ respectively, under the symmetry-breaking protocol, following a sudden quench of the flux from $\phi/\phi_{0}\!=\!1/6$ to $\phi/\phi_{0}\!=\!3/12$ at selected times $\tau$. During the dynamics, the momentum distribution oscillates from an anyonized one \textbf{(a)} with angular momentum $\ell \!=\! 1/6$  to another anyonic one \textbf{(d)}  with $\ell \!=\! 2/6$ , recovering its original anyonized profile in \textbf{(g)}. Intermediate times \textbf{(b)}, \textbf{(c)}, \textbf{(e)} and  \textbf{(f)} depict hybridized contributions from both states. Results obtained from exact diagonalization of the Fermi–Hubbard model with interactions $U/J$ = 1000, $N_{p}\!=\!6$ ,  on  $N_{s}\!=\!13$ sites, impurity barrier strength $\lambda_{\downarrow}/J \!=\! 0.1$ and hopping amplitude $J$.}
    \label{fig:dyna_var1}
\end{figure}

\noindent Dynamical anyonization is induced by suddenly quenching the flux to an avoided crossing opened by a localized barrier. In the main text, we considered a symmetry-breaking barrier and demonstrated the transition from the $\ell=0$ branch to the $\ell=1/6$ branch. Here, we present the two remaining cases, $\ell=1/6\leftrightarrow2/6$ and $\ell=2/6\leftrightarrow3/6$ --Figs.~\ref{fig:dyna_var1} and~\ref{fig:dyna_var2}. In each case, the impurity momentum distribution evolves from the profile associated with one anyonized branch to that of the other and subsequently returns to its initial form. This oscillatory evolution demonstrates coherent dynamical anyonization between states characterized by different exchange phases.\\

\noindent The same protocol also works with a symmetry-preserving barrier, as shown in Fig.~\ref{fig:symmdyna}. The system can again oscillate between the two momentum-distribution profiles, although the transfer occurs over longer times. The symmetry-preserving barrier has two main disadvantages. First, it can only couple parabolas belonging to the same spin-symmetry sector~\cite{polo2026static}. Second, a larger barrier strength is required to open a resolvable gap at the crossing. This difference follows from the form of the barrier terms. A symmetry-preserving barrier acts on the total local density,
\begin{equation}
\mathcal{H}_{b}^{\mathrm{SP}}=\lambda\left(n_{j_0,\uparrow}+n_{j_0,\downarrow}\right)=\lambda n_{j_0},
\end{equation}
and therefore perturbs only the charge sector. In the strongly interacting regime, the charge distribution is comparatively rigid, making the matrix element between the crossing branches small. \\
\begin{figure}[ht!]
    \centering
    \includegraphics[width=\linewidth]{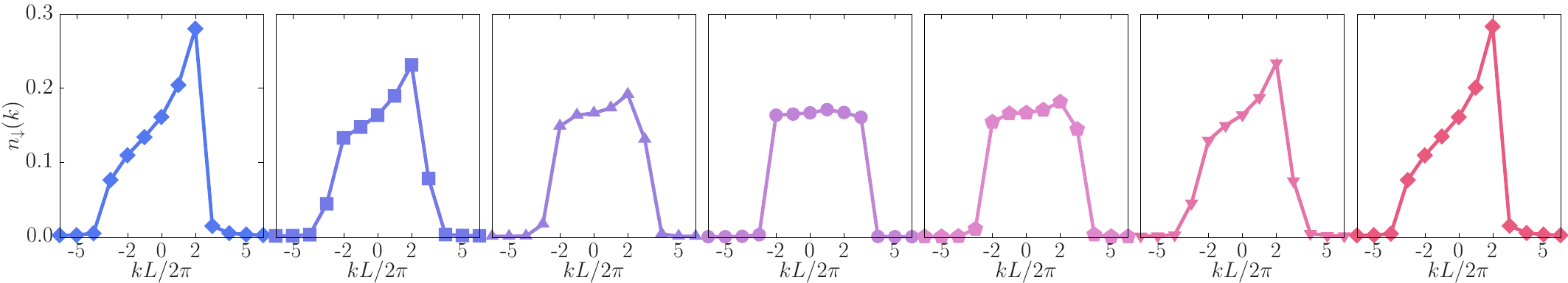}
    \put(-438,78){(\textbf{a})}
    \put(-488,78){$\frac{\tau J}{\hbar} = 0$}
    \put(-370,78){(\textbf{b})}
    \put(-418,78){$\frac{\tau J}{\hbar} = 141$}
    \put(-297,78){(\textbf{c})}
    \put(-350,78){$\frac{\tau J}{\hbar} = 211$}
    \put(-228,78){(\textbf{d})}
    \put(-280,78){$\frac{\tau J}{\hbar} = 291$}
    \put(-158,78){(\textbf{e})}
    \put(-210,78){$\frac{\tau J}{\hbar} = 370$}
    \put(-88,78){(\textbf{f})}
    \put(-138,78){$\frac{\tau J}{\hbar} = 470$}
    \put(-18,78){(\textbf{g})}
    \put(-70,78){$\frac{\tau J}{\hbar} \!=\! 600$}
    \caption{Dynamical anyonization from $\ell = 2/6$ to $\ell = 3/6$. Time evolved momentum distribution $n_{\downarrow}(k)$  as a function of the wavevector $k$, in units of $1/L$ and $2 \pi/L$ respectively, under the symmetry-breaking protocol, following a sudden quench of the flux from $\phi/\phi_{0}\!=\!2/6$ to $\phi/\phi_{0}\!=\!5/12$ at given times $\tau$. During the dynamics, the momentum distribution oscillates from an anyonized one \textbf{(a)} with angular momentum $\ell \!=\! 2/6$  to the fermionized one \textbf{(d)}  with $\ell \!=\! 3/6$ ,  its anyonized profile in \textbf{(g)}. Intermediate times \textbf{(b)},\textbf{(c)}, \textbf{(e)} and \textbf{(f)} depict hybridized contributions from both states. Results obtained from exact diagonalization of the Fermi–Hubbard model with interactions $U/J$ = 1000, $N_{p}\!=\!6$ ,  on  $N_{s}\!=\!13$ sites, impurity barrier strength $\lambda_{\downarrow}/J \!=\! 0.1$ and hopping amplitude $J$.}
    \label{fig:dyna_var2}
\end{figure}
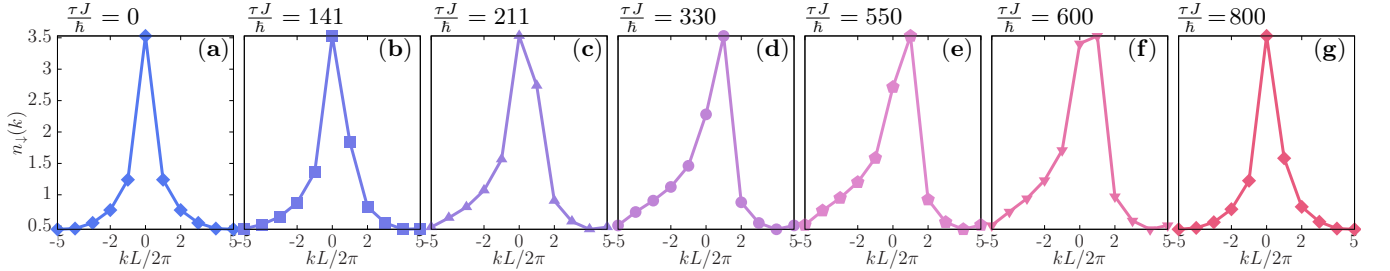
\begin{figure}[ht!]
    \centering
    \includegraphics[width=\linewidth]{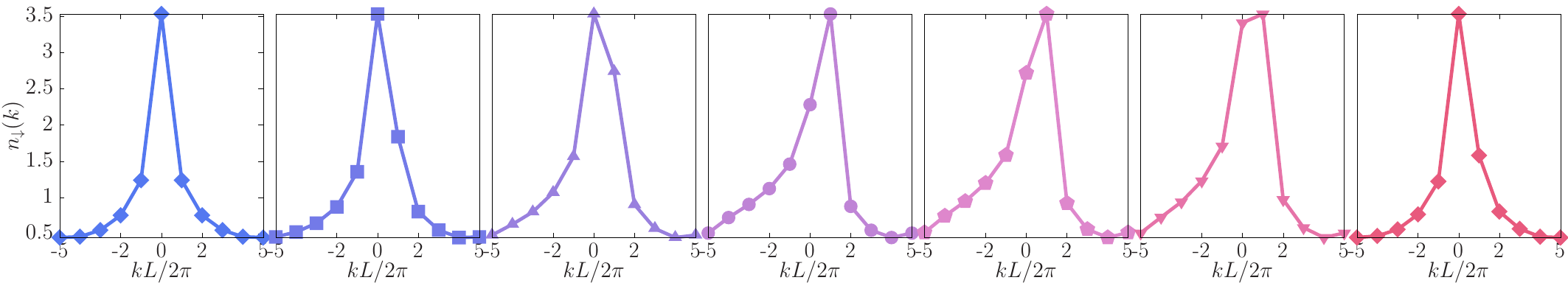}
    \put(-438,80){(\textbf{a})}
    \put(-488,92){$\frac{\tau J}{\hbar} = 0$}
    \put(-370,80){(\textbf{b})}
    \put(-420,92){$\frac{\tau J}{\hbar} = 141$}
    \put(-297,80){(\textbf{c})}
    \put(-350,92){$\frac{\tau J}{\hbar} = 211$}
    \put(-228,80){(\textbf{d})}
    \put(-280,92){$\frac{\tau J}{\hbar} = 330$}
    \put(-158,80){(\textbf{e})}
    \put(-210,92){$\frac{\tau J}{\hbar} = 550$}
    \put(-88,80){(\textbf{f})}
    \put(-138,92){$\frac{\tau J}{\hbar} = 600$}
    \put(-18,80){(\textbf{g})}
    \put(-70,92){$\frac{\tau J}{\hbar} \!=\! 800$}
    \caption{Dynamical anyonization from $\ell = 0$ to $\ell = 1/6$ with a symmetry-preserving barrier. Time evolved momentum distribution $n_{\downarrow}(k)$  as a function of the wavevector $k$, in units of $1/L$ and $2 \pi/L$ respectively, under the symmetry-breaking protocol, following a sudden quench of the flux from $\phi/\phi_{0}\!=\!0$ to $\phi/\phi_{0}\!=\! 1/12$ at given times $\tau$. During the dynamics, the momentum distribution oscillates from the bosonized one \textbf{(a)} with angular momentum $\ell \!=\! 0$  to the anyonized one \textbf{(d)}  with $\ell \!=\! 1/6$,  its original bosonized profile in \textbf{(g)}. Intermediate times \textbf{(b)},\textbf{(c)}, \textbf{(e)} and \textbf{(f)} depict hybridized contributions from both states.  Results obtained from exact diagonalization of the Fermi–Hubbard model with interactions $U/J$ = 1000, $N_{p}\!=\!6$ ,  on  $N_{s}\!=\!10$ sites, impurity barrier strength $\lambda_{\downarrow}/J=\lambda_{\uparrow}/J \!=\! 5$ and hopping amplitude $J$.}
    \label{fig:symmdyna}
\end{figure}

\noindent By contrast, a barrier acting on only one component  can be written as
\begin{equation}
\mathcal{H}_{b}^{\mathrm{SB}}=\lambda n_{j_0,\uparrow}=\frac{\lambda}{2}n_{j_0}+\lambda S_{j_0}^{z},
\end{equation}
by recalling the definition of the spin-z component $ S_j^{z}=(n_{j,\uparrow} - n_{j,\downarrow})/2$.
The local spin term breaks the SU(2) symmetry and acts directly on the spin configurations, allowing states from different spin-symmetry sectors to hybridize. It therefore opens a larger avoided-crossing gap at weaker barrier strength, leading to faster oscillations between the corresponding anyonized states.

\subsection{Conditions for true anyonization}

\subsubsection{Ring geometry}

\noindent Exact anyonization relies on the periodic geometry of the ring, which endows the spin sector with cyclic translational symmetry. The necklace states are eigenstates of the full cyclic permutation operator $P_{1\rightarrow N_{p}}$ and can be chosen as simultaneous eigenstates of the effective spin Hamiltonian only when $[H_{\mathrm{spin}},P_{1\rightarrow N_p}]=0.$ Here, we show that this condition is satisfied by the periodic spin chain but not by its open counterpart. \\

\noindent Let $\mathcal{T}\equiv P_{1\rightarrow N_{p}}$ denote the cyclic permutation operator, whose action on a general spin configuration is
\begin{equation}
\mathcal{T}|\sigma_1,\sigma_2,\ldots,\sigma_{N_p}\rangle = |\sigma_{N_p},\sigma_1,\ldots,\sigma_{N_p-1}\rangle .
\end{equation}
It shifts every spin by one position, $j\rightarrow j+1(mod \hspace{0.3mm} N_{p})$ and can be expressed as $\mathcal{T}=P_{1,2}P_{2,3}\cdots P_{N_{p}-1,N_{p}}$, where the rightmost operator acts first. In the single-impurity basis $|j\rangle=|\uparrow\cdots\uparrow\downarrow_j\uparrow\cdots\uparrow\rangle$, its action reduces to
\begin{equation}
\mathcal{T}|j\rangle=|j+1\rangle,\qquad |N_p+1\rangle\equiv|1\rangle,
\end{equation}
and hence $\mathcal{T}=\sum_{j=1}^{N_{p}}|j+1\rangle\langle j|$. For a ring of $N_{p}$ spins, the periodic permutation Hamiltonian is
\begin{equation}
H_{\mathrm{ring}} =t\sum_{j=1}^{N_{p}}P_{j,j+1}, \qquad P_{N_{p},N_{p}+1}\equiv P_{N_{p},1},
\end{equation}
where $t$ is the spin-exchange coupling. Under cyclic translation, a nearest-neighbour permutation transforms as $\mathcal{T} P_{j,j+1}\mathcal{T}^{-1}=P_{j+1,j+2}$,
with all indices understood modulo $N_{p}$. Therefore, the commutator  becomes
\begin{align}
[H_{\mathrm{ring}},\mathcal{T}] =t
\left(
\sum_{j=1}^{N_{p}}P_{j,j+1} - \sum_{j=1}^{N_{p}}P_{j+1,j+2} \right)\mathcal{T}.
\end{align}
Periodicity makes the two sums identical under a re-labelling of the index, yielding $[H_{\mathrm{ring}},\mathcal{T}]=0$. So, the cyclic quantum number $n$ is conserved, allowing the spin eigenstates to assume the necklace spin-wave form. \\

\noindent For an open chain, $H_{\mathrm{open}}=t\sum_{j=1}^{N_{p}-1}P_{j,j+1}$,
the corresponding commutator is
\begin{align}
[H_{\mathrm{open}},\mathcal{T}] = 
t\left(
\sum_{j=1}^{N_{p}-1}P_{j,j+1}
\sum_{j=1}^{N_{p}-1}P_{j+1,j+2}
\right)\mathcal T = t\left(P_{1,2}-P_{N_p,1}\right)\mathcal{T}.
\end{align}
The boundary terms break cyclic translational symmetry, so the spin eigenstates can no longer be labelled by a conserved cyclic quantum number. This prevents the uniform phase progression between successive impurity positions required for the exact anyonic correspondence.

\subsubsection{Artificial gauge field}

\noindent In our setup, the flux threading the ring couples identically to the spin-up and spin-down particles, a necessary requirement for exact anyonization. If the flux acts on only one component, or has different strengths for the two, the system may still exhibit anyonic signatures, but the correspondence with a genuine anyonic wavefunction is no longer exact. To clarify this distinction, we consider the Bethe Ansatz solution of a two-component fermionic mixture~\cite{shastry1990twisted} subject to component-dependent fluxes $\phi_{\uparrow}$ and $\phi_{\downarrow}$.
\begin{figure}[h!]
    \centering
    \includegraphics[width=\linewidth]{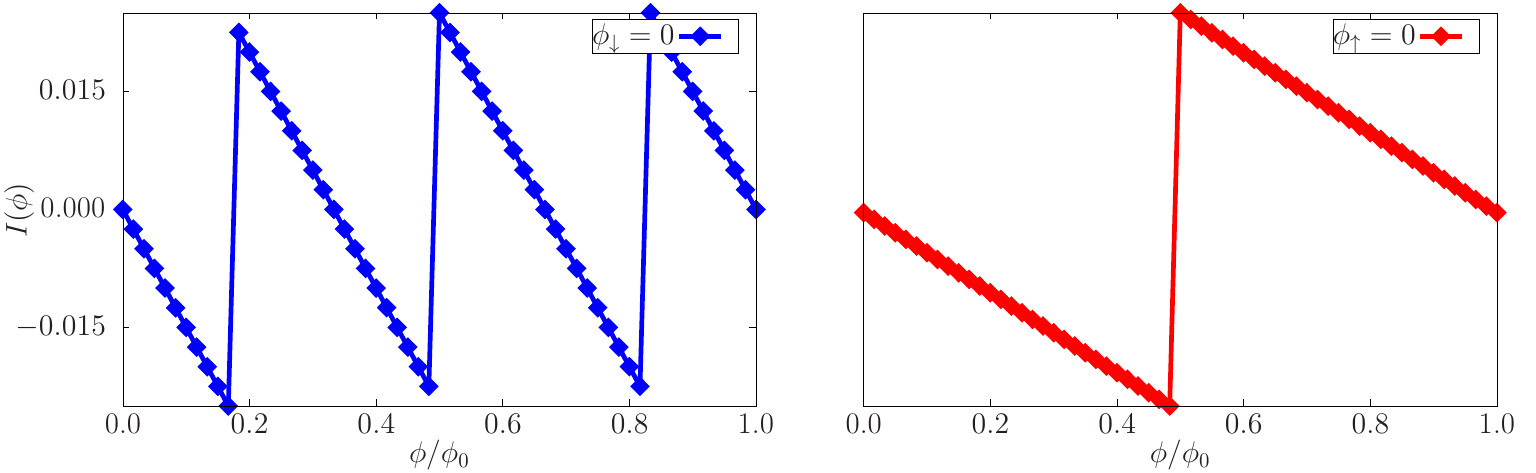}
    \put(-465,140){(\textbf{a})}
    \put(-220,140){(\textbf{b})}
\caption{Persistent current under a species-selective flux. Persistent current $I(\phi)$ as a function of the artificial flux $\phi/\phi_0$ for $N_p={4}$ two-component fermions with a single spin-down impurity, where $\phi_0$ is the bare flux quantum. In panel \textbf{(a)}, the flux acts only on the spin-up particles, $\phi_{\downarrow}=0$, producing $N_{p}-1$ fractionalized current peaks within one bare flux period. In panel \textbf{(b)}, only the spin-down impurity experiences the flux, $\phi_{\uparrow}=0$, and the current retains the usual period $\phi_{0}$. The results were obtained by exact diagonalization of the Fermi-Hubbard model on a ring of $N_{s}=15$ sites at interaction strength $U/J=1000$, where $J$ is the hopping amplitude.}
    \label{fig:diffflux}
\end{figure}
\noindent  In the limit of infinite repulsion, the charge quasimomenta take the form
\begin{equation}
k_{j}=\frac{1}{L}\left[2\pi I_{j}+\frac{2\pi\sum_{\alpha=1}^{M}J_{\alpha}
+N_{\uparrow}\phi_{\uparrow}+N_{\downarrow}\phi_{\downarrow}}{N_{p}}\right],
\end{equation}
where $N_{p}=N_{\uparrow}+N_{\downarrow}$, $M=N_{\downarrow}$, and $\{I_j\}$ and $\{J_\alpha\}$ are the charge and spin quantum numbers, respectively. This expression shows that the flux contribution to the charge sector depends on the number of particles experiencing each flux. In the single-impurity limit, we have that $N_{\downarrow}=1$ and $N_{\uparrow}=N_{p}-1$. If only the spin-down impurity experiences the flux, its contribution to each quasimomentum is $\phi_{\downarrow}/N_{p}$. A change $\sum_{\alpha}J_{\alpha}\rightarrow\sum_{\alpha}J_{\alpha}-1$ therefore compensates a flux variation $\Delta\phi_{\downarrow}=2\pi$, so the usual flux period is not reduced --Fig.~\ref{fig:diffflux}\textbf{(a)}. If instead only the spin-up background feels the flux, the corresponding contribution is $\frac{(N_{p}-1)\phi_{\uparrow}}{N_{p}}$, and the same change in the spin quantum numbers compensates a smaller flux variation, $\Delta\phi_{\uparrow}=\frac{2\pi}{N_{p}-1}$. This produces a reduced degree of fractionalization compared with the case of a common flux, $\phi_{\uparrow}=\phi_{\downarrow}\equiv\phi$, for which the flux contribution is simply $\phi$ and the compensating variation is $\Delta\phi=\frac{2\pi}{N_{p}}$ --Fig.~\ref{fig:diffflux}\textbf{(b)}. \\

\begin{figure}[h!]
    \centering
    \includegraphics[width=\linewidth]{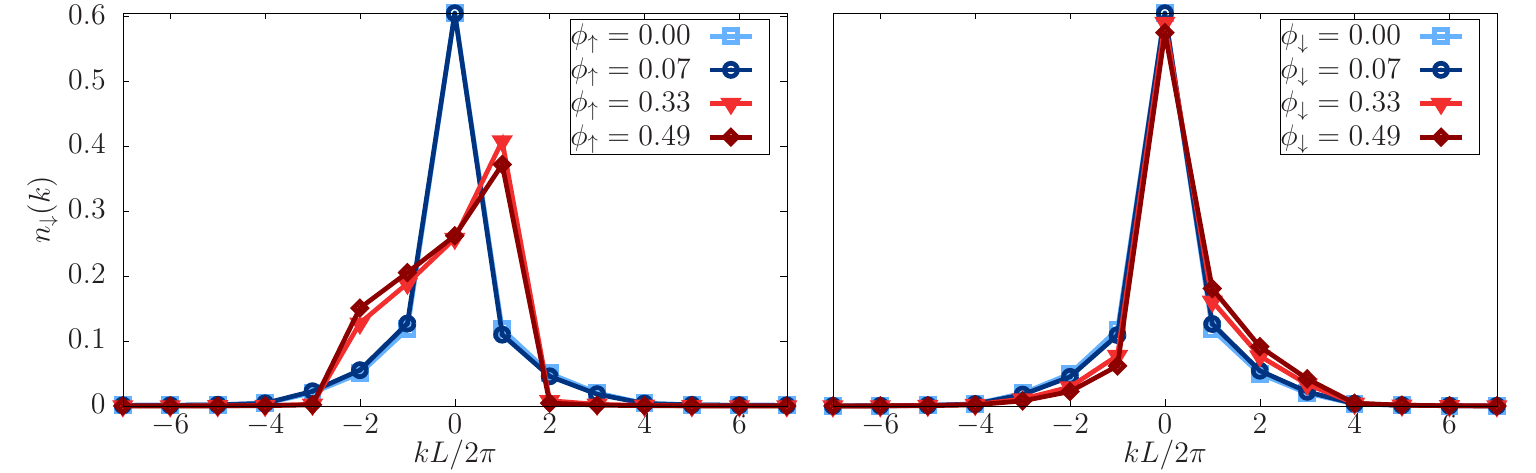}
    \put(-465,140){(\textbf{a})}
    \put(-220,140){(\textbf{b})}
\caption{Impurity momentum distribution under a species-selective flux. Momentum distribution of the impurity $n_{\downarrow}(k)$, units of $1/L$, as a function of the wavevectors $k$ in units of $2\pi/L$ for $N_{p}=4$ two-component fermions with a single spin-down impurity, where $\phi_0$ is the bare flux quantum. The distributions are shown for several flux values acting only on \textbf{(a)} the spin-up background, and \textbf{(b)} the spin-down impurity. The chosen fluxes correspond to the different angular momentum parabolas selected when both components experience the same flux. Panel \textbf{(a)} displays two qualitatively distinct families of profiles, with bosonized and anyonized character, whereas panel \textbf{(b)} exhibits only one. Shades of the same color identify flux values associated with the same angular momentum in the $\phi_{\uparrow}\neq 0$ case. Unlike the common-flux case, however, these profiles are not fixed by the discrete angular momentum alone but evolve continuously with the applied flux. The results were obtained by exact diagonalization of the Fermi-Hubbard model on a ring of $N_{s}=15$ sites at $U/J=1000$, where $J$ is the hopping amplitude.}
    \label{fig:diffflux2}
\end{figure}

\noindent However, the charge equation alone does not reveal how a species-dependent flux modifies the spin wavefunction. Such behaviour can be gleaned from the nested Bethe equation for the spin rapidities. In the infinitely repulsive limit, the latter reduces to the Bethe equation of a twisted XXX spin chain,
\begin{equation}
e^{i\delta\phi}\left(\frac{\Lambda_{\alpha}+i/2}{\Lambda_{\alpha}-i/2}\right)^{N_{p}}=\prod_{\substack{\beta=1\\\beta\neq\alpha}}^{M}\frac{\Lambda_{\alpha}-\Lambda_{\beta}+i}{\Lambda_{\alpha}-\Lambda_{\beta}-i},
\qquad\delta\phi\equiv\phi_{\downarrow}-\phi_{\uparrow},
\end{equation}
where $\Lambda_{\alpha}$ are the rescaled spin rapidities. The relative twist $e^{i\delta\phi}$ arises because transporting a spin-down particle around the ring produces the phase $e^{i\phi_{\downarrow}}$, whereas the spin-up reference background acquires $e^{i\phi_{\uparrow}}$. The spin sector therefore depends on the ratio of these phases, $e^{i(\phi_{\downarrow}-\phi_{\uparrow})}$, rather than on their common part. In the single-impurity limit, $M=1$, the product on the right-hand side is absent. Defining the spin-wave momentum $q$ through $e^{iq}=\frac{\Lambda+i/2}{\Lambda-i/2}$, one obtains
\begin{equation}
    q=\frac{2\pi n-\delta\phi}{N_{p}},
\end{equation}
so that the spin sector amplitudes satisfy $a_j\propto e^{-iqj}$. When both components experience the same flux, $\delta\phi=0$, the exchange phases are restricted to the discrete necklace values $q=2\pi n/N_{p}$; varying the common flux can therefore only select among different integer values of $n$. By contrast, when the two species experience different fluxes, $\delta\phi\neq0$ and $q$ varies continuously even for a fixed $n$. The statistical phase of the impurity wavefunction consequently evolves continuously, explaining why a component-dependent flux produces continuous, rather than discrete anyonization --Fig.~\ref{fig:diffflux2}\textbf{(a)} and \textbf{(b)}.

\subsubsection{Statistical transmutation: odd versus even particle numbers}

\noindent As discussed previously, the correspondence between the impurity momentum distributions of bosonic and fermionic mixtures depends on the parity of the particle number. For even $N_{p}$, exact statistical transmutation occurs: the spin-down impurity exhibits the same observables in both mixtures, independently of the underlying bosonic or fermionic statistics. For odd $N_{p}$, the transmutation is only partial. Although the impurity in each mixture still matches its corresponding Bose- or Fermi-mapped anyonic system, the impurity observables of the bosonic and fermionic mixtures no longer coincide with one another. \\

\noindent First we note that, whereas anyons can be tuned continuously from the bosonized to the fermionized regime for any particle number, this is not always possible for the mixtures. Their fractionalized angular momenta take the discrete values $\ell=\frac{n}{N_{p}}$. For odd $N_{p}$, no integer $n$ gives $\ell=1/2$, and the spectrum therefore contains no exactly fermionized parabola. The system can evolve from the bosonized regime through intermediate anyonized states, but it cannot complete the transmutation into a fermionized state. The full boson$\rightarrow$anyon$\rightarrow$fermion sequence is therefore possible only for even particle numbers. \\

\noindent The same parity effect can be understood from the quasimomenta $k_{j}$ entering the orbital wavefunction. For odd $N_{p}$, the ground-state configurations of both bosonic and fermionic mixtures are formed from integer quantum numbers. However for even $N_{p}$, the bosonic configuration is built from half-odd integers, whereas the fermionic one remains integer valued. Consequently, the shift produced by the spin quantum numbers acts in opposite directions for the two statistics: the shift that takes the fermionic mixture from the fermionized to the bosonized regime takes the bosonic mixture through the reverse sequence. Therefore, this lines up with corresponding Bose- and Fermi-anyon mappings that are offset by a phase $\pi$.

\subsubsection{Single impurity}

\noindent Exact anyonization is restricted to the single-impurity limit because every spin configuration is then specified solely by the position of the impurity in the ordered particle chain. All configurations belong to a single necklace, and each exchange of the impurity with a background particle multiplies the spin amplitude by the same phase. The spin wavefunction therefore takes the simple form $a_{q,j}\propto e^{-iqj}$, allowing the phase $q$ to be identified directly with a unique anyonic statistical angle. With multiple impurities, the spin configurations also depend on their relative positions and are generally distributed among several distinct necklaces. Cyclic symmetry fixes the amplitudes within each necklace but does not relate the coefficients of different necklaces. Therefore, the full spin wavefunction can  no longer be described by a single exchange phase or mapped exactly onto an anyonic wavefunction with one statistical angle. Multi-impurity systems may still display signatures of anyonic behaviour, but they do not exhibit the genuine anyonization found in the single-impurity limit.

\subsection{Traces of anyonization in interference dynamics}

\noindent Our proposed implementation is based on ultracold atomic gases, where the angular momentum in flux-threaded ring geometries can be measured through spiral interferograms
(see e.g.  ~\cite{amico2022colloquium,polo2025persistent} and references therein).  Two protocols are commonly used. In the homodyne protocol, the atomic cloud undergoes time-of-flight expansion and the resulting spatial density distribution is imaged. In the self-heterodyne protocol, the expanding ring-shaped gas interferes with a non-rotating gas placed at its center, which acts as a phase reference and reveals the phase profile of the system. Since the interference signals are directly connected to the one-body correlator and the corresponding momentum distribution, we expect anyonization to leave characteristic signatures in both protocols~\cite{chetcuti2023interference,chetcuti2025interferometric}. In particular, changes induced by the anyonic exchange phase in the component-resolved momentum distributions are inherited by the resulting interferograms.\\

\noindent \textit{Homodyne protocol:} This corresponds to the standard time-of-flight procedure: the gas is released from the confining potential at $t=0$ and allowed to expand freely. At sufficiently long expansion times, the measured spatial distribution maps onto the initial momentum distribution through $\mathbf{k}=m\mathbf{r}/(\hbar t)$. The total time-of-flight distribution is then
\begin{equation}\label{eq:momdist}
n_{\mathrm{TOF}}(\mathbf{k})=|w(\mathbf{k})|^{2}\sum_{\alpha=1}^{N}\sum_{j,l=1}^{N_s}
e^{i\mathbf{k}\cdot(\mathbf{r}_{l}-\mathbf{r}_{j})}\left\langle c_{l,\alpha}^{\dagger}c_{j,\alpha}\right\rangle,
\end{equation}
where $w(\mathbf{k})$ is the Fourier transform of Wannier function, $\mathbf{r}_{j}$ denotes the position of lattice site $j$ in the plane of the ring, and $c_{j,\alpha}^{\dagger}$ creates a particle of component $\alpha$ at that site. Therefore, the sum over $\alpha$  gives the TOF momentum distribution of the full mixture. \\
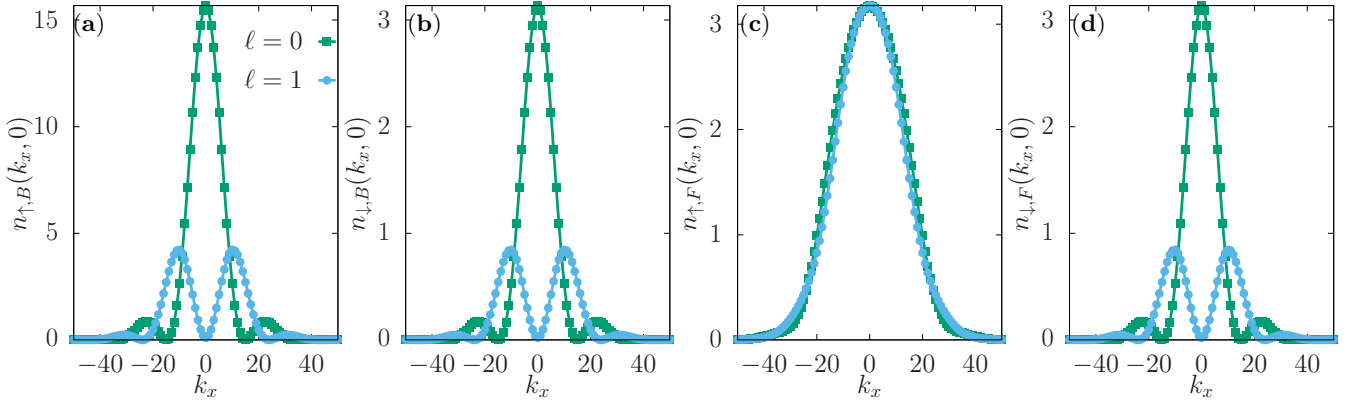
\begin{figure}[h!]
    \centering
    \includegraphics[width=\linewidth]{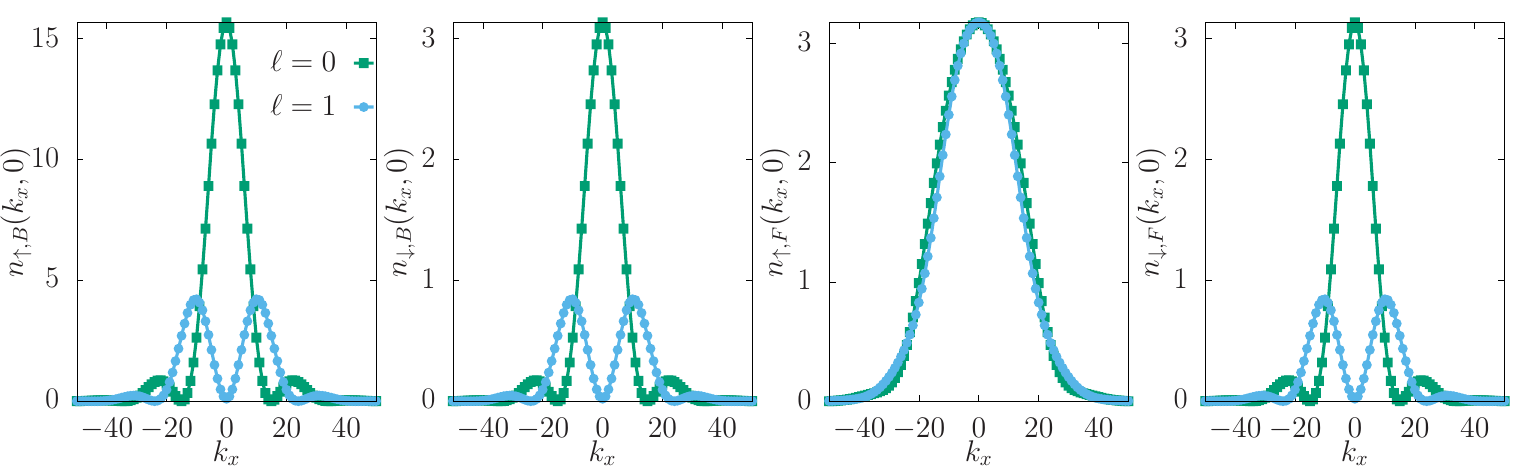}%
    \put(-485,140){(\textbf{a})}
    \put(-359,140){(\textbf{b})}
    \put(-234,140){(\textbf{c})}
    \put(-109,140){(\textbf{d})}
\caption{Cross section of the time-of-flight momentum distribution $n(k_x,0)$ for non-interacting bosonic (left pair) and fermionic (right pair) mixtures with $N_p=6$ particles, $N_{\uparrow}=5$ and $N_{\downarrow}=1$, at different angular momenta per particle $\ell$. Within each pair, the left and right sub-panels show the momentum distributions of the spin-up background and the spin-down impurity, respectively. The results were obtained by exact diagonalization of the corresponding Bose- and Fermi-Hubbard models on a ring of $N_{s}=20$ sites in the non-interacting regime.}
    \label{fig:TOFzero}
\end{figure}

\noindent A circulating state produces a central hole in the time-of-flight distribution, whose width increases with the angular momentum, or winding number, $\ell$~\cite{polo2025persistent}. For non-interacting particles, this behaviour can be understood from the occupation of the lattice-momentum modes $\eta$. A central hole appears when the zero-momentum mode $\eta=0$ is unoccupied. Since $\ell = \frac{1}{N_{p}}\sum_{\{\eta\}}\eta$, increasing $\ell$ shifts the momentum distribution away from the compact configuration centered around $\eta=0$ in the absence of flux. The appearance of this feature depends strongly on the particle statistics. For non-interacting bosons, all particles can occupy the same momentum mode, and a central hole therefore appears as soon as the system enters a current-carrying state with non-zero winding. By contrast, for fermions, the Pauli exclusion principle produces a Fermi sphere containing several momentum modes. Even though the systems acquires a current, $n_{\mathrm{TOF}}(\mathbf{k})$ remains peaked as long as the mode $\eta=0$ is occupied. Although this distribution can carry a finite current, the central peak remains present as long as the mode $\eta=0$ is occupied. A hole appears only after the Fermi sphere has been displaced by half, requiring a winding number of order $\ell=\lceil N_p/2\rceil$. Consequently, unlike in the bosonic case, the presence and size of the hole do not directly reflect the magnitude of the fermionic current. \\

\noindent This distinction is visible in the component-resolved distributions shown in Fig.~\ref{fig:TOFzero}. For the bosonic mixture, a central hole appears as soon as the system acquires angular momentum. In the fermionic mixture, however, the two components display different profiles. The spin-down impurity behaves similarly to the bosonic case, developing a hole immediately upon entering a current-carrying state. The spin-up background instead remains peaked at zero momentum over the range considered because its Fermi-sphere distribution continues to contain the $\eta=0$ mode.
\begin{figure}[h!]
    \centering
    \includegraphics[width=\linewidth]{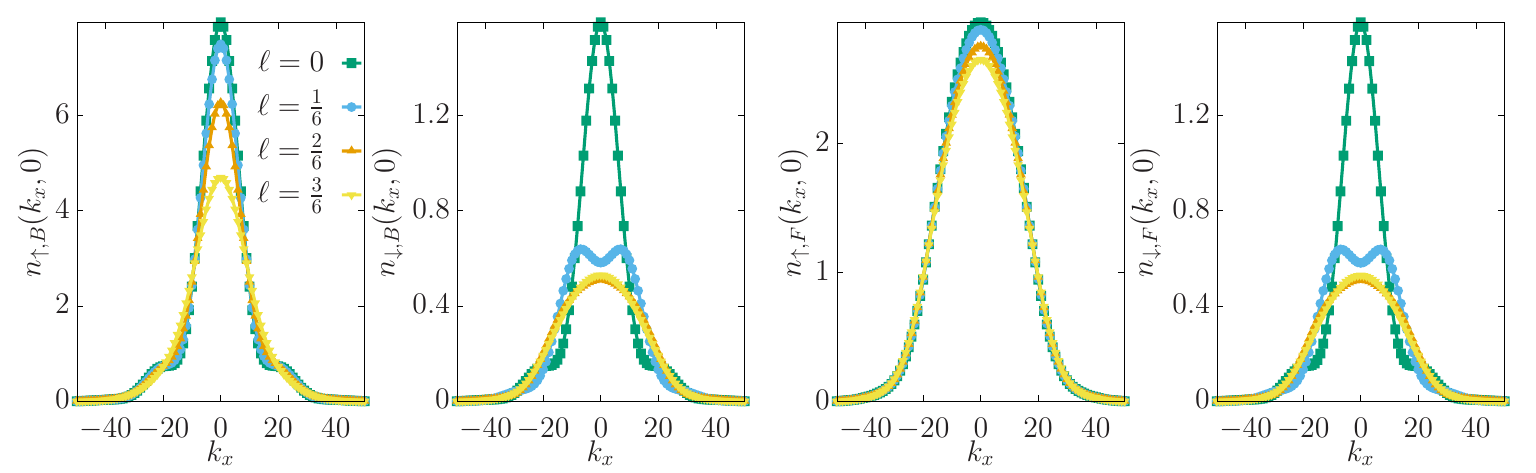}%
    \put(-485,140){(\textbf{a})}
    \put(-359,140){(\textbf{b})}
    \put(-230,140){(\textbf{c})}
    \put(-100,140){(\textbf{d})}
\caption{Cross section of the time-of-flight momentum distribution $n(k_x,0)$ for strongly interacting bosonic (left pair) and fermionic (right pair) mixtures with $N_p=6$ particles, $N_{\uparrow}=5$ and $N_{\downarrow}=1$, at different angular momenta per particle $\ell$. Within each pair, the left and right sub-panels show the momentum distributions of the spin-up background and the spin-down impurity, respectively. The results were obtained by exact diagonalization of the corresponding Bose- and Fermi-Hubbard models on a ring of $N_{s}=20$ sites subjected to interaction strengths $U/J=1000$ with $J$ being the hopping amplitude.}
    \label{fig:TOFinter}
\end{figure}
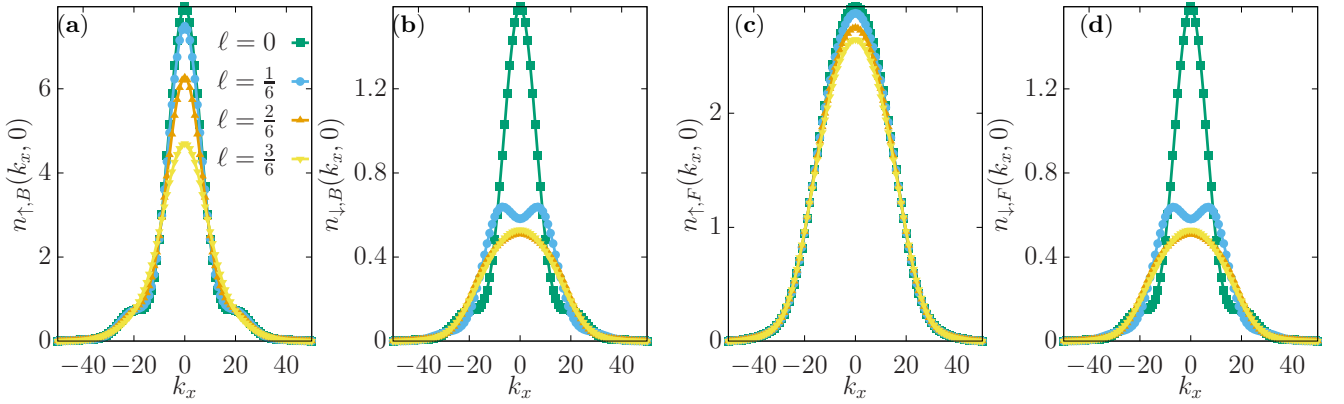

\noindent At strong repulsive interactions, the distinction between bosons and fermions becomes less pronounced, as the bosonic system develops fermion-like exclusion effects. In the presence of a flux, spin correlations lead to angular momentum fractionalization in both systems, with the angular momentum per particle changing in steps of $\frac{1}{N_{p}}$. Since each fractionalized state produces only a small displacement of the occupied momentum modes, the TOF distribution no longer develops a fully open central hole. Instead, a shallow depression appears at its center, and only after a larger flux has been applied~\cite{chetcuti2023interference}. Both features occur in the spin-up and spin-down distributions shown in Fig.~\ref{fig:TOFinter}. Nevertheless, a difference associated with the underlying particle statistics remains: the central peak decreases monotonically with flux for bosons, whereas its height alternates between successive fractionalized states for fermions. The clearest signature of anyonization appears in the spin-down distribution, which alternates between a shallow central depression and a fully peaked profile.\\

\noindent \textit{Self-heterodyne protocol:} The ring-shaped gas and a condensate placed at its center are released simultaneously and allowed to co-expand. The central condensate acts as a phase reference, and its interference with the expanding ring produces an interferogram in each experimental realization. Theoretically, this pattern is described through the density-density correlator
\begin{equation}
G(\mathbf{r},\mathbf{r}',t)=\sum_{\alpha,\beta=1}^{N}\left\langle n_{\alpha}(\mathbf{r},t)n_{\beta}(\mathbf{r}',t)\right\rangle,
\end{equation}
where $n_{\alpha}(\mathbf{r},t)=\psi_{\alpha}^{\dagger}(\mathbf{r},t) \psi_{\alpha}(\mathbf{r},t)$, $\psi_{\alpha}=\psi_{C,\alpha}+\psi_{R,\alpha}$ with
$\psi_{C,\alpha}$ and $\psi_{R,\alpha}$ being the field operators associated with the central condensate and the ring, respectively. The interference pattern is contained in the cross-terms:
\begin{equation}\label{eq:interference}
G_{R,C}(\mathbf{r},\mathbf{r}',t)=\sum_{\alpha=1}^{N}\sum_{j,l=1}^{N_{s}}
I_{j,l}(\mathbf{r},\mathbf{r}',t)\left[\delta_{jl}+\chi\left\langle
c_{l,\alpha}^{\dagger}c_{j,\alpha}\right\rangle\right],
\end{equation}
where $\chi=+1$ for bosons and $\chi=-1$ for fermions. The expansion kernel is
\begin{equation}
I_{j,l}(\mathbf{r},\mathbf{r}',t)=w_{C}(\mathbf{r}',t)w_{C}^{*}(\mathbf{r},t)w_{l}^{*}(\mathbf{r}',t)w_{j}(\mathbf{r},t),
\end{equation}
with $w_{C}(\mathbf{r},t)$ denoting the freely expanded wavefunction of the central condensate and $w_{j}(\mathbf{r},t)$ that of a particle initially localized at ring site $j$. The matrix element $\langle c_{l,\alpha}^{\dagger}c_{j,\alpha}\rangle$ is the one-body correlator of component $\alpha$ in the ring before release. 
\begin{figure}[h!]
    \centering
    \includegraphics[width=\linewidth]{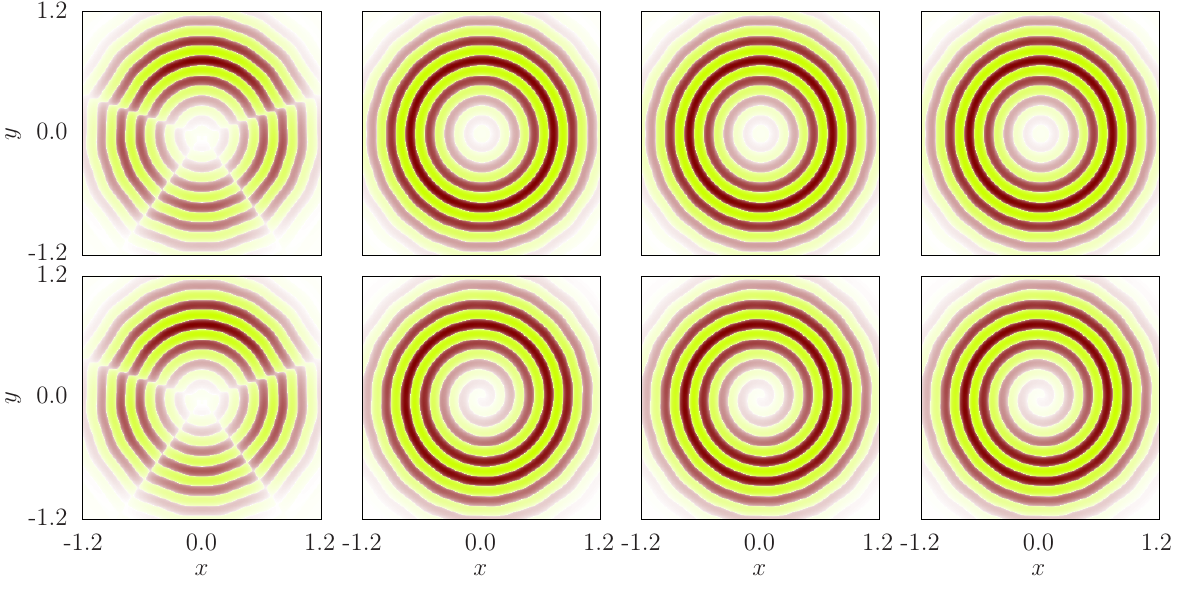}%
\caption{Real part of the cross-term correlator $\mathrm{Re}[G_{R,C}(\mathbf{x},\mathbf{x}';t)]$ for strongly repulsive fermionic (left block, first and second column) 
and bosonic (right block, third and fourth column) mixtures with $N_{p}=6$ particles, $N_{\uparrow}=5$ and $N_{\downarrow}=1$, on a ring of $N_{s}=20$ sites at interaction strength $U/J=0$. The top and bottom rows correspond to angular momenta $\ell=0$ and $\ell=1$, respectively. Within each block, the left and right 
panels show the interferograms of the spin-up background and the spin-down impurity, highlighting their distinct interference features. The results were obtained by exact diagonalization of the corresponding Fermi- and Bose-Hubbard models for hopping amplitude $J$. The Wannier functions are modeled as two-dimensional time-evolved Gaussians with ring radius $R=1.5$, initial width $\sigma=0.15$, and expansion time $\omega_0t=1.5$, while the reference point is fixed at $\mathbf{x}'=(R,0)$. To enhance the visibility of the interference features, the color scale represents $\operatorname{sgn}(Y)|Y|^{1/4}$, where $Y=\mathrm{Re}[G_{R,C}(\mathbf{x},\mathbf{x}';t)]$.}
    \label{fig:interzero}
\end{figure}

\begin{figure}[h!]
    \centering
    \includegraphics[width=0.95\linewidth]{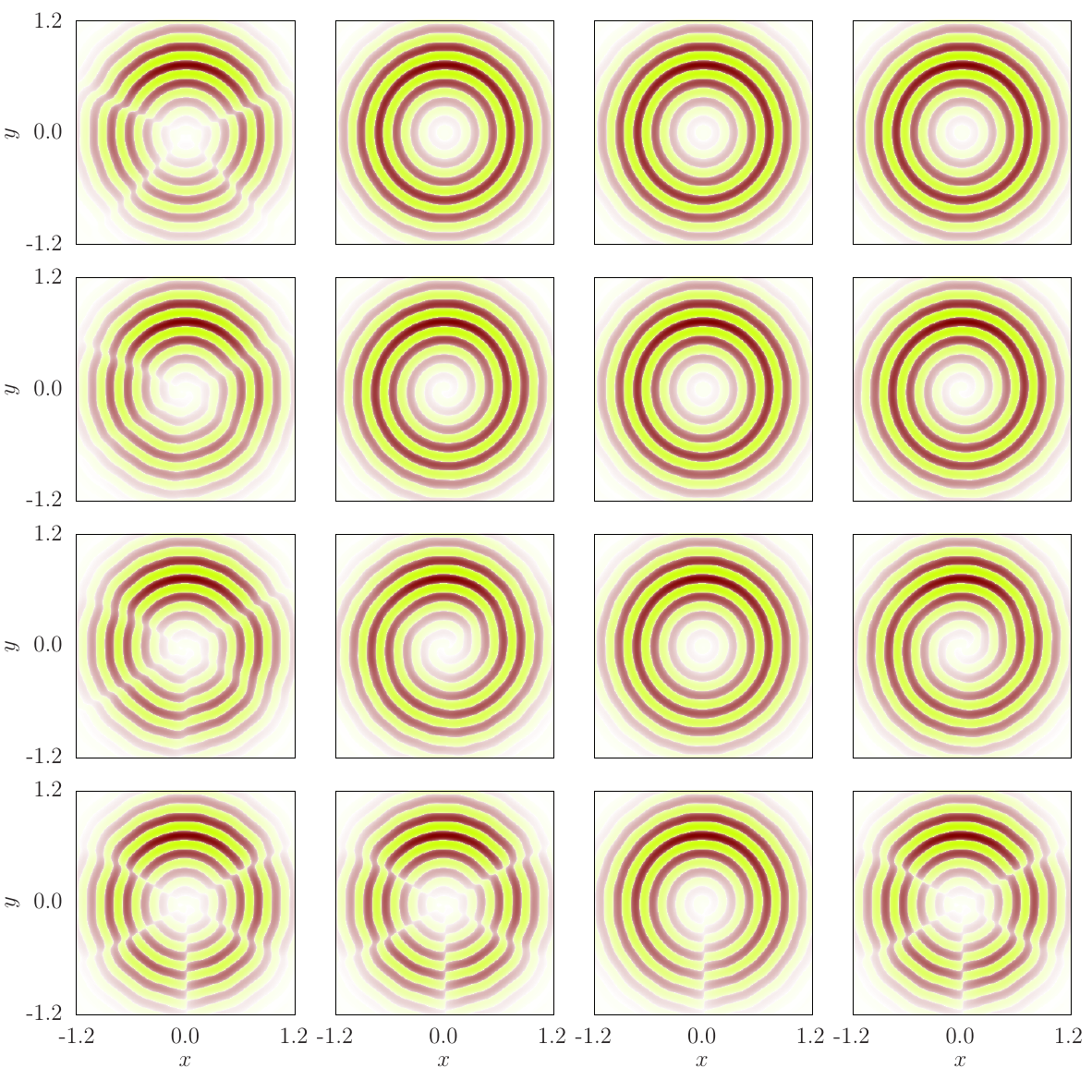}
\caption{Real part of the cross-term correlator $\mathrm{Re}[G_{R,C}(\mathbf{x},\mathbf{x}';t)]$ for strongly repulsive fermionic (left block, first and second column) and bosonic (right block, third and fourth column) mixtures with $N_p=6$ particles, $N_{\uparrow}=5$ and $N_{\downarrow}=1$, on a ring of $N_s=20$ sites at $U/J=1000$. From top to bottom, the four rows correspond to angular momenta per particle $\ell=0$, $1/6$, $2/6$, and $3/6$, respectively. Within each block, the left and right panels show the interferograms of the spin-up background and the spin-down impurity, highlighting their distinct interference features. The results were obtained by exact diagonalization of the corresponding Fermi- and Bose-Hubbard models for hopping amplitude $J$. The Wannier functions are modeled as two-dimensional time-evolved Gaussians with ring radius $R=1.5$, initial width $\sigma=0.15$, and expansion time $\omega_0t=1.5$, while the reference point is fixed at $\mathbf{x}'=(R,0)$. To enhance the visibility of the interference features, the color scale represents $\operatorname{sgn}(Y)|Y|^{1/4}$, where $Y=\mathrm{Re}[G_{R,C}(\mathbf{x},\mathbf{x}';t)]$.}
    \label{fig:interinter}
\end{figure}

\noindent In the self-heterodyne protocol, a circulating state produces a spiral interference pattern. The number of spiral arms is set by the winding number, while their orientation indicates the direction of the current. As in the homodyne protocol, a clear spiral pattern for fermions appears only after the occupied Fermi sea has been displaced by approximately half its width. Another important feature of the interferogram is the presence of radial dislocations, which appear as white lines segmenting the interference fringes and have recently been related to the nodes of the many-body wavefunction. For $N_{p}$ identical fermions, the conditional one-particle wavefunction contains $N_p-1$ nodes and therefore produces the same number of dislocations, whereas these features are absent for non-interacting bosons. In the single-impurity mixture considered here, the number of nodes: the conditional wavefunction of a spin-up particle contains $N_{\uparrow}-1$ nodes arising from the remaining spin-up particles, while that of the spin-down impurity contains no such nodes. The same difference is observed in the component-resolved interferograms --Fig.~\ref{fig:interzero}. \\

\noindent Going to the strongly repulsive regime, the number and orientation of the dislocations in the self-heterodyne interferograms reflect the fractionalized angular momentum of the system, allowing the bosonized, fermionized, and anyonized regimes to be distinguished --Fig.~\ref{fig:interinter}. On the bosonized parabola at $\ell=0$, the spin-up component of the fermionic mixture retains the $N_{\uparrow}-1$ dislocations found in the non-interacting system, whereas the spin-down impurity has none. By contrast, neither component of the bosonic mixture displays any dislocations. On reaching the fermionized parabola at $\ell=1/2$, both components of the fermionic mixture exhibit $N_p-1$ dislocations: the number in the spin-up component increases from $N_{\uparrow}-1$ to $N_p-1$, while the impurity interferogram develops the same pattern. In the bosonic mixture, the spin-down component likewise displays $N_p-1$ dislocations, while the spin-up species exhibits only one, independently of the total particle number. \\

\noindent In the intermediate anyonized states, the spin-down interferograms contain no clear dislocations but instead develop deformed spiral patterns for both bosonic and fermionic mixtures. These spirals appear even though the fermionic momentum distribution has not been displaced sufficiently to produce them in the non-interacting case. This reflects the fractionalized regime, in which the angular-momentum shift is distributed differently between the two components rather than arising from a rigid displacement of the entire Fermi sea. The spin-up component of the fermionic mixture also develops spirals, but they rotate in the opposite direction and contain a number of dislocations that varies between fractionalized parabolas. For the bosonic mixture, the spin-up interferograms remain as deformed concentric rings up to the $\ell=1/2$ parabola; beyond this point, they also develop counter-rotating spirals.

\subsection{Anyonic persistent currents}

\noindent Through the one-particle wavefunction and the one-body correlation function, we established a direct correspondence between the spin amplitudes of the mixture and the anyonic statistical phase. This correspondence suggests that, when subjected to an effective magnetic flux, the ground-state energy of a system of $N_{p}$ anyons should exhibit a reduced flux periodicity, or fractionalization, with the statistical phase playing the role of the spin amplitudes in the mixture. In what follows, we demonstrate this behaviour by calculating the ground-state energy $E_{0}(\phi)$ and the associated persistent current, $I(\phi)=-\frac{\partial E_{0}(\phi)}{\partial\phi}$. \\

\noindent To investigate this behaviour, we consider the one-dimensional anyon--Hubbard model~\cite{keilmann2011statistically,greschner2015anyon,bonkhoff2025anyonic},
\begin{equation}
H=-J\sum_{j=1}^{N_{s}}\left(a_{j}^{\dagger}a_{j+1}+a_{j+1}^{\dagger}a_{j} \right)+\frac{U}{2}\sum_{j=1}^{N_{s}}n_{j}(n_{j}-1),
\end{equation}
where periodic boundary conditions, $a_{N_{s}+1}\equiv a_1$, are imposed. Here, $a_{j}^{\dagger}$ and $a_{j}$ are the anyonic creation and annihilation operators, $n_{j}=a_{j}^{\dagger}a_{j}$ is the local number operator, and $N_s$ is the number of lattice sites. The parameters $J$ and $U$ denote the hopping and interaction strengths, respectively. The anyonic statistics are encoded in the generalized commutation relations
\begin{align}
a_{j}a_{l}^{\dagger}=e^{-i\theta\,\operatorname{sgn}(j-l)}a_{l}^{\dagger}a_{j}+\delta_{jl}, \hspace{10mm} a_{j}a_{l}
=e^{i\theta\,\operatorname{sgn}(j-l)}a_{l}a_{j},
\end{align}
where $\theta$ is the statistical angle. The limits $\theta=0$ and $\theta=\pi$ correspond to bosons and pseudo-fermions, respectively. In the latter case, the particles obey fermionic exchange statistics between different sites, while multiple occupation of the same site remains allowed unless the hard-core limit $U\rightarrow\infty$ is imposed. \\

\noindent Through the fractional Jordan--Wigner transformation~\cite{keilmann2011statistically},
\begin{equation}
a_{j}=b_{j}\exp\left(i\theta\sum_{l<j}n_{l}\right),\qquad a_{j}^{\dagger}=\exp\left(-i\theta\sum_{l<j}n_{l}
    \right)b_{j}^{\dagger},
\end{equation}
where $b_j^{\dagger}$ and $b_j$ are ordinary bosonic operators, the bulk hopping terms acquire an occupation-dependent phase. For periodic boundary conditions, the resulting Hamiltonian is
\begin{align}
H=-J\sum_{j=1}^{N_s-1}\left(b_{j}^{\dagger}b_{j+1}e^{i\theta n_j}
+e^{-i\theta n_{j}}b_{j+1}^{\dagger}b_{j}\right) -J\left[ e^{-i\theta(N_{p}-n_{N_s})}b_{N_s}^{\dagger}b_{1}+b_{1}^{\dagger}b_{N_{s}}
e^{i\theta(N_{p}-n_{N_{s}})}\right]+\frac{U}{2}\sum_{j=1}^{N_{s}}n_{j}(n_{j}-1),
\end{align}
where $N_{p}=\sum_{j} n_{j}$ is the conserved total particle number. In the bulk, hopping from $j+1$ to $j$ therefore acquires the phase $e^{i\theta n_{j}}$, while the reverse process carries its Hermitian conjugate. The additional phase in the boundary term accounts for the Jordan-Wigner string accumulated when a particle crosses the boundary of the ring. \\

\noindent An artificial magnetic flux $\phi$ is introduced through the Peierls substitution~\cite{peierls1933zur}. Defining the total Aharonov--Bohm phase as $\Phi=\frac{2\pi\phi}{N_{s}\phi_0}$, where $\phi_0$ is the bare flux quantum, the Bose-mapped Hamiltonian becomes
\begin{align}
H=-J\sum_{j=1}^{N_{s}-1}\left(e^{-i\Phi}b_{j}^{\dagger}b_{j+1}e^{i\theta n_{j}}+e^{i\Phi}e^{-i\theta n_{j}}b_{j+1}^{\dagger}b_{j}\right)
-J\left[e^{-i\Phi}e^{-i\theta(N_{p}-n_{N_{s}})}b_{N_{s}}^{\dagger}b_{1}
+e^{i\varphi}b_{1}^{\dagger}b_{N_{s}}e^{i\theta(N_{p}-n_{N_{s}})}\right]
    +\frac{U}{2}\sum_{j=1}^{N_{s}}n_{j}(n_{j}-1).
\end{align}
The persistent current follows from the Hellmann--Feynman theorem,
\begin{equation}
I(\phi)=-\frac{\partial E_0(\phi)}{\partial\phi}=-\left\langle\frac{\partial H}{\partial\phi}\right\rangle_{\mathrm{GS}}.
\end{equation}
Introducing $T_{j}=b_{j}^{\dagger}b_{j+1}e^{i\theta n_{j}}$ and $T_{\mathrm{B}}=e^{-i\theta(N_{p}-n_{N_{s}})}b_{N_{s}}^{\dagger}b_{1}$, the current can be written compactly as
\begin{equation}
I(\phi)= -\frac{2i\pi J}{N_{s}\phi_{0}}\left\langle e^{-i\varphi}
\left(\sum_{j=1}^{N_{s}-1}T_{j}+T_{\mathrm{B}}\right) -e^{i\varphi}\left(\sum_{j=1}^{N_{s}-1}T_{j}^{\dagger}+T_{\mathrm{B}}^{\dagger}\right)\right\rangle_{\mathrm{GS}}.
\end{equation}
For $\theta=0$, the occupation-dependent phases disappear and the standard persistent-current expression on a ring is recovered.
\begin{figure}[h!]
    \centering
    \includegraphics[width=0.95\linewidth]{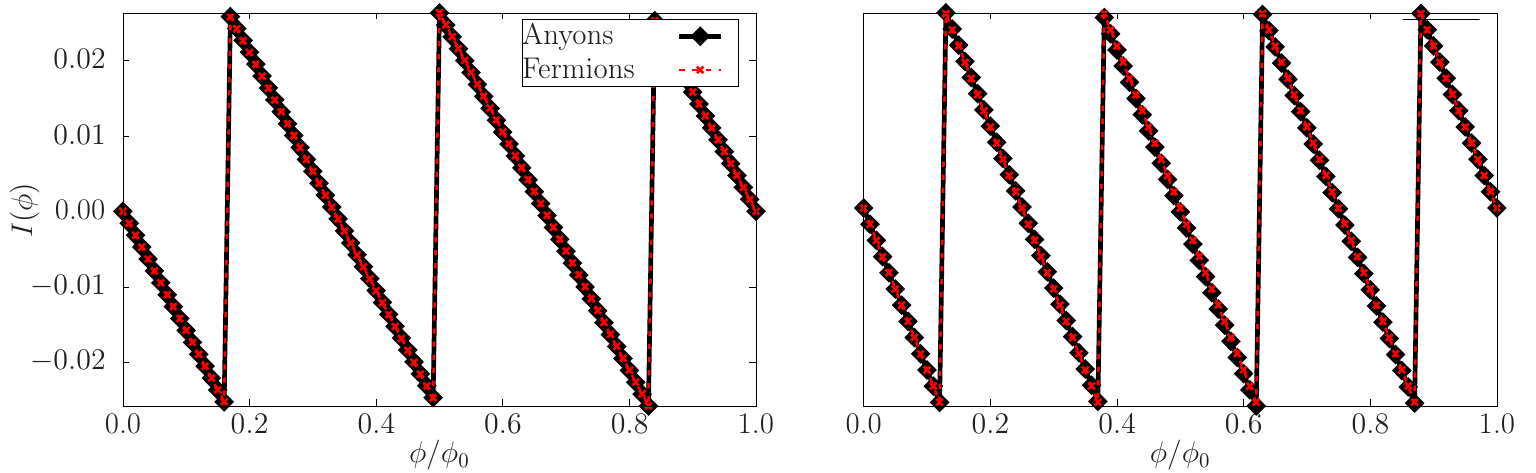}%
    \put(-445,135){(\textbf{a})}
    \put(-207,135){(\textbf{b})}
\caption{Comparison of the persistent current $I(\phi)$ as a function of the artificial gauge field $\phi$ in anyonic (black) and two-component fermionic (red) mixtures. Depending on the number of particles present in the system, $N_{p}=3$ and $N_{p}=4$ in the left/right panels respectively, the persistent current displays a reduced periodicity of $\phi_{0}/N_{p}$ with $\phi_{0}$ being the bare flux quantum. Results obtained with exact diagonalization of the anyon Bose-Hubbard model and SU(2) Fermi-Hubbard model for a ring of $N_{s}=15$ sites with $U/J=1000$ where $U$ is the interaction strength and $J$ is the hopping amplitude.}
    \label{fig:percurr}
\end{figure}

\begin{figure}[h!]
    \centering
    \includegraphics[width=0.95\linewidth]{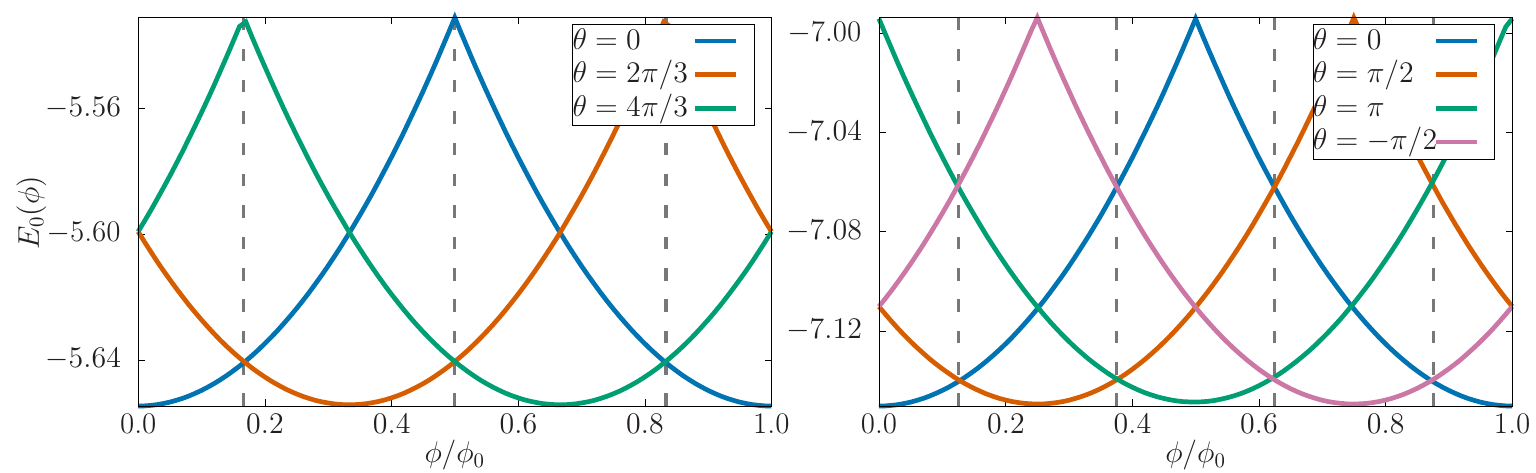}%
\caption{Ground-state energy $E_{0}$ as a function of the effective flux $\phi$ for a system of $N_{p}=3$ and $N_{p}=4$ anyons in the left and right panels respectively for various values of the statistical angle $\theta$. As the flux threading the system increases, there are energy level crossings between the ground- and excited states characterized by different $\theta$. Consequently, the period of the system is no longer given by the elementary flux quantum $\phi_{0}$ but is renormalized to $\phi_{0}/N_{p}$ (indicated by grey dotted lines). Results obtained with exact diagonalization of the anyon Bose-Hubbard model for a ring of $N_{s}=15$ sites with $U/J=1000$ where $U$ is the interaction strength and $J$ is the hopping amplitude.}
    \label{fig:ener}
\end{figure}

\noindent Figure~\ref{fig:percurr}\textbf{(a)} shows the ground-state energy as a function of the effective magnetic flux $\phi/\phi_{0}$ for the discrete statistical angles $\theta=2\pi n/N_{p}$. Each value of $\theta$ defines a distinct energy branch (see Fig.~\ref{fig:ener}), and to minimize the ground-state energy there are crossings between the different branches. The first occurs at $\phi/\phi_{0}=1/(2N_{p})$, with subsequent crossings separated by $1/N_{p}$, resulting in the reduced flux period $\phi_{0}/N_{p}$. These crossings directly mirror those between states belonging to different spin eigenstates of the quantum mixture, with the anyonic statistical phase playing the role of the spin ampltidues. The corresponding persistent current, shown in Fig.~\ref{fig:percurr}\textbf{(b)}, consequently exhibits the characteristic sawtooth profile with the same fractionalized period. Remarkably, the persistent currents of the anyonic system and the quantum mixture coincide exactly. Unlike the one-particle wavefunction and the one-body correlator, for which the exact correspondence applies specifically to the spin-down impurity, the persistent current involves the entire mixture, including both the spin-up background and the spin-down impurity. Since the flux affects both components in the same way, the persistent current measures the response of the whole system rather than that of the impurity alone. Therefore, its agreement with the anyonic current shows that the anyonic behaviour is also present in the collective motion of the mixture, even though not all spin-resolved quantities necessarily coincide.


\begin{thebibliography}{69}%
\makeatletter
\providecommand \@ifxundefined [1]{%
 \@ifx{#1\undefined}
}%
\providecommand \@ifnum [1]{%
 \ifnum #1\expandafter \@firstoftwo
 \else \expandafter \@secondoftwo
 \fi
}%
\providecommand \@ifx [1]{%
 \ifx #1\expandafter \@firstoftwo
 \else \expandafter \@secondoftwo
 \fi
}%
\providecommand \natexlab [1]{#1}%
\providecommand \enquote  [1]{``#1''}%
\providecommand \bibnamefont  [1]{#1}%
\providecommand \bibfnamefont [1]{#1}%
\providecommand \citenamefont [1]{#1}%
\providecommand \href@noop [0]{\@secondoftwo}%
\providecommand \href [0]{\begingroup \@sanitize@url \@href}%
\providecommand \@href[1]{\@@startlink{#1}\@@href}%
\providecommand \@@href[1]{\endgroup#1\@@endlink}%
\providecommand \@sanitize@url [0]{\catcode `\\12\catcode `\$12\catcode
  `\&12\catcode `\#12\catcode `\^12\catcode `\_12\catcode `\%12\relax}%
\providecommand \@@startlink[1]{}%
\providecommand \@@endlink[0]{}%
\providecommand \url  [0]{\begingroup\@sanitize@url \@url }%
\providecommand \@url [1]{\endgroup\@href {#1}{\urlprefix }}%
\providecommand \urlprefix  [0]{URL }%
\providecommand \Eprint [0]{\href }%
\providecommand \doibase [0]{http://dx.doi.org/}%
\providecommand \selectlanguage [0]{\@gobble}%
\providecommand \bibinfo  [0]{\@secondoftwo}%
\providecommand \bibfield  [0]{\@secondoftwo}%
\providecommand \translation [1]{[#1]}%
\providecommand \BibitemOpen [0]{}%
\providecommand \bibitemStop [0]{}%
\providecommand \bibitemNoStop [0]{.\EOS\space}%
\providecommand \EOS [0]{\spacefactor3000\relax}%
\providecommand \BibitemShut  [1]{\csname bibitem#1\endcsname}%
\let\auto@bib@innerbib\@empty
\bibitem [{\citenamefont {Dirac}(1926)}]{dirac1926theory}%
  \BibitemOpen
  \bibfield  {author} {\bibinfo {author} {\bibfnamefont {P.~A.~M.}\
  \bibnamefont {Dirac}},\ }\href {\doibase 10.1098/rspa.1926.0133} {\bibfield
  {journal} {\bibinfo  {journal} {Proceedings of the Royal Society of London.
  Series A}\ }\textbf {\bibinfo {volume} {112}},\ \bibinfo {pages} {661}
  (\bibinfo {year} {1926})}\BibitemShut {NoStop}%
\bibitem [{\citenamefont {Leinaas}\ and\ \citenamefont
  {Myrheim}(1977)}]{leinaas1977theory}%
  \BibitemOpen
  \bibfield  {author} {\bibinfo {author} {\bibfnamefont {J.~M.}\ \bibnamefont
  {Leinaas}}\ and\ \bibinfo {author} {\bibfnamefont {J.}~\bibnamefont
  {Myrheim}},\ }\href {\doibase 10.1007/BF02727953} {\bibfield  {journal}
  {\bibinfo  {journal} {Il Nuovo Cimento B (1971-1996)}\ }\textbf {\bibinfo
  {volume} {37}},\ \bibinfo {pages} {1} (\bibinfo {year} {1977})}\BibitemShut
  {NoStop}%
\bibitem [{\citenamefont {Wilczek}(1982)}]{wilczek1982quantum}%
  \BibitemOpen
  \bibfield  {author} {\bibinfo {author} {\bibfnamefont {F.}~\bibnamefont
  {Wilczek}},\ }\href {\doibase 10.1103/PhysRevLett.49.957} {\bibfield
  {journal} {\bibinfo  {journal} {Physical Review Letters}\ }\textbf {\bibinfo
  {volume} {49}},\ \bibinfo {pages} {957} (\bibinfo {year} {1982})}\BibitemShut
  {NoStop}%
\bibitem [{\citenamefont {Khare}(2005)}]{khare2005fractional}%
  \BibitemOpen
  \bibfield  {author} {\bibinfo {author} {\bibfnamefont {A.}~\bibnamefont
  {Khare}},\ }\href {\doibase 10.1142/5752} {\emph {\bibinfo {title}
  {{Fractional Statistics and Quantum Theory}}}},\ \bibinfo {edition} {2nd}\
  ed.\ (\bibinfo  {publisher} {World Scientific},\ \bibinfo {year}
  {2005})\BibitemShut {NoStop}%
\bibitem [{\citenamefont {Kundu}(1999)}]{kundu1999exact}%
  \BibitemOpen
  \bibfield  {author} {\bibinfo {author} {\bibfnamefont {A.}~\bibnamefont
  {Kundu}},\ }\href {\doibase 10.1103/PhysRevLett.83.1275} {\bibfield
  {journal} {\bibinfo  {journal} {Physical Review Letters}\ }\textbf {\bibinfo
  {volume} {83}},\ \bibinfo {pages} {1275} (\bibinfo {year}
  {1999})}\BibitemShut {NoStop}%
\bibitem [{\citenamefont {Girardeau}(2006)}]{girardeau2006anyon}%
  \BibitemOpen
  \bibfield  {author} {\bibinfo {author} {\bibfnamefont {M.~D.}\ \bibnamefont
  {Girardeau}},\ }\href {\doibase 10.1103/PhysRevLett.97.100402} {\bibfield
  {journal} {\bibinfo  {journal} {Physical Review Letters}\ }\textbf {\bibinfo
  {volume} {97}},\ \bibinfo {pages} {100402} (\bibinfo {year}
  {2006})}\BibitemShut {NoStop}%
\bibitem [{\citenamefont {Calabrese}\ and\ \citenamefont
  {Mintchev}(2007)}]{calabrese2007correlation}%
  \BibitemOpen
  \bibfield  {author} {\bibinfo {author} {\bibfnamefont {P.}~\bibnamefont
  {Calabrese}}\ and\ \bibinfo {author} {\bibfnamefont {M.}~\bibnamefont
  {Mintchev}},\ }\href {\doibase 10.1103/PhysRevB.75.233104} {\bibfield
  {journal} {\bibinfo  {journal} {Physical Review B}\ }\textbf {\bibinfo
  {volume} {75}},\ \bibinfo {pages} {233104} (\bibinfo {year}
  {2007})}\BibitemShut {NoStop}%
\bibitem [{\citenamefont {Batchelor}\ \emph {et~al.}(2006)\citenamefont
  {Batchelor}, \citenamefont {Guan},\ and\ \citenamefont
  {Oelkers}}]{batchelor2006one}%
  \BibitemOpen
  \bibfield  {author} {\bibinfo {author} {\bibfnamefont {M.~T.}\ \bibnamefont
  {Batchelor}}, \bibinfo {author} {\bibfnamefont {X.-W.}\ \bibnamefont {Guan}},
  \ and\ \bibinfo {author} {\bibfnamefont {N.}~\bibnamefont {Oelkers}},\ }\href
  {\doibase 10.1103/PhysRevLett.96.210402} {\bibfield  {journal} {\bibinfo
  {journal} {Physical Review Letters}\ }\textbf {\bibinfo {volume} {96}},\
  \bibinfo {pages} {210402} (\bibinfo {year} {2006})}\BibitemShut {NoStop}%
\bibitem [{\citenamefont {Batchelor}\ \emph {et~al.}(2007)\citenamefont
  {Batchelor}, \citenamefont {Guan},\ and\ \citenamefont
  {He}}]{batchelor2007bethe}%
  \BibitemOpen
  \bibfield  {author} {\bibinfo {author} {\bibfnamefont {M.~T.}\ \bibnamefont
  {Batchelor}}, \bibinfo {author} {\bibfnamefont {X.-W.}\ \bibnamefont {Guan}},
  \ and\ \bibinfo {author} {\bibfnamefont {J.-S.}\ \bibnamefont {He}},\ }\href
  {\doibase 10.1088/1742-5468/2007/03/P03007} {\bibfield  {journal} {\bibinfo
  {journal} {Journal of Statistical Mechanics: Theory and Experiment}\ }\textbf
  {\bibinfo {volume} {2007}},\ \bibinfo {pages} {P03007} (\bibinfo {year}
  {2007})}\BibitemShut {NoStop}%
\bibitem [{\citenamefont {Santachiara}\ and\ \citenamefont
  {Calabrese}(2008)}]{santachiara2008one}%
  \BibitemOpen
  \bibfield  {author} {\bibinfo {author} {\bibfnamefont {R.}~\bibnamefont
  {Santachiara}}\ and\ \bibinfo {author} {\bibfnamefont {P.}~\bibnamefont
  {Calabrese}},\ }\href {\doibase 10.1088/1742-5468/2008/06/P06005} {\bibfield
  {journal} {\bibinfo  {journal} {Journal of Statistical Mechanics: Theory and
  Experiment}\ }\textbf {\bibinfo {volume} {2008}},\ \bibinfo {pages} {P06005}
  (\bibinfo {year} {2008})}\BibitemShut {NoStop}%
\bibitem [{\citenamefont {Amico}\ \emph {et~al.}(1998)\citenamefont {Amico},
  \citenamefont {Osterloh},\ and\ \citenamefont {Eckern}}]{amico1998one}%
  \BibitemOpen
  \bibfield  {author} {\bibinfo {author} {\bibfnamefont {L.}~\bibnamefont
  {Amico}}, \bibinfo {author} {\bibfnamefont {A.}~\bibnamefont {Osterloh}}, \
  and\ \bibinfo {author} {\bibfnamefont {U.}~\bibnamefont {Eckern}},\ }\href
  {\doibase 10.1103/PhysRevB.58.R1703} {\bibfield  {journal} {\bibinfo
  {journal} {Physical Review B}\ }\textbf {\bibinfo {volume} {58}},\ \bibinfo
  {pages} {R1703(R)} (\bibinfo {year} {1998})}\BibitemShut {NoStop}%
\bibitem [{\citenamefont {Osterloh}\ \emph {et~al.}(2000)\citenamefont
  {Osterloh}, \citenamefont {Amico},\ and\ \citenamefont
  {Eckern}}]{osterloh2000bethe}%
  \BibitemOpen
  \bibfield  {author} {\bibinfo {author} {\bibfnamefont {A.}~\bibnamefont
  {Osterloh}}, \bibinfo {author} {\bibfnamefont {L.}~\bibnamefont {Amico}}, \
  and\ \bibinfo {author} {\bibfnamefont {U.}~\bibnamefont {Eckern}},\ }\href
  {\doibase 10.1088/0305-4470/33/9/101} {\bibfield  {journal} {\bibinfo
  {journal} {Journal of Physics A: Mathematical and General}\ }\textbf
  {\bibinfo {volume} {33}},\ \bibinfo {pages} {L87} (\bibinfo {year}
  {2000})}\BibitemShut {NoStop}%
\bibitem [{\citenamefont {Greschner}\ and\ \citenamefont
  {Santos}(2015)}]{greschner2015anyon}%
  \BibitemOpen
  \bibfield  {author} {\bibinfo {author} {\bibfnamefont {S.}~\bibnamefont
  {Greschner}}\ and\ \bibinfo {author} {\bibfnamefont {L.}~\bibnamefont
  {Santos}},\ }\href {\doibase 10.1103/PhysRevLett.115.053002} {\bibfield
  {journal} {\bibinfo  {journal} {Physical Review Letters}\ }\textbf {\bibinfo
  {volume} {115}},\ \bibinfo {pages} {053002} (\bibinfo {year}
  {2015})}\BibitemShut {NoStop}%
\bibitem [{\citenamefont {Bonkhoff}\ \emph {et~al.}(2025)\citenamefont
  {Bonkhoff}, \citenamefont {J\"agering}, \citenamefont {Hu}, \citenamefont
  {Pelster}, \citenamefont {Eggert},\ and\ \citenamefont
  {Schneider}}]{bonkhoff2025anyonic}%
  \BibitemOpen
  \bibfield  {author} {\bibinfo {author} {\bibfnamefont {M.}~\bibnamefont
  {Bonkhoff}}, \bibinfo {author} {\bibfnamefont {K.}~\bibnamefont
  {J\"agering}}, \bibinfo {author} {\bibfnamefont {S.}~\bibnamefont {Hu}},
  \bibinfo {author} {\bibfnamefont {A.}~\bibnamefont {Pelster}}, \bibinfo
  {author} {\bibfnamefont {S.}~\bibnamefont {Eggert}}, \ and\ \bibinfo {author}
  {\bibfnamefont {I.}~\bibnamefont {Schneider}},\ }\href {\doibase
  10.1103/7n1c-vq2p} {\bibfield  {journal} {\bibinfo  {journal} {Physical
  Review Letters}\ }\textbf {\bibinfo {volume} {135}},\ \bibinfo {pages}
  {036601} (\bibinfo {year} {2025})}\BibitemShut {NoStop}%
\bibitem [{\citenamefont {Forte}(1992)}]{forte1992quantum}%
  \BibitemOpen
  \bibfield  {author} {\bibinfo {author} {\bibfnamefont {S.}~\bibnamefont
  {Forte}},\ }\href {\doibase 10.1103/RevModPhys.64.193} {\bibfield  {journal}
  {\bibinfo  {journal} {Reviews of Modern Physics}\ }\textbf {\bibinfo {volume}
  {64}},\ \bibinfo {pages} {193} (\bibinfo {year} {1992})}\BibitemShut
  {NoStop}%
\bibitem [{\citenamefont {Stern}(2008)}]{stern2008anyons}%
  \BibitemOpen
  \bibfield  {author} {\bibinfo {author} {\bibfnamefont {A.}~\bibnamefont
  {Stern}},\ }\href {\doibase https://doi.org/10.1016/j.aop.2007.10.008}
  {\bibfield  {journal} {\bibinfo  {journal} {Annals of Physics}\ }\textbf
  {\bibinfo {volume} {323}},\ \bibinfo {pages} {204} (\bibinfo {year}
  {2008})},\ \bibinfo {note} {january Special Issue 2008}\BibitemShut {NoStop}%
\bibitem [{\citenamefont {Greiter}\ and\ \citenamefont
  {Wilczek}(2024)}]{greiter2024fractional}%
  \BibitemOpen
  \bibfield  {author} {\bibinfo {author} {\bibfnamefont {M.}~\bibnamefont
  {Greiter}}\ and\ \bibinfo {author} {\bibfnamefont {F.}~\bibnamefont
  {Wilczek}},\ }\href {\doibase 10.1146/annurev-conmatphys-040423-014045}
  {\bibfield  {journal} {\bibinfo  {journal} {Annual Review of Condensed Matter
  Physics}\ }\textbf {\bibinfo {volume} {15}},\ \bibinfo {pages} {131–157}
  (\bibinfo {year} {2024})}\BibitemShut {NoStop}%
\bibitem [{\citenamefont {Wilczek}(1990)}]{wilczek1990fractional}%
  \BibitemOpen
  \bibfield  {author} {\bibinfo {author} {\bibfnamefont {F.}~\bibnamefont
  {Wilczek}},\ }\href {\doibase 10.1142/0961} {\emph {\bibinfo {title}
  {{Fractional Statistics and Anyon Superconductivity}}}}\ (\bibinfo
  {publisher} {World Scientific},\ \bibinfo {year} {1990})\BibitemShut
  {NoStop}%
\bibitem [{\citenamefont {Nayak}\ \emph {et~al.}(2008)\citenamefont {Nayak},
  \citenamefont {Simon}, \citenamefont {Stern}, \citenamefont {Freedman},\ and\
  \citenamefont {Das~Sarma}}]{nayak2008non}%
  \BibitemOpen
  \bibfield  {author} {\bibinfo {author} {\bibfnamefont {C.}~\bibnamefont
  {Nayak}}, \bibinfo {author} {\bibfnamefont {S.~H.}\ \bibnamefont {Simon}},
  \bibinfo {author} {\bibfnamefont {A.}~\bibnamefont {Stern}}, \bibinfo
  {author} {\bibfnamefont {M.}~\bibnamefont {Freedman}}, \ and\ \bibinfo
  {author} {\bibfnamefont {S.}~\bibnamefont {Das~Sarma}},\ }\href {\doibase
  10.1103/RevModPhys.80.1083} {\bibfield  {journal} {\bibinfo  {journal}
  {Reviews of Modern Physics}\ }\textbf {\bibinfo {volume} {80}},\ \bibinfo
  {pages} {1083} (\bibinfo {year} {2008})}\BibitemShut {NoStop}%
\bibitem [{\citenamefont {Arovas}\ \emph {et~al.}(1984)\citenamefont {Arovas},
  \citenamefont {Schrieffer},\ and\ \citenamefont
  {Wilczek}}]{arovas1984fractional}%
  \BibitemOpen
  \bibfield  {author} {\bibinfo {author} {\bibfnamefont {D.}~\bibnamefont
  {Arovas}}, \bibinfo {author} {\bibfnamefont {J.~R.}\ \bibnamefont
  {Schrieffer}}, \ and\ \bibinfo {author} {\bibfnamefont {F.}~\bibnamefont
  {Wilczek}},\ }\href {\doibase 10.1103/PhysRevLett.53.722} {\bibfield
  {journal} {\bibinfo  {journal} {Physical Review Letters}\ }\textbf {\bibinfo
  {volume} {53}},\ \bibinfo {pages} {722} (\bibinfo {year} {1984})}\BibitemShut
  {NoStop}%
\bibitem [{\citenamefont {Kitaev}(2003)}]{kitaev2003fault}%
  \BibitemOpen
  \bibfield  {author} {\bibinfo {author} {\bibfnamefont {A.}~\bibnamefont
  {Kitaev}},\ }\href {\doibase https://doi.org/10.1016/S0003-4916(02)00018-0}
  {\bibfield  {journal} {\bibinfo  {journal} {Annals of Physics}\ }\textbf
  {\bibinfo {volume} {303}},\ \bibinfo {pages} {2} (\bibinfo {year}
  {2003})}\BibitemShut {NoStop}%
\bibitem [{\citenamefont {Bartolomei}\ \emph {et~al.}(2020)\citenamefont
  {Bartolomei}, \citenamefont {Kumar}, \citenamefont {Bisognin}, \citenamefont
  {Marguerite}, \citenamefont {Berroir}, \citenamefont {Bocquillon},
  \citenamefont {Plaçais}, \citenamefont {Cavanna}, \citenamefont {Dong},
  \citenamefont {Gennser}, \citenamefont {Jin},\ and\ \citenamefont
  {Fève}}]{bartolomei2020fractional}%
  \BibitemOpen
  \bibfield  {author} {\bibinfo {author} {\bibfnamefont {H.}~\bibnamefont
  {Bartolomei}}, \bibinfo {author} {\bibfnamefont {M.}~\bibnamefont {Kumar}},
  \bibinfo {author} {\bibfnamefont {R.}~\bibnamefont {Bisognin}}, \bibinfo
  {author} {\bibfnamefont {A.}~\bibnamefont {Marguerite}}, \bibinfo {author}
  {\bibfnamefont {J.-M.}\ \bibnamefont {Berroir}}, \bibinfo {author}
  {\bibfnamefont {E.}~\bibnamefont {Bocquillon}}, \bibinfo {author}
  {\bibfnamefont {B.}~\bibnamefont {Plaçais}}, \bibinfo {author}
  {\bibfnamefont {A.}~\bibnamefont {Cavanna}}, \bibinfo {author} {\bibfnamefont
  {Q.}~\bibnamefont {Dong}}, \bibinfo {author} {\bibfnamefont {U.}~\bibnamefont
  {Gennser}}, \bibinfo {author} {\bibfnamefont {Y.}~\bibnamefont {Jin}}, \ and\
  \bibinfo {author} {\bibfnamefont {G.}~\bibnamefont {Fève}},\ }\href
  {\doibase 10.1126/science.aaz5601} {\bibfield  {journal} {\bibinfo  {journal}
  {Science}\ }\textbf {\bibinfo {volume} {368}},\ \bibinfo {pages} {173}
  (\bibinfo {year} {2020})}\BibitemShut {NoStop}%
\bibitem [{\citenamefont {Kwan}\ \emph {et~al.}(2024)\citenamefont {Kwan},
  \citenamefont {Segura}, \citenamefont {Li}, \citenamefont {Kim},
  \citenamefont {Gorshkov}, \citenamefont {Eckardt}, \citenamefont
  {Bakkali-Hassani},\ and\ \citenamefont {Greiner}}]{kwan2024realization}%
  \BibitemOpen
  \bibfield  {author} {\bibinfo {author} {\bibfnamefont {J.}~\bibnamefont
  {Kwan}}, \bibinfo {author} {\bibfnamefont {P.}~\bibnamefont {Segura}},
  \bibinfo {author} {\bibfnamefont {Y.}~\bibnamefont {Li}}, \bibinfo {author}
  {\bibfnamefont {S.}~\bibnamefont {Kim}}, \bibinfo {author} {\bibfnamefont
  {A.~V.}\ \bibnamefont {Gorshkov}}, \bibinfo {author} {\bibfnamefont
  {A.}~\bibnamefont {Eckardt}}, \bibinfo {author} {\bibfnamefont
  {B.}~\bibnamefont {Bakkali-Hassani}}, \ and\ \bibinfo {author} {\bibfnamefont
  {M.}~\bibnamefont {Greiner}},\ }\href {\doibase 10.1126/science.adi3252}
  {\bibfield  {journal} {\bibinfo  {journal} {Science}\ }\textbf {\bibinfo
  {volume} {386}},\ \bibinfo {pages} {1055} (\bibinfo {year}
  {2024})}\BibitemShut {NoStop}%
\bibitem [{\citenamefont {Haldane}(1991)}]{haldane1991fractional}%
  \BibitemOpen
  \bibfield  {author} {\bibinfo {author} {\bibfnamefont {F.~D.~M.}\
  \bibnamefont {Haldane}},\ }\href {\doibase 10.1103/PhysRevLett.67.937}
  {\bibfield  {journal} {\bibinfo  {journal} {Physical Review Letters}\
  }\textbf {\bibinfo {volume} {67}},\ \bibinfo {pages} {937} (\bibinfo {year}
  {1991})}\BibitemShut {NoStop}%
\bibitem [{\citenamefont {Ha}(1995)}]{ha1995fractional}%
  \BibitemOpen
  \bibfield  {author} {\bibinfo {author} {\bibfnamefont {Z.}~\bibnamefont
  {Ha}},\ }\href {\doibase https://doi.org/10.1016/0550-3213(94)00537-O}
  {\bibfield  {journal} {\bibinfo  {journal} {Nuclear Physics B}\ }\textbf
  {\bibinfo {volume} {435}},\ \bibinfo {pages} {604} (\bibinfo {year}
  {1995})}\BibitemShut {NoStop}%
\bibitem [{\citenamefont {Scopa}\ \emph {et~al.}(2020)\citenamefont {Scopa},
  \citenamefont {Piroli},\ and\ \citenamefont {Calabrese}}]{scopa2020one}%
  \BibitemOpen
  \bibfield  {author} {\bibinfo {author} {\bibfnamefont {S.}~\bibnamefont
  {Scopa}}, \bibinfo {author} {\bibfnamefont {L.}~\bibnamefont {Piroli}}, \
  and\ \bibinfo {author} {\bibfnamefont {P.}~\bibnamefont {Calabrese}},\ }\href
  {\doibase 10.1088/1742-5468/abaed1} {\bibfield  {journal} {\bibinfo
  {journal} {Journal of Statistical Mechanics: Theory and Experiment}\ }\textbf
  {\bibinfo {volume} {2020}},\ \bibinfo {pages} {093103} (\bibinfo {year}
  {2020})}\BibitemShut {NoStop}%
\bibitem [{\citenamefont {Santachiara}\ \emph {et~al.}(2007)\citenamefont
  {Santachiara}, \citenamefont {Stauffer},\ and\ \citenamefont
  {Cabra}}]{santachiara2007entanglement}%
  \BibitemOpen
  \bibfield  {author} {\bibinfo {author} {\bibfnamefont {R.}~\bibnamefont
  {Santachiara}}, \bibinfo {author} {\bibfnamefont {F.}~\bibnamefont
  {Stauffer}}, \ and\ \bibinfo {author} {\bibfnamefont {D.~C.}\ \bibnamefont
  {Cabra}},\ }\href {\doibase 10.1088/1742-5468/2007/05/L05003} {\bibfield
  {journal} {\bibinfo  {journal} {Journal of Statistical Mechanics: Theory and
  Experiment}\ }\textbf {\bibinfo {volume} {2007}},\ \bibinfo {pages} {L05003}
  (\bibinfo {year} {2007})}\BibitemShut {NoStop}%
\bibitem [{\citenamefont {Pâţu}\ \emph {et~al.}(2007)\citenamefont {Pâţu},
  \citenamefont {Korepin},\ and\ \citenamefont {Averin}}]{patu2007correlation}%
  \BibitemOpen
  \bibfield  {author} {\bibinfo {author} {\bibfnamefont {O.~I.}\ \bibnamefont
  {Pâţu}}, \bibinfo {author} {\bibfnamefont {V.~E.}\ \bibnamefont {Korepin}},
  \ and\ \bibinfo {author} {\bibfnamefont {D.~V.}\ \bibnamefont {Averin}},\
  }\href {\doibase 10.1088/1751-8113/40/50/004} {\bibfield  {journal} {\bibinfo
   {journal} {Journal of Physics A: Mathematical and Theoretical}\ }\textbf
  {\bibinfo {volume} {40}},\ \bibinfo {pages} {14963} (\bibinfo {year}
  {2007})}\BibitemShut {NoStop}%
\bibitem [{\citenamefont {Hao}\ \emph {et~al.}(2008)\citenamefont {Hao},
  \citenamefont {Zhang},\ and\ \citenamefont {Chen}}]{hao2008ground}%
  \BibitemOpen
  \bibfield  {author} {\bibinfo {author} {\bibfnamefont {Y.}~\bibnamefont
  {Hao}}, \bibinfo {author} {\bibfnamefont {Y.}~\bibnamefont {Zhang}}, \ and\
  \bibinfo {author} {\bibfnamefont {S.}~\bibnamefont {Chen}},\ }\href {\doibase
  10.1103/PhysRevA.78.023631} {\bibfield  {journal} {\bibinfo  {journal}
  {Physical Review A}\ }\textbf {\bibinfo {volume} {78}},\ \bibinfo {pages}
  {023631} (\bibinfo {year} {2008})}\BibitemShut {NoStop}%
\bibitem [{\citenamefont {Hao}\ \emph {et~al.}(2009)\citenamefont {Hao},
  \citenamefont {Zhang},\ and\ \citenamefont {Chen}}]{hao2009ground}%
  \BibitemOpen
  \bibfield  {author} {\bibinfo {author} {\bibfnamefont {Y.}~\bibnamefont
  {Hao}}, \bibinfo {author} {\bibfnamefont {Y.}~\bibnamefont {Zhang}}, \ and\
  \bibinfo {author} {\bibfnamefont {S.}~\bibnamefont {Chen}},\ }\href {\doibase
  10.1103/PhysRevA.79.043633} {\bibfield  {journal} {\bibinfo  {journal}
  {Physical Review A}\ }\textbf {\bibinfo {volume} {79}},\ \bibinfo {pages}
  {043633} (\bibinfo {year} {2009})}\BibitemShut {NoStop}%
\bibitem [{\citenamefont {Gamayun}\ \emph {et~al.}(2020)\citenamefont
  {Gamayun}, \citenamefont {Lychkovskiy},\ and\ \citenamefont
  {Zvonarev}}]{gamayun2020zero}%
  \BibitemOpen
  \bibfield  {author} {\bibinfo {author} {\bibfnamefont {O.}~\bibnamefont
  {Gamayun}}, \bibinfo {author} {\bibfnamefont {O.}~\bibnamefont
  {Lychkovskiy}}, \ and\ \bibinfo {author} {\bibfnamefont {M.}~\bibnamefont
  {Zvonarev}},\ }\href {\doibase 10.21468/SciPostPhys.8.4.053} {\bibfield
  {journal} {\bibinfo  {journal} {SciPost Physics}\ }\textbf {\bibinfo {volume}
  {8}},\ \bibinfo {pages} {053} (\bibinfo {year} {2020})}\BibitemShut {NoStop}%
\bibitem [{\citenamefont {Gamayun}\ \emph {et~al.}(2024)\citenamefont
  {Gamayun}, \citenamefont {Quinn}, \citenamefont {Bidzhiev},\ and\
  \citenamefont {Zvonarev}}]{gamayun2024emergence}%
  \BibitemOpen
  \bibfield  {author} {\bibinfo {author} {\bibfnamefont {O.}~\bibnamefont
  {Gamayun}}, \bibinfo {author} {\bibfnamefont {E.}~\bibnamefont {Quinn}},
  \bibinfo {author} {\bibfnamefont {K.}~\bibnamefont {Bidzhiev}}, \ and\
  \bibinfo {author} {\bibfnamefont {M.~B.}\ \bibnamefont {Zvonarev}},\ }\href
  {\doibase 10.1103/PhysRevA.109.012209} {\bibfield  {journal} {\bibinfo
  {journal} {Physical Review A}\ }\textbf {\bibinfo {volume} {109}},\ \bibinfo
  {pages} {012209} (\bibinfo {year} {2024})}\BibitemShut {NoStop}%
\bibitem [{\citenamefont {Dhar}\ \emph {et~al.}(2025)\citenamefont {Dhar},
  \citenamefont {Wang}, \citenamefont {Horvath}, \citenamefont {Vashisht},
  \citenamefont {Zeng}, \citenamefont {Zvonarev}, \citenamefont {Goldman},
  \citenamefont {Guo}, \citenamefont {Landini},\ and\ \citenamefont
  {N{\"a}gerl}}]{dhar2025observing}%
  \BibitemOpen
  \bibfield  {author} {\bibinfo {author} {\bibfnamefont {S.}~\bibnamefont
  {Dhar}}, \bibinfo {author} {\bibfnamefont {B.}~\bibnamefont {Wang}}, \bibinfo
  {author} {\bibfnamefont {M.}~\bibnamefont {Horvath}}, \bibinfo {author}
  {\bibfnamefont {A.}~\bibnamefont {Vashisht}}, \bibinfo {author}
  {\bibfnamefont {Y.}~\bibnamefont {Zeng}}, \bibinfo {author} {\bibfnamefont
  {M.~B.}\ \bibnamefont {Zvonarev}}, \bibinfo {author} {\bibfnamefont
  {N.}~\bibnamefont {Goldman}}, \bibinfo {author} {\bibfnamefont
  {Y.}~\bibnamefont {Guo}}, \bibinfo {author} {\bibfnamefont {M.}~\bibnamefont
  {Landini}}, \ and\ \bibinfo {author} {\bibfnamefont {H.-C.}\ \bibnamefont
  {N{\"a}gerl}},\ }\href {\doibase 10.1038/s41586-025-09016-9} {\bibfield
  {journal} {\bibinfo  {journal} {Nature}\ }\textbf {\bibinfo {volume} {642}},\
  \bibinfo {pages} {53} (\bibinfo {year} {2025})}\BibitemShut {NoStop}%
\bibitem [{\citenamefont {Wang}\ \emph {et~al.}(2025)\citenamefont {Wang},
  \citenamefont {Vashisht}, \citenamefont {Guo}, \citenamefont {Dhar},
  \citenamefont {Landini}, \citenamefont {N\"agerl},\ and\ \citenamefont
  {Goldman}}]{wang2025anyonization}%
  \BibitemOpen
  \bibfield  {author} {\bibinfo {author} {\bibfnamefont {B.}~\bibnamefont
  {Wang}}, \bibinfo {author} {\bibfnamefont {A.}~\bibnamefont {Vashisht}},
  \bibinfo {author} {\bibfnamefont {Y.}~\bibnamefont {Guo}}, \bibinfo {author}
  {\bibfnamefont {S.}~\bibnamefont {Dhar}}, \bibinfo {author} {\bibfnamefont
  {M.}~\bibnamefont {Landini}}, \bibinfo {author} {\bibfnamefont {H.-C.}\
  \bibnamefont {N\"agerl}}, \ and\ \bibinfo {author} {\bibfnamefont
  {N.}~\bibnamefont {Goldman}},\ }\href {\doibase 10.1103/2np8-mp39} {\bibfield
   {journal} {\bibinfo  {journal} {Physical Review Letters}\ }\textbf {\bibinfo
  {volume} {135}},\ \bibinfo {pages} {253403} (\bibinfo {year}
  {2025})}\BibitemShut {NoStop}%
\bibitem [{\citenamefont {Aupetit-Diallo}\ \emph {et~al.}(2025)\citenamefont
  {Aupetit-Diallo}, \citenamefont {Pecci}, \citenamefont {Volosniev},
  \citenamefont {Albert}, \citenamefont {Minguzzi},\ and\ \citenamefont
  {Vignolo}}]{aupetit2025necklace}%
  \BibitemOpen
  \bibfield  {author} {\bibinfo {author} {\bibfnamefont {G.}~\bibnamefont
  {Aupetit-Diallo}}, \bibinfo {author} {\bibfnamefont {G.}~\bibnamefont
  {Pecci}}, \bibinfo {author} {\bibfnamefont {A.~G.}\ \bibnamefont
  {Volosniev}}, \bibinfo {author} {\bibfnamefont {M.}~\bibnamefont {Albert}},
  \bibinfo {author} {\bibfnamefont {A.}~\bibnamefont {Minguzzi}}, \ and\
  \bibinfo {author} {\bibfnamefont {P.}~\bibnamefont {Vignolo}},\ }\href
  {\doibase 10.21468/SciPostPhysCore.8.1.022} {\bibfield  {journal} {\bibinfo
  {journal} {SciPost Physics Core}\ }\textbf {\bibinfo {volume} {8}},\ \bibinfo
  {pages} {022} (\bibinfo {year} {2025})}\BibitemShut {NoStop}%
\bibitem [{\citenamefont {Gaudin}(1967)}]{gaudin1967un}%
  \BibitemOpen
  \bibfield  {author} {\bibinfo {author} {\bibfnamefont {M.}~\bibnamefont
  {Gaudin}},\ }\href {\doibase https://doi.org/10.1016/0375-9601(67)90193-4}
  {\bibfield  {journal} {\bibinfo  {journal} {Physics Letters A}\ }\textbf
  {\bibinfo {volume} {24}},\ \bibinfo {pages} {55} (\bibinfo {year}
  {1967})}\BibitemShut {NoStop}%
\bibitem [{\citenamefont {Yang}(1967)}]{yang1967some}%
  \BibitemOpen
  \bibfield  {author} {\bibinfo {author} {\bibfnamefont {C.~N.}\ \bibnamefont
  {Yang}},\ }\href {\doibase 10.1103/PhysRevLett.19.1312} {\bibfield  {journal}
  {\bibinfo  {journal} {Physical Review Letters}\ }\textbf {\bibinfo {volume}
  {19}},\ \bibinfo {pages} {1312} (\bibinfo {year} {1967})}\BibitemShut
  {NoStop}%
\bibitem [{\citenamefont {Oelkers}\ \emph {et~al.}(2006)\citenamefont
  {Oelkers}, \citenamefont {Batchelor}, \citenamefont {Bortz},\ and\
  \citenamefont {Guan}}]{oelkers2006bethe}%
  \BibitemOpen
  \bibfield  {author} {\bibinfo {author} {\bibfnamefont {N.}~\bibnamefont
  {Oelkers}}, \bibinfo {author} {\bibfnamefont {M.~T.}\ \bibnamefont
  {Batchelor}}, \bibinfo {author} {\bibfnamefont {M.}~\bibnamefont {Bortz}}, \
  and\ \bibinfo {author} {\bibfnamefont {X.-W.}\ \bibnamefont {Guan}},\ }\href
  {\doibase 10.1088/0305-4470/39/5/005} {\bibfield  {journal} {\bibinfo
  {journal} {Journal of Physics A: Mathematical and General}\ }\textbf
  {\bibinfo {volume} {39}},\ \bibinfo {pages} {1073} (\bibinfo {year}
  {2006})}\BibitemShut {NoStop}%
\bibitem [{sup()}]{suppmat}%
  \BibitemOpen
  \href@noop {} {}\bibinfo {note} {{See Supplemental Material which includes
  Refs. FILL for details.}}\BibitemShut {Stop}%
\bibitem [{\citenamefont {Ogata}\ and\ \citenamefont
  {Shiba}(1990)}]{ogata1990bethe}%
  \BibitemOpen
  \bibfield  {author} {\bibinfo {author} {\bibfnamefont {M.}~\bibnamefont
  {Ogata}}\ and\ \bibinfo {author} {\bibfnamefont {H.}~\bibnamefont {Shiba}},\
  }\href {\doibase 10.1103/PhysRevB.41.2326} {\bibfield  {journal} {\bibinfo
  {journal} {Physical Review B}\ }\textbf {\bibinfo {volume} {41}},\ \bibinfo
  {pages} {2326} (\bibinfo {year} {1990})}\BibitemShut {NoStop}%
\bibitem [{\citenamefont {Deuretzbacher}\ \emph {et~al.}(2008)\citenamefont
  {Deuretzbacher}, \citenamefont {Fredenhagen}, \citenamefont {Becker},
  \citenamefont {Bongs}, \citenamefont {Sengstock},\ and\ \citenamefont
  {Pfannkuche}}]{deuretzbacher2008exact}%
  \BibitemOpen
  \bibfield  {author} {\bibinfo {author} {\bibfnamefont {F.}~\bibnamefont
  {Deuretzbacher}}, \bibinfo {author} {\bibfnamefont {K.}~\bibnamefont
  {Fredenhagen}}, \bibinfo {author} {\bibfnamefont {D.}~\bibnamefont {Becker}},
  \bibinfo {author} {\bibfnamefont {K.}~\bibnamefont {Bongs}}, \bibinfo
  {author} {\bibfnamefont {K.}~\bibnamefont {Sengstock}}, \ and\ \bibinfo
  {author} {\bibfnamefont {D.}~\bibnamefont {Pfannkuche}},\ }\href {\doibase
  10.1103/PhysRevLett.100.160405} {\bibfield  {journal} {\bibinfo  {journal}
  {Physical Review Letters}\ }\textbf {\bibinfo {volume} {100}},\ \bibinfo
  {pages} {160405} (\bibinfo {year} {2008})}\BibitemShut {NoStop}%
\bibitem [{\citenamefont {Deuretzbacher}\ \emph {et~al.}(2016)\citenamefont
  {Deuretzbacher}, \citenamefont {Becker},\ and\ \citenamefont
  {Santos}}]{deuretzbacher2016momentum}%
  \BibitemOpen
  \bibfield  {author} {\bibinfo {author} {\bibfnamefont {F.}~\bibnamefont
  {Deuretzbacher}}, \bibinfo {author} {\bibfnamefont {D.}~\bibnamefont
  {Becker}}, \ and\ \bibinfo {author} {\bibfnamefont {L.}~\bibnamefont
  {Santos}},\ }\href {\doibase 10.1103/PhysRevA.94.023606} {\bibfield
  {journal} {\bibinfo  {journal} {Physical Review A}\ }\textbf {\bibinfo
  {volume} {94}},\ \bibinfo {pages} {023606} (\bibinfo {year}
  {2016})}\BibitemShut {NoStop}%
\bibitem [{\citenamefont {Osterloh}\ \emph {et~al.}(2023)\citenamefont
  {Osterloh}, \citenamefont {Polo}, \citenamefont {Chetcuti},\ and\
  \citenamefont {Amico}}]{osterloh2023exact}%
  \BibitemOpen
  \bibfield  {author} {\bibinfo {author} {\bibfnamefont {A.}~\bibnamefont
  {Osterloh}}, \bibinfo {author} {\bibfnamefont {J.}~\bibnamefont {Polo}},
  \bibinfo {author} {\bibfnamefont {W.~J.}\ \bibnamefont {Chetcuti}}, \ and\
  \bibinfo {author} {\bibfnamefont {L.}~\bibnamefont {Amico}},\ }\href
  {\doibase 10.21468/SciPostPhys.15.1.006} {\bibfield  {journal} {\bibinfo
  {journal} {SciPost Physics}\ }\textbf {\bibinfo {volume} {15}},\ \bibinfo
  {pages} {006} (\bibinfo {year} {2023})}\BibitemShut {NoStop}%
\bibitem [{\citenamefont {Girardeau}(1960)}]{girardeau1960relationship}%
  \BibitemOpen
  \bibfield  {author} {\bibinfo {author} {\bibfnamefont {M.}~\bibnamefont
  {Girardeau}},\ }\href {\doibase 10.1063/1.1703687} {\bibfield  {journal}
  {\bibinfo  {journal} {Journal of Mathematical Physics}\ }\textbf {\bibinfo
  {volume} {1}},\ \bibinfo {pages} {516} (\bibinfo {year} {1960})}\BibitemShut
  {NoStop}%
\bibitem [{\citenamefont {Amico}\ \emph {et~al.}(2021)\citenamefont {Amico},
  \citenamefont {Boshier}, \citenamefont {Birkl}, \citenamefont {Minguzzi},
  \citenamefont {Miniatura}, \citenamefont {Kwek}, \citenamefont {Aghamalyan},
  \citenamefont {Ahufinger}, \citenamefont {Anderson}, \citenamefont {Andrei}
  \emph {et~al.}}]{amico2021roadmap}%
  \BibitemOpen
  \bibfield  {author} {\bibinfo {author} {\bibfnamefont {L.}~\bibnamefont
  {Amico}}, \bibinfo {author} {\bibfnamefont {M.}~\bibnamefont {Boshier}},
  \bibinfo {author} {\bibfnamefont {G.}~\bibnamefont {Birkl}}, \bibinfo
  {author} {\bibfnamefont {A.}~\bibnamefont {Minguzzi}}, \bibinfo {author}
  {\bibfnamefont {C.}~\bibnamefont {Miniatura}}, \bibinfo {author}
  {\bibfnamefont {L.-C.}\ \bibnamefont {Kwek}}, \bibinfo {author}
  {\bibfnamefont {D.}~\bibnamefont {Aghamalyan}}, \bibinfo {author}
  {\bibfnamefont {V.}~\bibnamefont {Ahufinger}}, \bibinfo {author}
  {\bibfnamefont {D.}~\bibnamefont {Anderson}}, \bibinfo {author}
  {\bibfnamefont {N.}~\bibnamefont {Andrei}},  \emph {et~al.},\ }\href
  {\doibase 10.1116/5.0026178} {\bibfield  {journal} {\bibinfo  {journal} {AVS
  Quantum Science}\ }\textbf {\bibinfo {volume} {3}},\ \bibinfo {pages}
  {039201} (\bibinfo {year} {2021})}\BibitemShut {NoStop}%
\bibitem [{\citenamefont {Amico}\ \emph {et~al.}(2022)\citenamefont {Amico},
  \citenamefont {Anderson}, \citenamefont {Boshier}, \citenamefont {Brantut},
  \citenamefont {Kwek}, \citenamefont {Minguzzi},\ and\ \citenamefont {von
  Klitzing}}]{amico2022colloquium}%
  \BibitemOpen
  \bibfield  {author} {\bibinfo {author} {\bibfnamefont {L.}~\bibnamefont
  {Amico}}, \bibinfo {author} {\bibfnamefont {D.}~\bibnamefont {Anderson}},
  \bibinfo {author} {\bibfnamefont {M.}~\bibnamefont {Boshier}}, \bibinfo
  {author} {\bibfnamefont {J.-P.}\ \bibnamefont {Brantut}}, \bibinfo {author}
  {\bibfnamefont {L.-C.}\ \bibnamefont {Kwek}}, \bibinfo {author}
  {\bibfnamefont {A.}~\bibnamefont {Minguzzi}}, \ and\ \bibinfo {author}
  {\bibfnamefont {W.}~\bibnamefont {von Klitzing}},\ }\href {\doibase
  10.1103/RevModPhys.94.041001} {\bibfield  {journal} {\bibinfo  {journal}
  {Reviews of Modern Physics}\ }\textbf {\bibinfo {volume} {94}},\ \bibinfo
  {pages} {041001} (\bibinfo {year} {2022})}\BibitemShut {NoStop}%
\bibitem [{\citenamefont {Dalibard}\ \emph {et~al.}(2011)\citenamefont
  {Dalibard}, \citenamefont {Gerbier}, \citenamefont
  {Juzeli\ifmmode~\bar{u}\else \={u}\fi{}nas},\ and\ \citenamefont
  {\"Ohberg}}]{dalibard2011colloquium}%
  \BibitemOpen
  \bibfield  {author} {\bibinfo {author} {\bibfnamefont {J.}~\bibnamefont
  {Dalibard}}, \bibinfo {author} {\bibfnamefont {F.}~\bibnamefont {Gerbier}},
  \bibinfo {author} {\bibfnamefont {G.}~\bibnamefont
  {Juzeli\ifmmode~\bar{u}\else \={u}\fi{}nas}}, \ and\ \bibinfo {author}
  {\bibfnamefont {P.}~\bibnamefont {\"Ohberg}},\ }\href {\doibase
  10.1103/RevModPhys.83.1523} {\bibfield  {journal} {\bibinfo  {journal}
  {Reviews of Modern Physics}\ }\textbf {\bibinfo {volume} {83}},\ \bibinfo
  {pages} {1523} (\bibinfo {year} {2011})}\BibitemShut {NoStop}%
\bibitem [{\citenamefont {Polo}\ \emph {et~al.}(2025)\citenamefont {Polo},
  \citenamefont {Chetcuti}, \citenamefont {Haug}, \citenamefont {Minguzzi},
  \citenamefont {Wright},\ and\ \citenamefont {Amico}}]{polo2025persistent}%
  \BibitemOpen
  \bibfield  {author} {\bibinfo {author} {\bibfnamefont {J.}~\bibnamefont
  {Polo}}, \bibinfo {author} {\bibfnamefont {W.}~\bibnamefont {Chetcuti}},
  \bibinfo {author} {\bibfnamefont {T.}~\bibnamefont {Haug}}, \bibinfo {author}
  {\bibfnamefont {A.}~\bibnamefont {Minguzzi}}, \bibinfo {author}
  {\bibfnamefont {K.}~\bibnamefont {Wright}}, \ and\ \bibinfo {author}
  {\bibfnamefont {L.}~\bibnamefont {Amico}},\ }\href {\doibase
  https://doi.org/10.1016/j.physrep.2025.06.003} {\bibfield  {journal}
  {\bibinfo  {journal} {Physics Reports}\ }\textbf {\bibinfo {volume} {1137}},\
  \bibinfo {pages} {1} (\bibinfo {year} {2025})},\ \bibinfo {note} {persistent
  currents in ultracold gases}\BibitemShut {NoStop}%
\bibitem [{\citenamefont {Leggett}(1991)}]{leggett1991dephasing}%
  \BibitemOpen
  \bibfield  {author} {\bibinfo {author} {\bibfnamefont {A.~J.}\ \bibnamefont
  {Leggett}},\ }\enquote {\bibinfo {title} {{Dephasing and Non-Dephasing
  Collisions in Nanostructures}},}\ in\ \href {\doibase
  10.1007/978-1-4899-3689-9_19} {\emph {\bibinfo {booktitle} {Granular
  Nanoelectronics}}},\ \bibinfo {editor} {edited by\ \bibinfo {editor}
  {\bibfnamefont {D.~K.}\ \bibnamefont {Ferry}}, \bibinfo {editor}
  {\bibfnamefont {J.~R.}\ \bibnamefont {Barker}}, \ and\ \bibinfo {editor}
  {\bibfnamefont {C.}~\bibnamefont {Jacoboni}}}\ (\bibinfo  {publisher}
  {Springer US},\ \bibinfo {address} {Boston, MA},\ \bibinfo {year} {1991})\
  pp.\ \bibinfo {pages} {297--311}\BibitemShut {NoStop}%
\bibitem [{\citenamefont {Yu}\ and\ \citenamefont
  {Fowler}(1992)}]{yu1992persistent}%
  \BibitemOpen
  \bibfield  {author} {\bibinfo {author} {\bibfnamefont {N.}~\bibnamefont
  {Yu}}\ and\ \bibinfo {author} {\bibfnamefont {M.}~\bibnamefont {Fowler}},\
  }\href {\doibase 10.1103/PhysRevB.45.11795} {\bibfield  {journal} {\bibinfo
  {journal} {Physical Review B}\ }\textbf {\bibinfo {volume} {45}},\ \bibinfo
  {pages} {11795} (\bibinfo {year} {1992})}\BibitemShut {NoStop}%
\bibitem [{\citenamefont {Chetcuti}\ \emph {et~al.}(2022)\citenamefont
  {Chetcuti}, \citenamefont {Haug}, \citenamefont {Kwek},\ and\ \citenamefont
  {Amico}}]{chetcuti2022persistent}%
  \BibitemOpen
  \bibfield  {author} {\bibinfo {author} {\bibfnamefont {W.~J.}\ \bibnamefont
  {Chetcuti}}, \bibinfo {author} {\bibfnamefont {T.}~\bibnamefont {Haug}},
  \bibinfo {author} {\bibfnamefont {L.-C.}\ \bibnamefont {Kwek}}, \ and\
  \bibinfo {author} {\bibfnamefont {L.}~\bibnamefont {Amico}},\ }\href
  {\doibase 10.21468/SciPostPhys.12.1.033} {\bibfield  {journal} {\bibinfo
  {journal} {SciPost Physics}\ }\textbf {\bibinfo {volume} {12}},\ \bibinfo
  {pages} {033} (\bibinfo {year} {2022})}\BibitemShut {NoStop}%
\bibitem [{\citenamefont {Pecci}\ \emph {et~al.}(2023)\citenamefont {Pecci},
  \citenamefont {Aupetit-Diallo}, \citenamefont {Albert}, \citenamefont
  {Vignolo},\ and\ \citenamefont {Minguzzi}}]{pecci2023persistent}%
  \BibitemOpen
  \bibfield  {author} {\bibinfo {author} {\bibfnamefont {G.}~\bibnamefont
  {Pecci}}, \bibinfo {author} {\bibfnamefont {G.}~\bibnamefont
  {Aupetit-Diallo}}, \bibinfo {author} {\bibfnamefont {M.}~\bibnamefont
  {Albert}}, \bibinfo {author} {\bibfnamefont {P.}~\bibnamefont {Vignolo}}, \
  and\ \bibinfo {author} {\bibfnamefont {A.}~\bibnamefont {Minguzzi}},\ }\href
  {\doibase 10.5802/crphys.157} {\bibfield  {journal} {\bibinfo  {journal}
  {Comptes Rendus. Physique}\ }\textbf {\bibinfo {volume} {24}},\ \bibinfo
  {pages} {87} (\bibinfo {year} {2023})}\BibitemShut {NoStop}%
\bibitem [{Note1()}]{Note1}%
  \BibitemOpen
  \bibinfo {note} {Equivalently, the algebra may be formulated utilizing
  bosonic anyonic commutation rules, which differ from the fermionic ones by a
  $\pi $ shift of the statistical parameter.}\BibitemShut {Stop}%
\bibitem [{\citenamefont {Chetcuti}\ \emph {et~al.}(2025)\citenamefont
  {Chetcuti}, \citenamefont {Minguzzi}, \citenamefont {Polo},\ and\
  \citenamefont {Amico}}]{chetcuti2025interferometric}%
  \BibitemOpen
  \bibfield  {author} {\bibinfo {author} {\bibfnamefont {W.~J.}\ \bibnamefont
  {Chetcuti}}, \bibinfo {author} {\bibfnamefont {A.}~\bibnamefont {Minguzzi}},
  \bibinfo {author} {\bibfnamefont {J.}~\bibnamefont {Polo}}, \ and\ \bibinfo
  {author} {\bibfnamefont {L.}~\bibnamefont {Amico}},\ }\href
  {https://arxiv.org/abs/2512.13795} {\enquote {\bibinfo {title}
  {Interferometric probe for the zeros of the many-body wavefunction},}\ }
  (\bibinfo {year} {2025}),\ \Eprint {http://arxiv.org/abs/2512.13795}
  {arXiv:2512.13795} \BibitemShut {NoStop}%
\bibitem [{Note2()}]{Note2}%
  \BibitemOpen
  \bibinfo {note} {This loss of statistical identity is confined to the
  impurity, as the momentum distributions of the majority component remain
  distinguishable in bosonic and fermionic mixtures.}\BibitemShut {Stop}%
\bibitem [{\citenamefont {Musolino}\ \emph {et~al.}(2024)\citenamefont
  {Musolino}, \citenamefont {Albert}, \citenamefont {Minguzzi},\ and\
  \citenamefont {Vignolo}}]{musolino2024symmetry}%
  \BibitemOpen
  \bibfield  {author} {\bibinfo {author} {\bibfnamefont {S.}~\bibnamefont
  {Musolino}}, \bibinfo {author} {\bibfnamefont {M.}~\bibnamefont {Albert}},
  \bibinfo {author} {\bibfnamefont {A.}~\bibnamefont {Minguzzi}}, \ and\
  \bibinfo {author} {\bibfnamefont {P.}~\bibnamefont {Vignolo}},\ }\href
  {\doibase 10.1103/PhysRevLett.133.183402} {\bibfield  {journal} {\bibinfo
  {journal} {Physical Review Letters}\ }\textbf {\bibinfo {volume} {133}},\
  \bibinfo {pages} {183402} (\bibinfo {year} {2024})}\BibitemShut {NoStop}%
\bibitem [{\citenamefont {Musolino}\ \emph {et~al.}(2025)\citenamefont
  {Musolino}, \citenamefont {Albert}, \citenamefont {Vignolo},\ and\
  \citenamefont {Minguzzi}}]{musolino2025symmetry}%
  \BibitemOpen
  \bibfield  {author} {\bibinfo {author} {\bibfnamefont {S.}~\bibnamefont
  {Musolino}}, \bibinfo {author} {\bibfnamefont {M.}~\bibnamefont {Albert}},
  \bibinfo {author} {\bibfnamefont {P.}~\bibnamefont {Vignolo}}, \ and\
  \bibinfo {author} {\bibfnamefont {A.}~\bibnamefont {Minguzzi}},\ }\href
  {\doibase 10.1103/bl58-878c} {\bibfield  {journal} {\bibinfo  {journal}
  {Physical Review A}\ }\textbf {\bibinfo {volume} {112}},\ \bibinfo {pages}
  {063308} (\bibinfo {year} {2025})}\BibitemShut {NoStop}%
\bibitem [{\citenamefont {Polo}\ \emph {et~al.}(2026)\citenamefont {Polo},
  \citenamefont {Chetcuti}, \citenamefont {Minguzzi}, \citenamefont
  {Osterloh},\ and\ \citenamefont {Amico}}]{polo2026static}%
  \BibitemOpen
  \bibfield  {author} {\bibinfo {author} {\bibfnamefont {J.}~\bibnamefont
  {Polo}}, \bibinfo {author} {\bibfnamefont {W.~J.}\ \bibnamefont {Chetcuti}},
  \bibinfo {author} {\bibfnamefont {A.}~\bibnamefont {Minguzzi}}, \bibinfo
  {author} {\bibfnamefont {A.}~\bibnamefont {Osterloh}}, \ and\ \bibinfo
  {author} {\bibfnamefont {L.}~\bibnamefont {Amico}},\ }\href {\doibase
  10.1103/tcmp-nf52} {\bibfield  {journal} {\bibinfo  {journal} {Physical
  Review Research}\ }\textbf {\bibinfo {volume} {8}},\ \bibinfo {pages}
  {L022015} (\bibinfo {year} {2026})}\BibitemShut {NoStop}%
\bibitem [{\citenamefont {Lunt}\ \emph
  {et~al.}(2024{\natexlab{a}})\citenamefont {Lunt}, \citenamefont {Hill},
  \citenamefont {Reiter}, \citenamefont {Preiss}, \citenamefont {Ga\l{}ka},\
  and\ \citenamefont {Jochim}}]{lunt2024engineering}%
  \BibitemOpen
  \bibfield  {author} {\bibinfo {author} {\bibfnamefont {P.}~\bibnamefont
  {Lunt}}, \bibinfo {author} {\bibfnamefont {P.}~\bibnamefont {Hill}}, \bibinfo
  {author} {\bibfnamefont {J.}~\bibnamefont {Reiter}}, \bibinfo {author}
  {\bibfnamefont {P.~M.}\ \bibnamefont {Preiss}}, \bibinfo {author}
  {\bibfnamefont {M.}~\bibnamefont {Ga\l{}ka}}, \ and\ \bibinfo {author}
  {\bibfnamefont {S.}~\bibnamefont {Jochim}},\ }\href {\doibase
  10.1103/PhysRevA.110.063315} {\bibfield  {journal} {\bibinfo  {journal}
  {Physical Review A}\ }\textbf {\bibinfo {volume} {110}},\ \bibinfo {pages}
  {063315} (\bibinfo {year} {2024}{\natexlab{a}})}\BibitemShut {NoStop}%
\bibitem [{\citenamefont {Lunt}\ \emph
  {et~al.}(2024{\natexlab{b}})\citenamefont {Lunt}, \citenamefont {Hill},
  \citenamefont {Reiter}, \citenamefont {Preiss}, \citenamefont {Ga\l{}ka},\
  and\ \citenamefont {Jochim}}]{lunt2024realization}%
  \BibitemOpen
  \bibfield  {author} {\bibinfo {author} {\bibfnamefont {P.}~\bibnamefont
  {Lunt}}, \bibinfo {author} {\bibfnamefont {P.}~\bibnamefont {Hill}}, \bibinfo
  {author} {\bibfnamefont {J.}~\bibnamefont {Reiter}}, \bibinfo {author}
  {\bibfnamefont {P.~M.}\ \bibnamefont {Preiss}}, \bibinfo {author}
  {\bibfnamefont {M.}~\bibnamefont {Ga\l{}ka}}, \ and\ \bibinfo {author}
  {\bibfnamefont {S.}~\bibnamefont {Jochim}},\ }\href {\doibase
  10.1103/PhysRevLett.133.253401} {\bibfield  {journal} {\bibinfo  {journal}
  {Physical Review Letters}\ }\textbf {\bibinfo {volume} {133}},\ \bibinfo
  {pages} {253401} (\bibinfo {year} {2024}{\natexlab{b}})}\BibitemShut
  {NoStop}%
\bibitem [{\citenamefont {Deguchi}\ \emph {et~al.}(2000)\citenamefont
  {Deguchi}, \citenamefont {Essler}, \citenamefont {Göhmann}, \citenamefont
  {Klümper}, \citenamefont {Korepin},\ and\ \citenamefont
  {Kusakabe}}]{deguchi2000thermodynamics}%
  \BibitemOpen
  \bibfield  {author} {\bibinfo {author} {\bibfnamefont {T.}~\bibnamefont
  {Deguchi}}, \bibinfo {author} {\bibfnamefont {F.}~\bibnamefont {Essler}},
  \bibinfo {author} {\bibfnamefont {F.}~\bibnamefont {Göhmann}}, \bibinfo
  {author} {\bibfnamefont {A.}~\bibnamefont {Klümper}}, \bibinfo {author}
  {\bibfnamefont {V.}~\bibnamefont {Korepin}}, \ and\ \bibinfo {author}
  {\bibfnamefont {K.}~\bibnamefont {Kusakabe}},\ }\href {\doibase
  https://doi.org/10.1016/S0370-1573(00)00010-7} {\bibfield  {journal}
  {\bibinfo  {journal} {Physics Reports}\ }\textbf {\bibinfo {volume} {331}},\
  \bibinfo {pages} {197} (\bibinfo {year} {2000})}\BibitemShut {NoStop}%
\bibitem [{\citenamefont {Korepin}\ \emph {et~al.}(1993)\citenamefont
  {Korepin}, \citenamefont {Bogoliubov},\ and\ \citenamefont
  {Izergin}}]{korepin1993quantum}%
  \BibitemOpen
  \bibfield  {author} {\bibinfo {author} {\bibfnamefont {V.~E.}\ \bibnamefont
  {Korepin}}, \bibinfo {author} {\bibfnamefont {N.~M.}\ \bibnamefont
  {Bogoliubov}}, \ and\ \bibinfo {author} {\bibfnamefont {A.~G.}\ \bibnamefont
  {Izergin}},\ }\enquote {\bibinfo {title} {{The Quantum Inverse Scattering
  Method}},}\ in\ \href@noop {} {\emph {\bibinfo {booktitle} {Quantum Inverse
  Scattering Method and Correlation Functions}}},\ \bibinfo {series and number}
  {Cambridge Monographs on Mathematical Physics}\ (\bibinfo  {publisher}
  {Cambridge University Press},\ \bibinfo {year} {1993})\ p.\ \bibinfo {pages}
  {115–136}\BibitemShut {NoStop}%
\bibitem [{\citenamefont {Essler}\ \emph {et~al.}(2005)\citenamefont {Essler},
  \citenamefont {Frahm}, \citenamefont {Göhmann}, \citenamefont {Klümper},\
  and\ \citenamefont {Korepin}}]{essler2005one}%
  \BibitemOpen
  \bibfield  {author} {\bibinfo {author} {\bibfnamefont {F.~H.~L.}\
  \bibnamefont {Essler}}, \bibinfo {author} {\bibfnamefont {H.}~\bibnamefont
  {Frahm}}, \bibinfo {author} {\bibfnamefont {F.}~\bibnamefont {Göhmann}},
  \bibinfo {author} {\bibfnamefont {A.}~\bibnamefont {Klümper}}, \ and\
  \bibinfo {author} {\bibfnamefont {V.~E.}\ \bibnamefont {Korepin}},\
  }\href@noop {} {\emph {\bibinfo {title} {{The One-Dimensional Hubbard
  Model}}}}\ (\bibinfo  {publisher} {Cambridge University Press},\ \bibinfo
  {year} {2005})\BibitemShut {NoStop}%
\bibitem [{Note3()}]{Note3}%
  \BibitemOpen
  \bibinfo {note} {This correspondence depends crucially on particle-number
  parity. Here, we restrict the discussion to even $N_{p}$ that will be
  addressed later.}\BibitemShut {Stop}%
\bibitem [{\citenamefont {Forrester}\ \emph {et~al.}(2003)\citenamefont
  {Forrester}, \citenamefont {Frankel}, \citenamefont {Garoni},\ and\
  \citenamefont {Witte}}]{forrester2003finite}%
  \BibitemOpen
  \bibfield  {author} {\bibinfo {author} {\bibfnamefont {P.~J.}\ \bibnamefont
  {Forrester}}, \bibinfo {author} {\bibfnamefont {N.~E.}\ \bibnamefont
  {Frankel}}, \bibinfo {author} {\bibfnamefont {T.~M.}\ \bibnamefont {Garoni}},
  \ and\ \bibinfo {author} {\bibfnamefont {N.~S.}\ \bibnamefont {Witte}},\
  }\href {\doibase 10.1103/PhysRevA.67.043607} {\bibfield  {journal} {\bibinfo
  {journal} {Physical Review A}\ }\textbf {\bibinfo {volume} {67}},\ \bibinfo
  {pages} {043607} (\bibinfo {year} {2003})}\BibitemShut {NoStop}%
\bibitem [{\citenamefont {Shastry}\ and\ \citenamefont
  {Sutherland}(1990)}]{shastry1990twisted}%
  \BibitemOpen
  \bibfield  {author} {\bibinfo {author} {\bibfnamefont {B.~S.}\ \bibnamefont
  {Shastry}}\ and\ \bibinfo {author} {\bibfnamefont {B.}~\bibnamefont
  {Sutherland}},\ }\href {\doibase 10.1103/PhysRevLett.65.243} {\bibfield
  {journal} {\bibinfo  {journal} {Physical Review Letters}\ }\textbf {\bibinfo
  {volume} {65}},\ \bibinfo {pages} {243} (\bibinfo {year} {1990})}\BibitemShut
  {NoStop}%
\bibitem [{\citenamefont {Chetcuti}\ \emph {et~al.}(2023)\citenamefont
  {Chetcuti}, \citenamefont {Osterloh}, \citenamefont {Amico},\ and\
  \citenamefont {Polo}}]{chetcuti2023interference}%
  \BibitemOpen
  \bibfield  {author} {\bibinfo {author} {\bibfnamefont {W.~J.}\ \bibnamefont
  {Chetcuti}}, \bibinfo {author} {\bibfnamefont {A.}~\bibnamefont {Osterloh}},
  \bibinfo {author} {\bibfnamefont {L.}~\bibnamefont {Amico}}, \ and\ \bibinfo
  {author} {\bibfnamefont {J.}~\bibnamefont {Polo}},\ }\href {\doibase
  10.21468/SciPostPhys.15.4.181} {\bibfield  {journal} {\bibinfo  {journal}
  {SciPost Physics}\ }\textbf {\bibinfo {volume} {15}},\ \bibinfo {pages} {181}
  (\bibinfo {year} {2023})}\BibitemShut {NoStop}%
\bibitem [{\citenamefont {Keilmann}\ \emph {et~al.}(2011)\citenamefont
  {Keilmann}, \citenamefont {Lanzmich}, \citenamefont {McCulloch},\ and\
  \citenamefont {Roncaglia}}]{keilmann2011statistically}%
  \BibitemOpen
  \bibfield  {author} {\bibinfo {author} {\bibfnamefont {T.}~\bibnamefont
  {Keilmann}}, \bibinfo {author} {\bibfnamefont {S.}~\bibnamefont {Lanzmich}},
  \bibinfo {author} {\bibfnamefont {I.}~\bibnamefont {McCulloch}}, \ and\
  \bibinfo {author} {\bibfnamefont {M.}~\bibnamefont {Roncaglia}},\ }\href
  {\doibase 10.1038/ncomms1353} {\bibfield  {journal} {\bibinfo  {journal}
  {Nature Communications}\ }\textbf {\bibinfo {volume} {2}} (\bibinfo {year}
  {2011}),\ 10.1038/ncomms1353}\BibitemShut {NoStop}%
\bibitem [{\citenamefont {Peierls}(1933)}]{peierls1933zur}%
  \BibitemOpen
  \bibfield  {author} {\bibinfo {author} {\bibfnamefont {R.}~\bibnamefont
  {Peierls}},\ }\href {\doibase 10.1007/BF01342591} {\bibfield  {journal}
  {\bibinfo  {journal} {Zeitschrift f{\"u}r Physik}\ }\textbf {\bibinfo
  {volume} {80}},\ \bibinfo {pages} {763} (\bibinfo {year} {1933})}\BibitemShut
  {NoStop}%
\end{thebibliography}
\end{document}